\documentclass[12pt]{report}
\usepackage{amssymb}
\usepackage{epsfig}

\newcommand{\half}{{\frac{1}{2}}}

\def\nicefrac#1/#2{\leavevmode\kern.1em
\raise.5ex\hbox{\the\scriptfont0 #1}\kern-.1em
/\kern-.15em\lower.25ex\hbox{\the\scriptfont0 #2}}

\newcommand{\EE}{{\cal E}}
\newcommand{\MM}{{\cal M}}

\newcommand{\OO}{{\cal O}}
\newcommand{\ct}{\eta}

\newcommand{\cd}{\cdot}

\newcommand{\al}{\alpha}
\renewcommand{\b}{\beta}
\newcommand{\de}{\delta}
\newcommand{\De}{\Delta}
\newcommand{\ep}{\epsilon}
\newcommand{\ga}{\gamma}
\newcommand{\Ga}{\Gamma}
\newcommand{\ka}{\kappa}
\newcommand{\io}{\iota}
\newcommand{\La}{\Lambda}
\newcommand{\la}{\lambda}
\newcommand{\Om}{\Omega}
\newcommand{\om}{\omega}
\newcommand{\si}{\sigma}
\newcommand{\Si}{\Sigma}
\renewcommand{\th}{\theta}
\newcommand{\Th}{\Theta}
\newcommand{\vth}{\vartheta}
\newcommand{\vph}{\varphi}
\newcommand{\vep}{\varepsilon}
\newcommand{\ra}{\rightarrow}
\newcommand{\bk}{{\bf k}}
\newcommand{\bx}{{\bf x}}

\newcommand{\bn}{{\bf n}}

\newcommand{\bp}{{\bf p}}
\newcommand{\bv}{{\bf v}}

\newcommand{\pr}{\prime}

\newcommand{\bm}[1]{\mbox{\boldmath $#1$}}

\newcommand{\be}{\begin{equation}}
\newcommand{\ee}{\end{equation}}
\newcommand{\bea}{\begin{eqnarray}}
\newcommand{\eea}{\end{eqnarray}}
\newcommand{\bean}{\begin{eqnarray*}}
\newcommand{\eean}{\end{eqnarray*}}
\newcommand{\dd}{\partial}

\newcommand{\gsim}{\stackrel{>}{\sim}}
\newcommand{\lsim}{\stackrel{<}{\sim}}

\newcommand{\lan}{\langle}
\newcommand{\ran}{\rangle}
\renewcommand{\r}{\right}
\renewcommand{\l}{\left}
\newcommand{\ie}{{\em i.e.}}
\newcommand{\eg}{{\em e.g.}}
\newcommand{\da}{\dot{a}}
\newcommand{\rad}{\mathrm{rad}}
\newcommand{\lo}{{(\mathrm{long})}}
\newcommand{\mr}{\mathrm}
\newcommand{\dec}{\mathrm{dec}}
\newcommand{\etal}{{\it et al.}}

\begin{document}

%\begin{titlepage}
\vspace*{1cm}
\centerline{\bf \huge The theory of CMB Anisotropies}
\vspace{1.5cm}
\begin{center}{\Large\bf  by  \vspace{1cm}\\ Ruth Durrer}\\
Universit\'e de Gen\`eve, D\'epartement de Physique Th\'eorique, \\
Quai E. Ansermet 24, 1211 Gen\`eve 4, Switzerland
\vspace{0.1cm}\\
and
\vspace{0.1cm}\\
School of Natural Sciences, Institute for Advanced Study\\
Einstein Drive, Princeton, NJ 08540, USA
\vspace{1cm}\\
\today
\end{center}
\vspace{5cm}
\begin{figure}[ht]
\begin{center}
\begin{minipage}{.7\linewidth}
\centering
\end{minipage}
\end{center}
\end{figure}
%\end{titlepage}
\newpage
\tableofcontents

\setcounter{page}{1}
%\pagenumbering{arabic}
\chapter{Introduction}
\thispagestyle{empty}

In this review I would like to show  the importance and the power
of measurements of anisotropies in the cosmic microwave background
(CMB).

CMB anisotropies  are so useful mainly because they are small: For a
given model, they can be calculated within linear
perturbation theory, to very good approximation. They are
influenced only little by the non-linear processes of galaxy
formation. This allows us to compute them very precisely (to about
1\%, which is high precision for present cosmological standards). For
given initial fluctuations, the result depends only on the
cosmological parameters. If we can measure CMB anisotropies to a
precision of, say 1\%, this allows us therefore to determine
cosmological parameters to about 1\%. An unprecedented possibility!
Consider that at present, after the work of two generations, {\em e.g.}
the Hubble parameter is known only to about 25\%, the baryon density
is known to about 10\% and the uncertainties in the dark
matter density, the cosmological constant and the space curvature are
even larger.

This somewhat too optimistic conclusion has however three caveats
which we want to mention before entering the subject of this review.
\begin{enumerate}
\item {\bf Initial conditions:}  The result depends on the model for 
the initial fluctuations. The simplest inflationary scenarios which lead 
to adiabatic perturbations, contain in general three to four free
parameters, like the ratio of  tensor to scalar perturbations
($r$) and the spectral index of the scalar and tensor perturbations
($n_S$ and $n_T$), so a few more parameters need to be
fitted additionally to the data.

More generic initial conditions allow for at least four additional 
isocurvature modes with arbitrary (anti-)correlations. The initial 
conditions are then given by a $5\times 5$ positive semi-definite matrix, 
and, in principle, several spectral indices~\cite{bucher,trotta}.
In most of this review we shall ignore this possibility and assume that 
initial perturbations are purely adiabatic. Even if isocurvature constributions
cannot be excluded, this most simple model is in good agreement with the 
present data.

If the perturbations are generated by active sources like, \eg,
topological defects, then the modeling is far more complicated,
and the analysis is too different to be included in this review.

\item {\bf Degeneracy:} Even though we can measure over $1000$ 
  independent modes
($C_\ell$'s) of the CMB anisotropy spectrum, there are certain
combinations of the cosmological parameters that lead to degeneracies
in the CMB spectrum. The result is, \eg, very sensitive to
the sum $\Om_\mr{matter}+\Om_\La$, but not to the difference
(``cosmic confusion'').
%{\bf maybe mention that SNIa can break this degeneracy}

\item {\bf Cosmic variance:} Since the fluctuations are created by 
 random processes,
we can only calculate expectation values. Yet we have only
one universe to take measurements  (``cosmic variance'').
For small--scale fluctuations we can in general assume
that the expectation value over ensembles of universes
is the same as a spatial average (a kind of ergodic hypothesis),
but for large scales we can't escape large statistical errors.
\end{enumerate}

\section{Friedmann-Lema\^\i tre universes}

Friedmann-Lema\^\i tre universes are homogeneous and
isotropic solutions of Einstein's equations. The hyper-surfaces
of constant time are homogeneous and isotropic, \ie, spaces
of constant curvature with metric 
$a^2(\eta) \gamma_{ij} dx^i dx^j$, where $\ga_{ij}$ is the
metric of a space with constant curvature $\ka$. This metric can be 
expressed in the form
\bea
\ga_{ij} dx^i dx^j &=& dr^2+\chi^2(r)\l(d\vth^2+sin^2\vth d\vph^2\r)\\
\chi^2(r) &=& \l\{ \begin{array}{lcl}
	r^2      & , & \ka=0 \\
	\sin^2r  & , & \ka=1 \\
	\sinh^2r & , & \ka=-1 , \end{array} \r.
\eea
where we have rescaled $a(\eta)$ such that $\ka=\pm1$ or $0$.
(With this normalization the scale factor  $a$ has the dimension of a 
length and $\eta$ and $r$ are dimensionless for $\ka \neq 0$.) The 
four-dimensional metric is then of the form
\be
g_{\mu\nu} dx^\mu dx^\nu = -a^2(\eta) d\eta^2
	+a^2(\eta) \ga_{ij} dx^i dx^j.
\ee
Here $\eta$ is called the {\em conformal time}.

 Einstein's equations reduce to ordinary
differential equations for the function $a(\eta)$
(with $\dot{}\equiv d/d\eta$):
\bea
\l(\frac{\da}{a}\r)^2+\ka &=& \frac{8\pi G}{3} a^2 \rho
	+ \frac{1}{3} \La a^2 \label{fl1} \\
\l(\frac{\da}{a}\r)^{\textstyle \cdot} &=& -{4\pi G\over 3} a^2 \l(\rho+3 p\r)
	+ \frac{1}{3} \La a^2 = \l(\frac{\ddot{a}}{a}\r)
	- \l(\frac{\da}{a}\r)^2 , \label{fl2}
\eea
where $\rho=-T^0_0$, $p=T^i_i$ (no sum!) and all other components of
the energy momentum tensor have to vanish by the requirement of
isotropy and homogeneity. $\La$ is the cosmological constant.

Energy momentum ``conservation'' (which is also a consequence of 
(\ref{fl1})
and (\ref{fl2}) due to the contracted Bianchi identity) reads
\be
\dot{\rho}=-3\l(\frac{\da}{a}\r) (\rho+p) . \label{emcons}
\ee

After these preliminaries (which we suppose to be known to the
audience) let us answer the following question: Given an object
with comoving diameter $\la$\footnote{or physical
size $a(\eta)\la=d$} at a redshift $z(\eta)=(a_0/a)-1$. 
Under which angle $\vth(\la,z)$ do we see
this object today and how does this angle depend on $\Om_\La$ and
$\Om_\ka$?

We define
\bea
\Om_m &=& \l(\frac{8\pi G\rho a^2}{3
\l(\frac{\da}{a}\r)^2}\r)_{\eta=\eta_0} \nonumber \\
\Om_\La &=& \left.\frac{\La a^2}
	{3 \l(\frac{\da}{a}\r)^2}\right|_{\eta=\eta_0} \\
\Om_\ka &=& \left.\frac{-\ka}
	{\l(\frac{\da}{a}\r)^2}\right|_{\eta=\eta_0} , \nonumber
\eea
where the index $_0$ indicates the value of a given variable today.
Friedmann's equation (\ref{fl1}) then requires
\be
1 = \Om_m + \Om_\La + \Om_\ka .
\ee

Back to our problem:
\begin{figure}[ht]
\begin{center}
\begin{minipage}{.7\linewidth}
\centering
\epsfig{figure=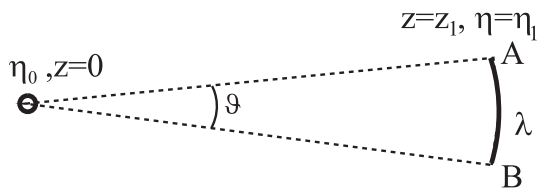,width=7cm}
\end{minipage}
\caption{{\small The two ends of the object emit a flash simultaneously
from $A$ and $B$ at $z_1$ which reaches us today.}}
\end{center}
\end{figure}
Without loss of generality we set $r=0$ at our position and thus
$r=r_1=\eta_0-\eta_1$ at the position of the flashes, $A$ and $B$ at
redshift $z_1$. If $\la$
denotes the comoving arc length between $A$ and $B$ we have
$\la=\chi(r_1)\vth=\chi(\eta_0-\eta_1)\vth$, \ie
\be
\vth = \frac{\la}{\chi(\eta_0-\eta_1)} .
\ee
It remains to calculate $(\eta_0-\eta_1)(z_1)$.

Note that in the case $\ka=0$ we can still normalize the scale factor
$a$ as we want, and it is  convenient to choose $a_0=1$, so
that comoving scales today become physical scales. However, for
$\ka\neq 0$, we have already normalized $a$ such that $\ka=\pm 1$
and $\chi=\sin r$ or $\sinh r$. We have in principle no normalization
constant left.

From the Friedmann equation we have
\be
\da^2 = \frac{8\pi G}{3} a^4 \rho + \frac{1}{3} \La a^4 -\ka a^2.
\ee
We assume that $\rho$ is a combination of ``dust'' (cold, non--relativistic
matter) with $p_d=0$ and radiation with $p_\rad=\nicefrac1/3 \rho_\rad$.

From (\ref{emcons}) we find that $\rho_\rad \propto a^{-4}$ and
$\rho_d \propto a^{-3}$. Therefore, with $H_0=\l(\frac{\da}{a^2}\r) (\eta_0)$,
we define
\bea
\frac{8\pi G}{3} a^4 \rho &=& H_0^2 \l(a_0^4\Om_\rad + \Om_d aa_0^3\r)\\
\frac{1}{3}\La a^4 &=& H_0^2 \Om_\La a^4 \\
-\ka a^2 &=& H_0^2 \Om_\ka a^2 a_0^2~.
\eea
The Friedmann equation then implies
\be
\frac{da}{d\eta} = H_0a_0^2\l(\Om_\rad + {a\over a_0}\Om_d+{a^4\over
a_0^4}\Om_\La+{a^2\over a_0^2}\Om_\ka\r)^\frac{1}{2} \ee
so that
\be
\eta_0-\eta_1 = \frac{1}{H_0 a_0}\int_0^{z_1}
\frac{dz}{\l[\Om_\rad (z+1)^4 +\Om_d (z+1)^3 +
\Om_\La+\Om_\ka (z+1)^2 \r]^\frac{1}{2}} \label{int}.
\ee
Here we have introduced the cosmological redshift $z+1= a_0/a$. 
(In principle we could of course also add other matter components
like, {\em e.g.} ``quintessence''~\cite{Steinhard}, which would lead
to a somewhat different form of the integral~(\ref{int}), but for
definiteness, we remain with dust, radiation and a cosmological
constant.)
 
In general, this integral has to be solved numerically. It determines
the angle $\vth(\la,z_1)$ under which an object with comoving size $\la$
at $z_1$ is seen.

On the other hand, the angular diameter distance to an object
of physical size $d$ seen under angle $\vth$ is given by
$\eta_0-\eta_1 = r_1 = \chi^{-1}\l({d\over a_1\vth}\r)$. If we are able to  
measure the redshift and the comoving angular diameter distance of a 
certain class of objects comparing with Eq.~(\ref{int}) allows in principle to
determine the parameters $\Om_m$, $\Om_\La$, $\Om_\ka$ and $H_0$.  
\begin{figure}[ht]
\begin{center}
\begin{minipage}{1.\linewidth}
\centering
\epsfig{figure=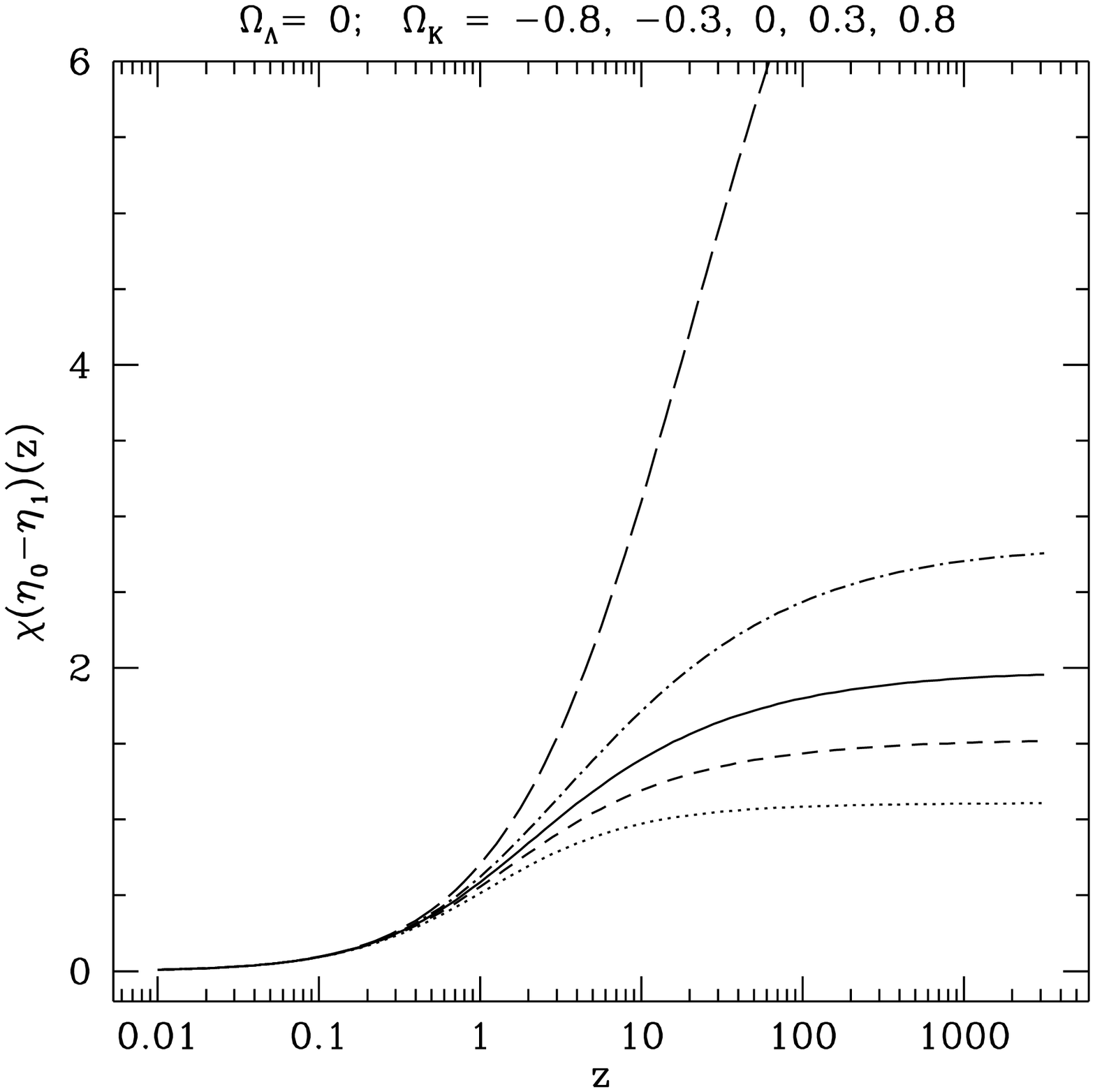,width=6.8cm} \quad
\epsfig{figure=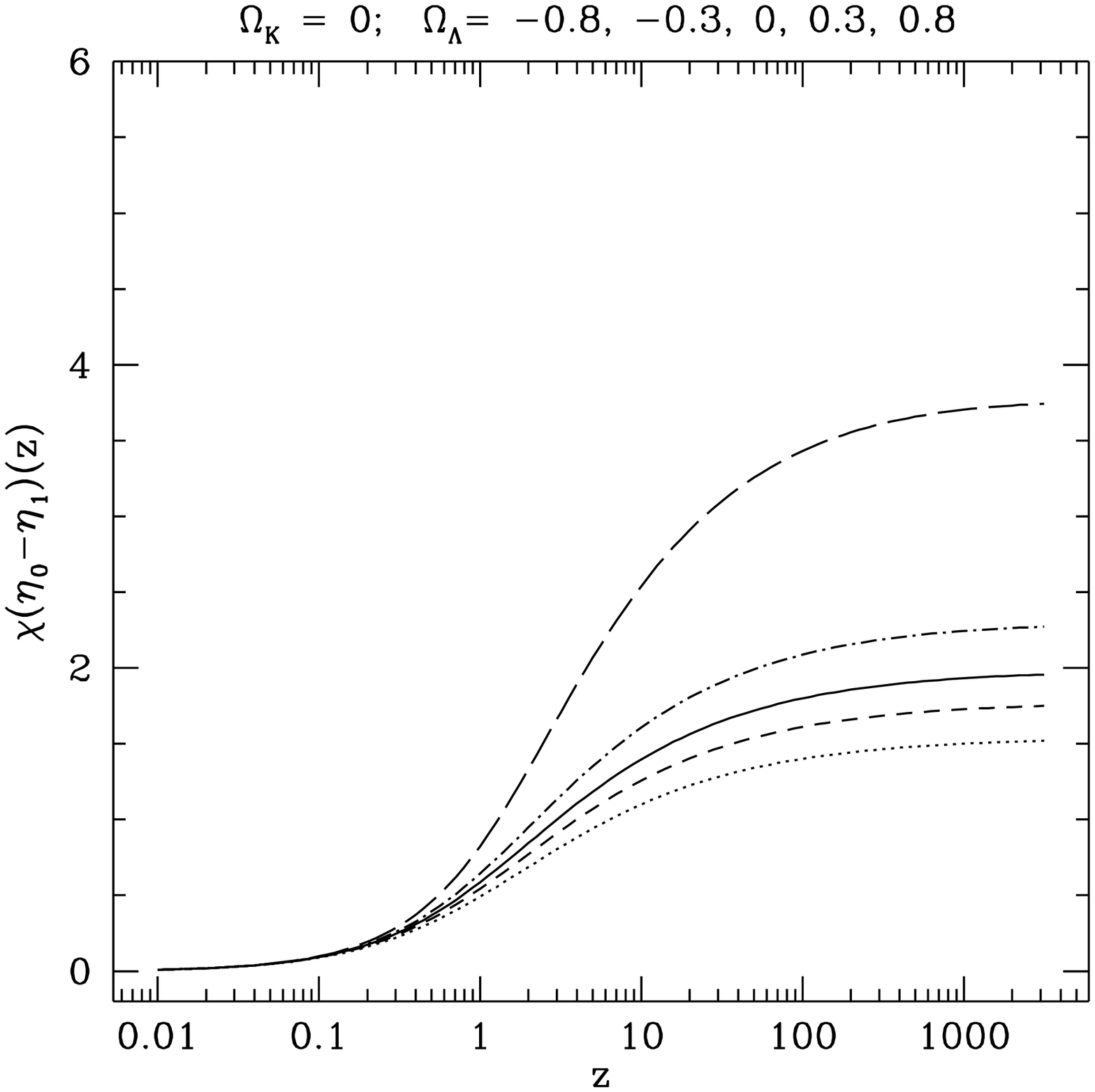,width=6.8cm}
\end{minipage}
\caption{{\small The function $\chi(\eta_0-\eta_1)$ as a function 
of the redshift $z$ for different values of the
cosmological parameters $\Om_\ka$ (left, with $\Om_\La$=0)
and $\Om_\La$ (right, with $\Om_\ka$=0), namely $-0.8$ [dotted], $-0.3$
[short--dashed], $0$ [solid], $0.3$ [dot--dashed], $0.8$ [long--dashed].}\label{chi}}
\end{center}
\end{figure}

% For $z_1\ll 1$, $(\eta_0-\eta_1)H_0 a_0$ is relatively small.
We have
$\frac{-\ka}{H_0^2 a_0^2} = \Om_\ka \Rightarrow 
H_0 a_0 = \frac{1}{\sqrt{|\Om_\ka|}}$ for $\Om_\ka \neq 0$.

Observationally we know $10^{-5}<\Om_\rad\leq 10^{-4}$ as well as
$0.1\le\Om_d\lsim 1 $, $|\Om_\La|\lsim 1$ and $|\Om_\ka|\lsim 1$.

If we are interested in small redshifts, $z_1\lsim 10$, we may safely
neglect $\Om_\rad$. In this region, Eq.~(\ref{int}) is very sensitive
to $\Om_\La$ and provides an excellent mean to constrain the
cosmological constant.

At high redshift, $z_1\gsim 1000$, neglecting radiation  is no longer a
 good approximation.

%{\bf stuff missing}...

We shall later need the opening angle of the {\em horizon} distance,
\bea
\vth_H(z_1)&=&\frac{\eta_1}{\chi(\eta_0-\eta_1)} ,\\
\eta_1&=&\frac{1}{H_0 a_0} \int_{z_1}^\infty 
\frac{dz}{\l[\Om_\rad (z+1)^4 +\Om_d (z+1)^3+\Om_\La+\Om_\ka (z+1)^2 \r]^\frac{1}{2}}.
\eea
(Clearly this integral diverges if $\Om_\rad=\Om_d=0$. This is exactly
what  happens during an inflationary period and leads there to the
solution of the horizon problem.)

\begin{figure}[ht]
\begin{center}
\begin{minipage}{1.\linewidth}
\centering
\epsfig{figure=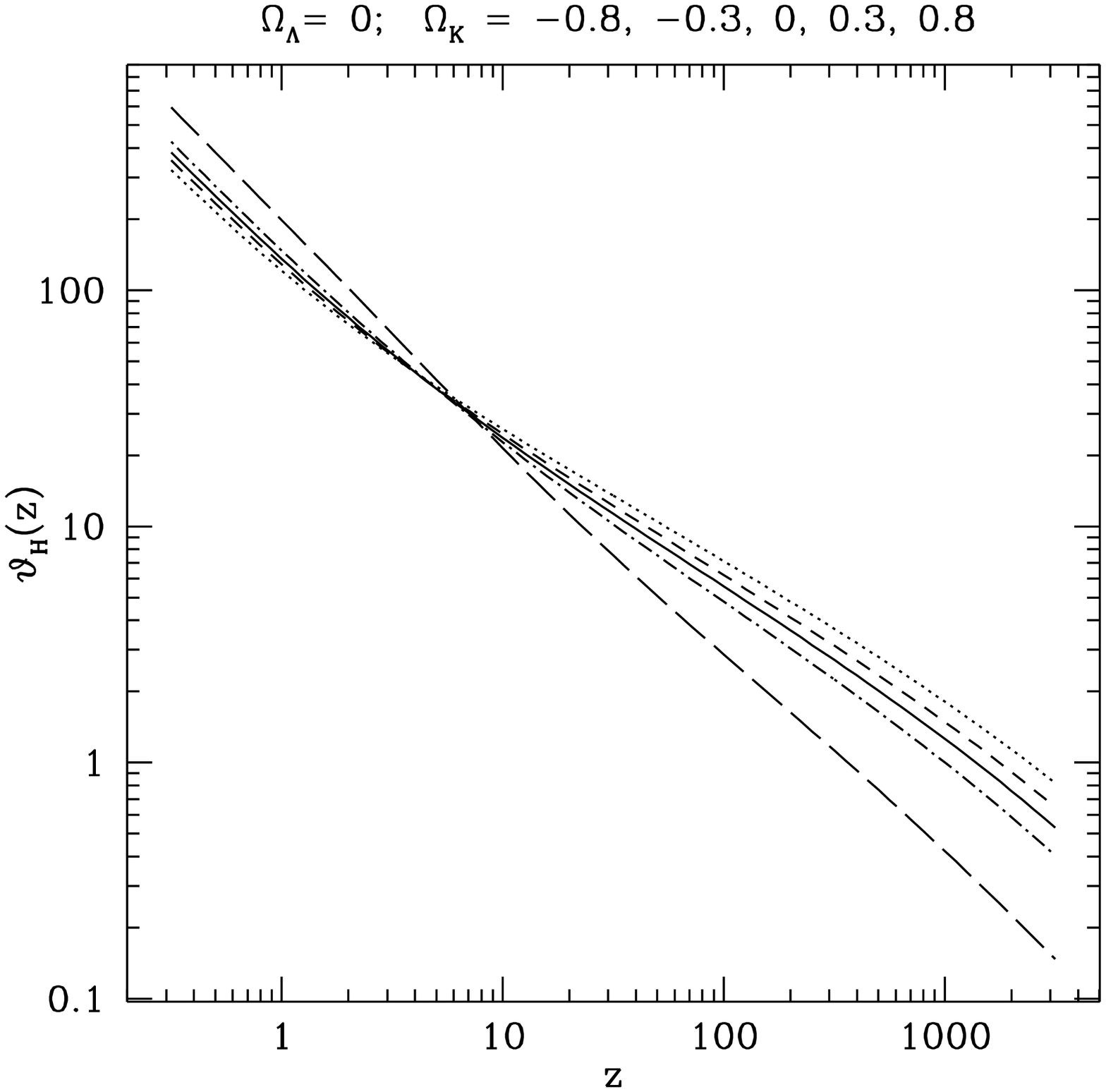,width=6.8cm} \quad
\epsfig{figure=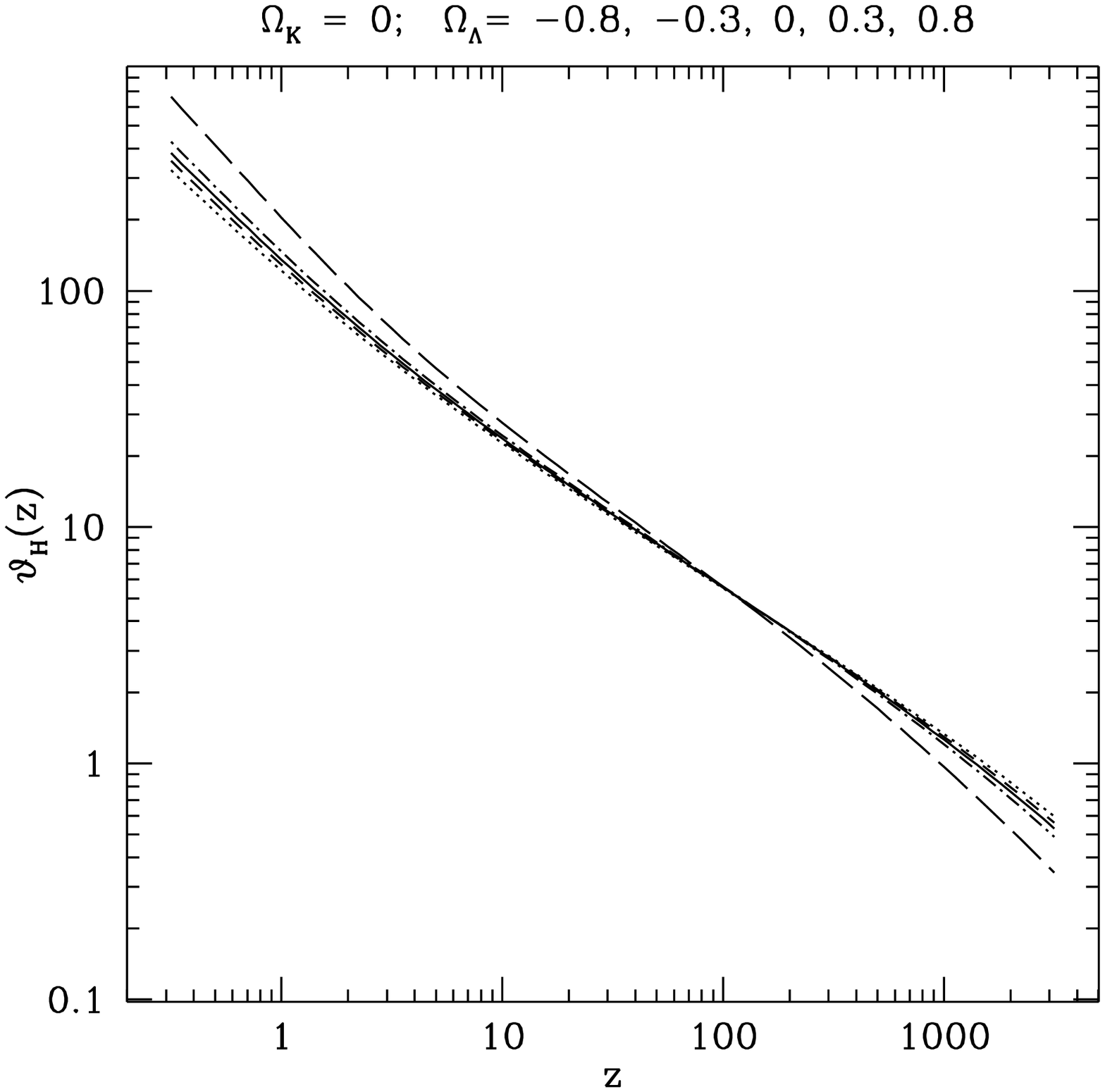,width=6.8cm}
\end{minipage}
\caption{{\small $\vth_H(z_1)$ (in degrees) for different values of the
cosmological parameters $\Om_\ka$ and $\Om_\La$ the line styles are as in 
Fig.~\ref{chi}.\label{theta}}}
\end{center}
\end{figure}

The value of the radiation density is well known. For photons plus three sorts of massless neutrinos we have
\[ \rho_\mr{rad}=7.94\times
	10^{-34}(T_0/2.737\mr{K})^4\mr{g}/\mr{cm}^3~. \]
This gives
\bea
\Om_\rad h^2 = 4.2\cdot10^{-5} (T_0/2.737\mr{K})^4~, \label{Omrad}\\
 H_0 = 100 h \frac{\mathrm{km}}{\mathrm{s Mpc}}~.
\eea

 Neglecting $\Om_\rad$, for
$\Om_\La=0$ and small curvature, $0< |\Om_\kappa|\ll \Om_d$ at high
enough redshift, $z_1\ge 10$, one has  $\eta_0-\eta_1 \simeq 
2 \sqrt{|\Om_\kappa|/ \Om_d}= 2/(H_0a_0 \sqrt{\Om_d}) $. This yields 
$\vth(\la,z_1) \simeq \sqrt{\Om_d}H_0a_0\la/2 = 
\half\sqrt{\Om_d}H_0\la_{phys}/(z_1+1)$,
where $\la_{phys}=a_1\la$ is the physical scale corresponding to comoving 
size $\la$.

\section{Recombination and the cosmic microwave background (CMB)}

During its expansion, the universe cools adiabatically. At early
times, it is dominated by a thermal radiation background with
$\rho=C/a^4=g_\mathrm{eff} a_\mr{SB} T^4$,\footnote{We will use
units with $\hbar=c=k_\mr{B}=1$ throughout this report. The
Stefan--Boltzmann constant is then given by
 $a_\mr{SB}=\pi^2 k_\mr{B}^4/(60 \hbar^3 c^2)=\pi^2/60$.}
and we find  that $T\propto a^{-1}$. Here $g_{eff}= n_b +7/8n_F$ is
the effective number of degrees of freedom, bosons counting as $1$ and
fermions counting as $7/8$ (see {\it e.g.}~\cite{Paddy}). At temperatures below
$0.5$MeV only neutrinos and photons are still relativistic leading to the
density parameter given in Eq.~(\ref{Omrad}).
( Neutrinos have a somewhat lower temperature than photons, $T_\nu
=(4/11)^{1/3}T$, since they have already dropped out of thermal
equilibrium at $T\simeq 1$MeV, before $e^{\pm}$ annihilation which therefore 
reheats the photons but not the neutrinos, see 
{\em e.g.}~\cite{Paddy,peebles}.)
 
The photons obey a Planck distribution,
\be
f(\om)=\frac{1}{e^{\om/T}-1}.
\ee

At a temperature of about $T\sim4000\mr{K}\sim0.4\mr{eV}$, the number
density of photons with energies above the hydrogen ionization
energy drops below the baryon density of the universe, and the
protons begin to (re-)combine to neutral hydrogen. (Helium has
already recombined earlier.) Photons and baryons are tightly
coupled before (re-)combination by non--relativistic Thomson
scattering of electrons. During recombination the free electron
density drops sharply and the mean free path of the photon
grows larger than the Hubble scale. At the temperature 
$T_\mr{dec}\sim3000\mr{K}$ (corresponding to the redshift
$z_\mr{dec}\simeq1100$ and the physical time 
$t_\mr{dec}=a_0 \eta_\mr{dec}\simeq10^5 \mr{years}$)
photons become {\em free} and the universe becomes
transparent.

After recombination, the photon distribution evolves according to
Liouville's equation (geodesic spray):
\be
p^0 \dd_\eta f - \Ga^i_{\mu\nu} p^\mu p^\nu \frac{\dd f}{\dd p^i}
\equiv L_{X_g}f = 0, \label{geosp}
\ee
where $i=1,2,3$. Since the photons are massless,
$|{\bf p}|^2=\sum_{i=1}^3p_ip^i= \om^2~~
(\om=ap^0)$. Isotropy of the distribution implies that $f$ depends on
$p^i$ only via $|{\bf p}|=\om$, and so 
\be
\frac{\dd f}{\dd p^i} = \frac{\dd \om}{\dd p^i} \frac{\dd f}{\dd \om}
=\frac{p^i}{\om} \frac{\dd f}{\dd \om}.
\ee
In a Friedmann universe (also if $\ka\neq0$!) we find for
$p^\mu p_\mu=-\om^2+\bp^2=0$ [exercise!]
\be
\Ga^i_{\mu\nu} p^\mu p^\nu p_i = -\om^3 \l(\frac{\da}{a^2}\r).
\ee

Inserting this result into (\ref{geosp}) leads to
\be
\dd_\eta f+\om \l(\frac{\da}{a}\r) \frac{\dd f}{\dd \om} =0,
\ee
which is satisfied by an arbitrary function $f=f(\om a)$. Hence the
distribution of free--streaming photons changes just
by redshifting the momenta. Therefore, setting  $T\propto a^{-1}$ even after
recombination, the blackbody shape of
the photon distribution remains unchanged.

Note however that after recombination the photons are no
longer in thermal equilibrium and the $T$ in the Planck
distribution is not a temperature in the thermo-dynamical sense
but merely a parameter in the photon distribution function.

\begin{figure}[ht]
\begin{center}
\begin{minipage}{1.\linewidth}
\centering
\epsfig{figure=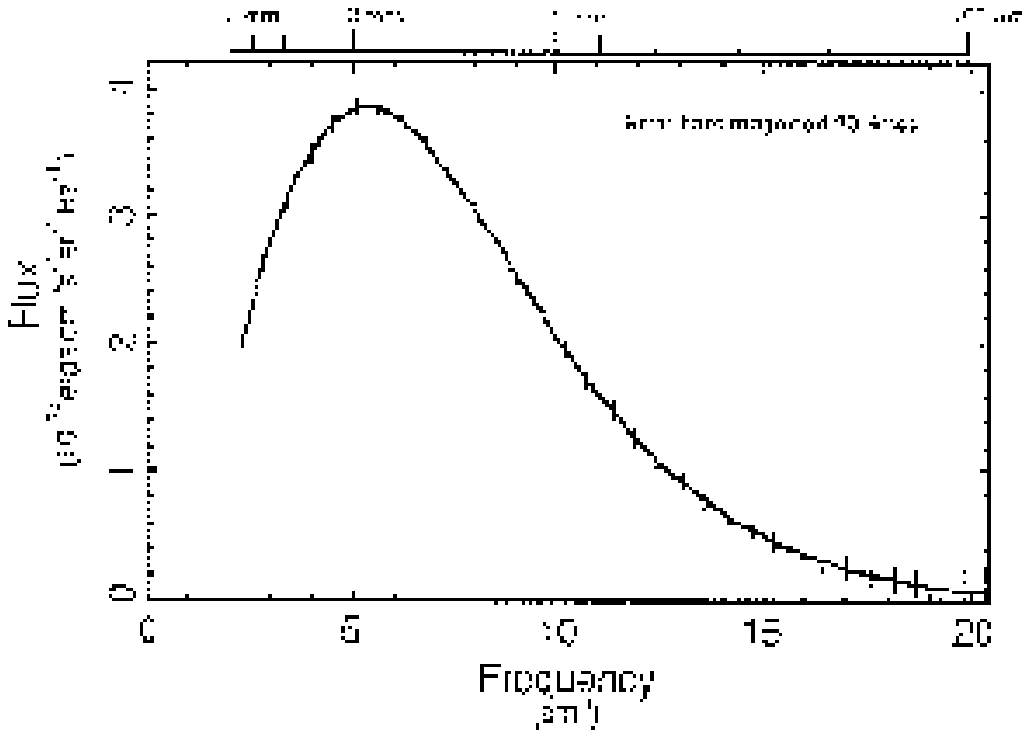,width=7cm} \quad
\epsfig{figure=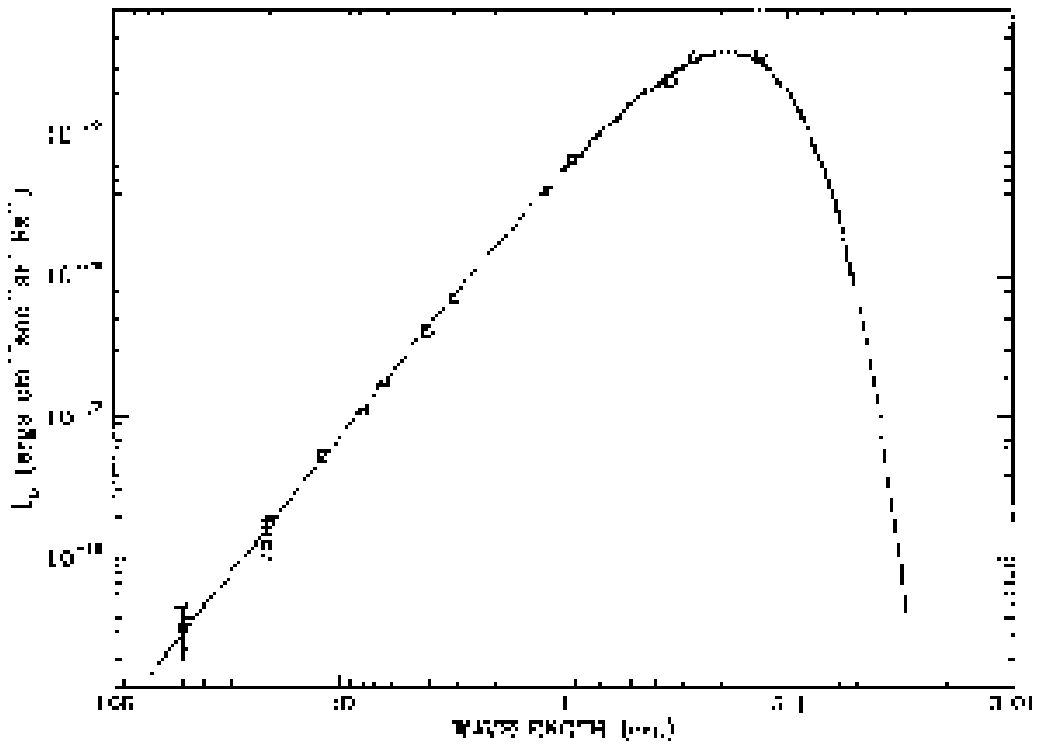,width=7cm}
\end{minipage}
\caption{{\small Spectrum of the cosmic background
radiation. The graph on the left shows the measurements
of the FIRAS experiment on COBE (the vertical bars), overlaid
by a blackbody spectrum at a temperature of 2.73 K. The error bars are
20 times magnified! The
image on the right shows a larger number of measurements.
The FIRAS data is represented by the fat line around the
peak of the spectrum (from Peebles~\cite{peebles}).\label{cobe}}}
\end{center}
\end{figure}

The blackbody spectrum of these cosmic photons which are called the 
``cosmic microwave background'' (CMB) is extremely well verified 
observationally (see Fig.~\ref{cobe}).
The limits on deviations are often parameterized in terms of
three parameters: The chemical potential $\mu$, the Compton
$y$ parameter (which quantifies a well defined change in the
spectrum arising from interactions with a non--relativistic
electron gas at a different temperature, see {\it
e.g.}~\cite{peebles}) and $Y_\mr{ff}$
(describing a contamination by free-free emission).

The present limits on these parameters are (at 95\% CL, \cite{PDG})
\be
|\mu| < 9\cdot 10^{-5}, \quad
|y| < 1.2 \cdot 10^{-5}, \quad
|Y_\mr{ff}| < 1.9 \cdot 10^{-5}.
\ee

The CMB Photons have not only a very thermal spectrum, but they
are also distributed very isotropically, apart from a dipole
which is (most probably) simply due to our motion relative to
the surface of last scattering:

An observer moving with velocity $\bv$ relative to a source
emitting a photon with proper momentum $\bp=-\om\bn$ sees
this photon redshifted with frequency
\be
\om'=\ga\om\l(1-\bn\bv\r) ,
\ee
where $\gamma={1\over \sqrt{1-v^2}}$ is the relativistic $\ga$-factor.
For an isotropic emission of photons coming from all directions $\bn$ this 
leads to a dipole anisotropy in first order in $\bv$. This
dipole anisotropy, which is of the order of 
\[  \l({\De T\over T}\r)_{\mr{dipole}}\simeq 10^{-3}  \]
 has already been discovered in the 70ties~\cite{Conklin,Henry}.
Interpreting it as due to our motion with respect to the last
scattering surface  implies a velocity for the solar-system
bary-center of $v=371\pm0.5\,\mr{km/s}$
at 68\% CL (\cite{PDG}).

The COBE\footnote{Cosmic Background Explorer, NASA satellite
launched 1990.} DMR experiment (Differential Microwave Radiometer)
has found fluctuations of
\be
\sqrt{\l\lan\l(\frac{\De T}{T}\r)^2\r\ran} \sim 10^{-5}
\ee
on all angular scales $\th \geq 7^\circ$~\cite{DMR}.
On smaller angular scales many experiments have found fluctuations
(we shall describe the experimental results in more detail later), but all 
of them are $\lsim 10^{-4}$.

As we shall see later, the CMB fluctuations
on large scales provide a measure for the deviation of the
geometry from the Friedmann-Lema\^\i tre one. The geometry
perturbations are thus small and we may calculate their effects
by {\em linear perturbation theory}. On smaller scales, $\De T/T$ reflects
the fluctuations in the energy density in the baryon/radiation
plasma prior to recombination. Their amplitude is just about right
to allow the  formation of the presently observed non--linear structures (like
galaxies, clusters, etc.) out of small initial fluctuations by
gravitational instability.

These findings strongly support the hypothesis which we assume here, 
namely that the large scale structure
(\ie~galaxy distribution) observed in the universe formed by gravitational
instability from relatively small ($\sim 10^{-4} - 10^{-5}$) initial
fluctuations. As we shall see, such initial fluctuations leave
an interesting ``fingerprint'' on the cosmic microwave background.

\chapter{Perturbation Theory}
The tool for the analysis of CMB anisotropies is cosmological
perturbation theory. We spend therefore some time on this
subject, especially on the fundamental level.

Once all the variables are defined, we will be rather brief in
 the derivation of the basic perturbation equations.
First of all, because these derivations are in general not very illuminating
and secondly because nowadays all of you can obtain them very
easily by setting
\be
g_{\mu\nu} = \bar{g}_{\mu\nu}+\vep a^2 h_{\mu\nu}
\ee
($\bar{g}_{\mu\nu}$ being the unperturbed Friedmann metric) and asking
Mathematic or Maple to calculate the Einstein Tensor using the
condition $\vep^2=0$. We conventionally set (absorbing the ``smallness''
parameter $\vep$ into $h_{\mu\nu}$)
\be \begin{array}{llll}
g_{\mu\nu} = \bar{g}_{\mu\nu} + a^2 h_{\mu\nu}, &
	\quad \bar{g}_{00}=-a^2, &
	\quad \bar{g}_{ij}=a^2 \ga_{ij}&
	\quad |h_{\mu\nu}| \ll 1 \\[.2cm]
T^\mu_\nu = \overline{T}^\mu_\nu + \th^\mu_\nu,&
	\quad \overline{T}^0_0 = -\bar{\rho},&
	\quad \overline{T}^i_j = \bar{p} \de^i_j&
	\quad |\th^\mu_\nu|/\bar{\rho} \ll 1 .
\end{array} \ee

\section{Gauge transformation, gauge invariance}

The first fundamental problem we want to discuss is the problem of
'choice of gauge' in cosmological perturbation theory:

For linear perturbation theory to apply, the spacetime manifold
$\MM$ with metric $g$ and the energy momentum tensor $T$ of the real,
observable universe must be in some sense close to a Friedmann universe,
\ie,
the manifold $\cal M$ with a Robertson--Walker metric $\bar{g}$ and a
homogeneous and isotropic energy momentum tensor $\overline{T}$. It
is an interesting, non--trivial unsolved problem how to construct
$\bar{g}$ and $\overline{T}$ from the physical fields $g$ and $T$ in
practice.  There are two main
difficulties: Spatial averaging procedures depend on the choice of a
hyper--surface of constant time
 and do not commute with derivatives, so that averaged
 fields $\bar{g}$ and $\overline{T}$
 will in general not satisfy Einstein's equations. Secondly, averaging
is in practice impossible over super--horizon scales.

Even though we cannot give a constructive prescription,
we now assume that there exists an averaging procedure which
 leads to a Friedmann
universe with spatially averaged tensor fields $\overline{Q}$, such that
the deviations
$(T_{\mu\nu}-\overline{T}_{\mu\nu})/\max_{\{\al\b\}}\{|\overline{T}_{\al\b}|\}$
and
$(g_{\mu\nu}-\overline{g}_{\mu\nu})/\max_{\{\al\b\}}\{\overline{g}_{\al\b}\}$
  are small, and $\bar{g}$ and $\overline{T}$ satisfy Friedmann's equations.
Let us call such an averaging procedure 'admissible'.
There may be many different admissible averaging procedures (e.g. over a
different hyper--surface) leading to slightly different Friedmann backgrounds.
%Let us denote another admissible average by $(\bar{g}_2,\overline{T}_2)$.
%The averaging procedures are isomorphic via an isomorphism
% $\phi$ on $\cal M$ which is close to unity:
%\bean \bar{g}_2 &=& \phi_*\bar{g}     \\
% \overline{T}_2 &=& \phi_*\overline{T} ~, \eean
%where $\phi_*(Q)$ denotes the push-forward of the tensor field $Q$ 
%under $\phi$.
But since $|g-\bar{g}|$ is small of order $\ep$, the difference of the
two Friedmann backgrounds must also be small of order  $\ep$ and we
can regard it as part of the perturbation. 

We consider now a fixed
admissible Friedmann background $(\bar{g}, \bar{T})$ as chosen. Since
the theory is invariant under diffeomorphisms (coordinate
transformations),  the perturbations are not unique. For an arbitrary
diffeomorphism $\phi$ and its pullback $\phi^*$, the two metrics
$g$ and $\phi^*(g)$ describe the same geometry. Since we have chosen
the background metric $\bar{g}$ we only allow diffeomorphisms which
leave  $\bar{g}$ invariant {\em i.e.} which deviate only in first
order form the identity. Such an 'infinitesimal' isomorphism 
can be represented as the infinitesimal flow of a vector field $X$,
$\phi = \phi_\ep^X$. Remember the definition of the flow: For the integral
curve $\ga_x(s)$ of $X$ with starting point $x$, i.e., $\ga_x(s=0)=x$
we have $\phi_s^X(x) = \ga_x(s)$. In terms of the vector field $X$, to
 first order in $\ep$, its pullback is then
of the form
\[ \phi^* = id +\ep L_X \]
($L_X$ denotes the Lie derivative in
direction $X$). The transformation $g\ra \phi^*(g)$  is
equivalent to $\bar g +\ep a^2 h \ra \bar g +\ep(a^2h+L_X\bar g)$,
{\em i.e.} under an 'infinitesimal coordinate transformation' the
metric perturbation $h$ transforms as
\be
 h \ra h + a^{-2}L_X\bar{g} ~. \label{2gt}
\ee
In the context of cosmological perturbation theory, infinitesimal
coordinate transformations are called 'gauge transformation'. The
perturbation of a arbitrary tensor field $Q=\bar Q +\ep Q^{(1)}$ obeys
the gauge transformation law
\be
 Q^{(1)}  \ra  Q^{(1)} + L_X\bar{Q}~.  \label{gaugeQ}
\ee

%The isomorphism $\phi$ 
%relation of the two averaging procedures is given by
%\bea \bar{g}_2 &=& \phi_*\bar{g} = \bar{g} -\ep L_{\!X}\bar{g}
%   + {\cal O}(\ep^2)  \label{2ggt} \\
% \overline{T}_2 &=& \phi_*\overline{T} = \overline{T} -\ep L_{\!X}
%\overline{T}   + {\cal O}(\ep^2) ~. \eea
%In the context of cosmological perturbation theory, the isomorphism
%$\phi$ is called a gauge transformation. And the choice of a background
%$(\bar{g},\overline{T})$ corresponds to a choice of gauge.
%The above relation is of course true for all averaged tensor fields
%$\overline{Q}$ and $\overline{Q}_2$.
%Separating $Q$ into a background component and a small perturbation,
%$ Q = \overline{Q} + \ep Q^{(1)}=  \overline{Q}_2 + \ep Q^{(2)}$ , we
%obtain the relation
%\be Q^{(2)} = Q^{(1)} +L_{\!X}\overline{Q}   \;. \label{2gt} \ee
Since every vector field $X$ generates a gauge
transformation $\phi = \phi_\ep^X$, we can conclude that only perturbations
 of tensor fields with  $ L_X\overline{Q}=0$ for all vector fields $X$,
i.e., with vanishing (or constant) 'background contribution' are gauge
invariant. This simple result is sometimes referred to as the
'Stewart Walker Lemma' \cite{StW}.

The gauge dependence of perturbations has caused many controversies in the
literature, since it is often difficult to extract the physical
meaning of gauge dependent perturbations, especially on super--horizon
scales. This has led to the development of gauge invariant
perturbation theory
which we are going to use throughout this review. The advantage of the
gauge--invariant formalism is that the variables used have simple geometric
and physical meanings and are not plagued by gauge modes.
Although the derivation requires somewhat more work, the final system
of perturbation equations is usually simple and well suited for numerical
treatment. We shall also see, that on sub-horizon scales, the
gauge invariant matter perturbations variables approach the usual, gauge
dependent ones. Since one of the gauge invariant geometrical perturbation
variables corresponds to the Newtonian potential, the Newtonian limit can be 
performed easily.

First we note that since all  relativistic
equations are covariant (i.e. can be written in the form $Q=0$ for some
tensor field $Q$), it is always possible to express the corresponding
perturbation equations in terms of gauge invariant variables
\cite{Ba,KS,Rfund}.

\section{Gauge invariant perturbation variables}

Since the $\{\eta=\mr{const}\}$ hyper-surfaces are homogeneous and
isotropic, it is sensible to perform a harmonic analysis: A
(spatial) tensor field $Q$ on these hyper-surfaces can be decomposed
into components with transform irreducibly under translations and
rotations. All such components evolve independently. For a scalar
quantity $f$ in the case $\ka=0$ this is nothing else than its
Fourier decomposition:
\be
f(\bx,\eta)=\int d^3\!k \hat{f}(\bk) e^{i\bk\bx}.
\ee
(The exponentials $Y_\bk(\bx)=e^{i\bk\bx}$ are the unitary irreducible
representations of the Euclidean translation group.) For $\ka=1$ such a
decomposition also exists, but the values $k$ are discrete,
$k^2=\ell(\ell+2)$ and for $\ka=-1$, they are bounded from below,
$k^2>1$. Of course, the functions $Y_\bk$ are different for $\ka\neq 0$.

They are always the complete orthogonal set of eigenfunctions of the
Laplacian,
\be
\Delta Y^{(S)}=-k^2 Y^{(S)} .
\ee

In addition, a tensorial variable (at fixed position $\bx$) can be
decomposed into irreducible components under the rotation
group $SO(3)$.

For a vector field, this is its decomposition
into a gradient and a rotation,
\be
V_i = \nabla_i \vph + B_i ,\label{vdec} %\quad \mr{where } B^i_{,i}=0.
\ee
where
\be
B^i_{|i}=0 ,
\ee
where we used $X_{|i}$ to denote the three--dimensional covariant
derivative of $X$.
$\vph$ is the spin 0 and ${\bf B}$ is the spin 1 component
of V.

For a symmetric tensor field we have
\be
H_{ij} = H_L \ga_{ij} +\l(\nabla_i\nabla_j-\frac{1}{3}\De\ga_{ij}\r)H_T
	+\frac{1}{2}\l(H^{(V)}_{i|j}+H^{(V)}_{j|i}\r) + H^{(T)}_{ij}, 
\label{tdec}
\ee
where
\be
H^{(V)|i}_i=H^{(T)^i}_i=H^{(T)^j}_{i|j}=0.
\ee
Here $H_L$ and $H_T$ are spin 0 components, $H^{(V)}_i$ is a
spin 1 component and $H^{(T)}_{ij}$ is a spin 2 component.

We shall not need higher tensors (or spinors).
As a basis for vector and tensor modes we use the vector and tensor
type eigenfunctions to the Laplacian,
\bea
 \Delta Y_j^{(V)} &=&-k^2 Y_j^{(V)} \\
\mbox{ and} && \nonumber\\
 \Delta Y_{ji}^{(T)} &=&-k^2 Y_{ji}^{(T)}~, 
\eea
where $Y_j^{(V)}$ is a transverse vector,  $Y_j^{(V)|j}=0$ and $
Y_{ji}^{(T)} $ is a symmetric transverse traceless tensor, $Y_j^{(T)j}=
Y_{ji}^{(T)|i}= 0$.

According to Eqs.~(\ref{vdec}) and (\ref{tdec}) we can construct scalar type
vectors and tensors and vector type tensors. To this goal we define
\bea
Y^{(S)}_j &\equiv& -k^{-1}Y^{(S)}_{|j} \\
Y^{(S)}_{ij} &\equiv& k^{-2}Y^{(S)}_{|ij} +{1\over 3}\ga_{ij}Y^{(S)} \\ 
Y^{(V)}_{ij} &\equiv& -{1\over 2k}(Y^{(V)}_{i|j} +Y^{(V)}_{j|i}) ~.
\eea
In the following we shall extensively use this decomposition
and write down the perturbation equations for a given mode
$k$. 

The decomposition of a vector field is then of the form
\be
B_i = B Y^{(S)}_i+B^{(V)} Y^{(V)}_i. \label{vdec2}
\ee
The decomposition of a scalar field is given by (compare~\ref{tdec})
\be 
H_{ij}=H_LY^{(S)} \ga_{ij} +H_TY^{(S)}_{ij} 
	+H^{(V)}Y^{(V)}_{ij} + H^{(T)}Y^{(T)}_{ij}, 
\label{tdec2}
\ee
where $B$, $B^{(V)}_i$, $H_L$, $H_T$, $H^{(V)}_i$ and $H^{(T)}_{ij}$ are 
functions of $\eta$ and $\bk$

\subsection{Metric perturbations}
Perturbations of the metric are of the form
\be
g_{\mu\nu} = \bar{g}_{\mu\nu} + a^2 h_{\mu\nu} .
\ee
We parameterize them as
\be
h_{\mu\nu} dx^\mu dx^\nu = -2 A d\eta^2 - 2B_i d\eta dx^i
	+2 H_{ij} dx^i dx^j, \label{defpert}
\ee
and we decompose the perturbation variables $B_i$ and $H_{ij}$ according  to
(\ref{vdec2}) and (\ref{tdec2}).

Let us consider the behavior of $h_{\mu\nu}$ under gauge
transformations. We set the vector field defining the gauge
transformation to
\be
{\bf X} = T\dd_\eta+L^i\dd_i .
\ee
Using simple identities from differential geometry like
$L_{\bf X}(df)=d(L_{\bf X} f)$ and \\ 
$(L_{\bf X}\ga)_{ij}=X_{i|j}+X_{j|i}$, we obtain 
\bea
L_{\bf X} \bar{g} &=& a^2\l[-2\l(\frac{\da}{a}T+\dot{T}\r)d\eta^2
	+2\l(\dot{L}_i-T_{,i}\r)d\eta dx^i \r. \nonumber \\ &&
 \l.    +\l(2\frac{\da}{a}T\ga_{ij}+L_{i|j}+L_{j|i}\r)dx^i dx^j\r].
\eea

Comparing this with (\ref{defpert}) and using (\ref{2gt}) we
obtain the following behavior of our perturbation variables under
gauge transformations (decomposing $L_i=LY^{(S)}_i+L^{(V)}Y^{(V)}_i$):
\bea
A &\rightarrow& A + \frac{\da}{a}T + \dot{T} \label{transA}\\
B &\rightarrow& B - \dot{L} -k T\\
B^{(V)} &\rightarrow& B^{(V)} - \dot{L}^{(V)}\\
H_L &\rightarrow& H_L + \frac{\da}{a}T + \frac{k}{3} L\\
H_T &\rightarrow& H_T - kL \label{transHT}\\
H^{(V)} &\rightarrow& H^{(V)} -k L^{(V)} \\
H^{(T)} &\rightarrow& H^{(T)} .
\eea
Two scalar and one vector variable can be brought to disappear
by gauge transformations.

One often chooses $kL=H_T$ and $T=B+\dot{L}$, so that the variables
$H_T$ and $B$ vanish. In this gauge (longitudinal gauge), scalar
perturbations of the metric are of the form ($H_T=B=0$):
\be
h_{\mu\nu}^{(S)} = -2 \Psi d\eta^2 + 2 \Phi \ga_{ij} dx^i dx^j .
\ee

$\Psi$ and $\Phi$ are the so called {\em Bardeen} potentials.
In general they are defined by
\bea
\Psi &=& A - \frac{\da}{a}k^{-1} \sigma -k^{-1}\dot{\sigma} \\
\Phi &=& H_L + \frac{1}{3} H_T -\frac{\da}{a}k^{-1} \sigma
\eea
with $\si=k^{-1}\dot{H}_T-B$. A short calculation using Eqs. (\ref{transA}) 
to (\ref{transHT}) shows that they
are gauge invariant.

For vector perturbations it is convenient to set $kL^{(V)}=H^{(V)}$
so that $H^{(V)}$ vanishes and we have
\be
h_{\mu\nu}^{(V)} dx^\mu dx^\nu = 2 \si^{(V)}Y^{(V)}_i d\eta dx^i.
 \label{vecg}\ee
We shall call this gauge the ``vector gauge''.
In general $\si^{(V)}=k^{-1}\dot{H}^{(V)}-B^{(V)}$ is gauge invariant\footnote{
$Y^{(V)}_{ij}\si^{(V)}$ is the shear of the hyper-surfaces
of constant time.}.

Clearly there are no tensorial (spin 2) gauge transformation
and hence $H_{ij}^{(T)}$ is gauge invariant.

\subsection{Perturbations of the energy momentum tensor}

Let $T^\mu_\nu=\overline{T}^\mu_\nu+\Th^\mu_\nu$ be the full
energy momentum tensor. We define its energy density $\rho$
and its energy flow 4-vector $u$ as the time-like eigenvalue
and eigenvector of $T^\mu_\nu$:
\be
T^\mu_\nu u^\nu = -\rho u^\mu , \quad u^2=-1.
\ee

We then define their perturbations by
\be
\rho=\bar{\rho}\l(1+\de\r), \quad u=u^0 \dd_t + u^i \dd_i .
\ee
$u^0$ is fixed by the normalization condition,
\be
u^0 = \frac{1}{a} (1-A) .
\ee
We further set
\be
u^i = \frac{1}{a} v^i = vY^{(S)i} + v^{(V)}Y^{(V)i}.
\ee

We define $P^\mu_\nu\equiv u^\mu u_\nu+\de^\mu_\nu$, the
projection tensor onto the part of tangent space normal to
$u$ and set the stress tensor
\be
\tau^{\mu\nu}=P^\mu_\al P^\nu_\b T^{\al\b} .
\ee

In the unperturbed case we have 
$\tau^0_0=0, \tau^i_j=\bar{p} \de^i_j$. Including 
 perturbations, to first order we still obtain
\be
\tau^0_0=\tau^0_i=\tau^i_0=0.
\ee

But $\tau^i_j$ contains in general perturbations. We set
\be
\tau^i_j = \bar{p}\l[\l(1+\Pi_L\r)\de^i_j+\Pi^i_j\r]
	, \quad \mr{with} \quad \Pi^i_i=0.
\ee
We decompose $\Pi^i_j$ as
\be
\Pi^i_j =  \Pi^{(S)}Y^{(S)\,i}_j
	+ \Pi^{(V)}Y^{(V)\,i}_j
	+ \Pi^{(T)} Y^{(T)\,i}_j.
\ee

We shall not derive the gauge transformation
properties in detail, but just state some results which can be obtained
as an exercise (see also~\cite{KS}):
\begin{itemize}
\item Of the variables defined above only the $\Pi^{(S,V,T)}$
are gauge invariant; they describe the anisotropic stress tensor,
$\Pi^\mu_\nu=\tau^\mu_\nu-\nicefrac1/3\tau^\al_\al\de^\mu_\nu$.
They are gauge invariant due to the Stewart--Walker
lemma, since $\bar{\Pi}=0$. For perfect fluids $\Pi^\mu_\nu=0$.
\item A second gauge invariant variable is
\be
\Ga=\pi_L-\frac{c_s^2}{w}\de,
\ee
where $c_s^2\equiv\dot{p}/\dot{\rho}$ is the adiabatic sound speed
and $w\equiv p/\rho$ is the enthalpy. One can show that $\Ga$ is
proportional to the divergence of the entropy flux of the
perturbations. Adiabatic perturbations are characterized
by $\Ga=0$.
\item Gauge invariant density and velocity perturbations
can be found by combining $\de$, $v$ and $v_i^{(V)}$ with
metric perturbations.
\end{itemize}

We shall use
\bea
V &\equiv& v - \frac{1}{k} \dot{H}_T = v^\lo \\[2mm]
D_g &\equiv& \de + 3(1+w) \l(H_L+\frac{1}{3} H_T\r)=\de^\lo+3(1+w)\Phi\\
D &\equiv& \de^\lo+3(1+w) \l(\frac{\da}{a}\r)\frac{V}{k}\\[2mm]
V^{(V)} &\equiv& v^{(V)} - \frac{1}{k}\dot{H}^{(V)} 
	= v^{(\mr{vec})}\\
\Om &\equiv& v^{(V)} - B^{(V)} =  v^{(\mr{vec})}-B^{(V)}\\
 \Om - V^{(V)} &=&\si^{(V)} .
\eea
Here $ v^\lo,~\de^\lo$ and $ v_i^{(\mr{vec})}$ are the velocity (and
density) perturbations in the longitudinal and vector gauge
respectively and $\si^{(V)}$ is the metric perturbation in vector
gauge (see Eq.~(\ref{vecg})).
 These variables
 can be  interpreted nicely in terms of gradients
of the energy density and the shear and vorticity of the velocity
field~\cite{Ellis}.

But we just want to show that on scales much smaller than the
Hubble scale, $k\eta\gg 1$, the metric perturbations are much
smaller than $\de$ and $v$ and we can thus ``forget them'' (which
will be important when comparing experimental results with
calculations in this formalism):

The perturbations of the Einstein tensor are given by
second derivatives of the metric perturbations. Einstein's
equations yield the following order of magnitude estimate:
\bea
\OO\l(\frac{\de T}{T}\r)
\underbrace{\OO\l(8\pi GT\r)}_{\OO\l(\frac{\da}{a}\r)^2=\OO\l(\eta^{-2}\r)}
&=& \OO\l(\frac{1}{\eta^2}h+\frac{k}{\eta}h+k^2 h\r)\\
\OO\l(\frac{\de T}{T}\r) &=& \OO\l(h+k\eta h+(k\eta)^2 h\r) .
\eea
For $k\eta\gg 1$ this gives $\OO(\de,v)=\OO\l(\frac{\de T}{T}\r)\gg\OO(h)$.
On sub-horizon scales the difference between $\de$, $\de^\lo$, $D_g$
and $D$ is negligible as well as the difference between $v$ and $V$
or $v^{(V)}$, $V^{(V)}$ and $\Om^{(V)}$.

Later we shall also need other perturbation variables like the
perturbation of the photon brightness (energy--integrated photon
distribution function), but we shall introduce them as we get
there and discuss some applications first.

%\clearemptydoublepage

\section{Basic perturbation equations}

As already announced, we do not derive Einstein's equations but just
write down those which we shall need later:

\subsection{Constraint equations}
\bea \l. \begin{array}{rclc}
4\pi G a^2 \rho D & = & (k^2 -3\ka)\Phi & (00)\\
4\pi G a^2 (\rho+p) V & = & k\l(\l(\frac{\da}{a}\r)\Psi-\dot{\Phi}\r) & (0i)
\end{array} \r\} && \mr{(scalar)}\label{scalcntr} \\
8\pi G a^2 (\rho+p)\Om = \frac{1}{2} \l(2\ka-k^2\r)\si^{(V)} \quad (0i)
\quad && \mr{(vector)} \label{veccntr}
\eea

\subsection{Dynamical equations}
\bea
-k^2 \l(\Phi+\Psi\r) &=& 8\pi G a^2 p\Pi^{(S)} 
 \quad\quad \mr{(scalar)} \label{scaldyn}\\
k\l(\dot{\si}^{(V)}+2\l(\frac{\da}{a}\r)\si^{(V)}\r) &=& 
8\pi G a^2 p \Pi^{(V)} 
	\quad\quad\mr{(vector)}\\
\ddot{H}^{(T)}
	+2\l(\frac{\da}{a}\r)\dot{H}^{(T)}
	+\l(2\ka+k^2\r)H^{(T)}  &=& 
8\pi G a^2 p \Pi_{ij}^{(T)}
\quad\quad \mr{(tensor)} \label{tensdyn}
\eea
There is a second dynamical scalar eqn., which is however complicated and 
not needed, since we may instead use one of the conservation eqns. below.
Note that for perfect fluids, where $\Pi^i_j\equiv0$, we have
$\Phi=-\Psi$, $\si^{(V)}\propto 1/a^2$ and $H$ obeys a damped
wave equation. The damping term can be neglected on small scales (over
short time periods) when
$\eta^{-2}\lsim 2\ka+k^2$, and $H_{ij}$ represents propagating
gravitational waves. For vanishing curvature, these are just the
sub-horizon scales, $k\eta\gsim 1$. For $\ka<0$, waves oscillate with
a somewhat smaller frequency, $\om=\sqrt{2\ka+k^2}$,
 while for $\ka>0$ the frequency is somewhat larger.

\subsection{Conservation equations}
\bea \hspace{-1cm}\l. \begin{array}{c}
\dot{D}_g+3\l(c_s^2-w\r)\l(\frac{\da}{a}\r)D_g
+(1+w)kV+3w\l(\frac{\da}{a}\r)\Ga=0 \\
\dot{V}+\l(\frac{\da}{a}\r)\l(1-3c_s^2\r)V = k\l(\Psi-3c_s^2\Phi\r)
+\frac{c_s^2 k}{1+w}D_g \\
 \hspace{2cm} +\frac{wk}{1+w}\l[\Ga-\frac{2}{3}\l(1-\frac{3\ka}{k^2}\r)\Pi\r]
\end{array} \r\} && \mr{(scalar)} \label{scalcons}\\
\hspace{-1cm}\dot{\Om}_i+\l(1-3c_s^2\r)\l(\frac{\da}{a}\r)\Om_i
=\frac{p}{2(\rho+p)}\l(k-\frac{2\ka}{k}\r)\Pi_i^{(V)} \qquad && \mr{(vector)}
\label{veccons}
\eea

\chapter{Simple applications}
We first discuss some simple applications which
will be important for the CMB. We could of course also write
(\ref{scalcons}) in terms of $D$, but we shall just work with
the relation
\be
D=D_g+3(1+w)\l(-\Phi+\l(\frac{\da}{a}\r)k^{-1}V\r) . \label{DDgrel}
\ee

\section{The pure dust fluid at $\ka=0, \La=0$}
We assume the dust to have $w=c_s^2=p=0$ and $\Pi=\Ga=0$.
The equations (\ref{scalcons}), (\ref{scaldyn}) and
(\ref{scalcntr}) then reduce to
\bea
\dot{D}_g&=&-kV\quad\mr{(energy~conservation~eqn.)} \label{dust1}\\
\dot{V}+\l(\frac{\da}{a}\r)V &=& k\Psi \quad
	\mr{(gravitational~acceleration~eqn.)}  \label{dust2}\\
\Phi &=& -\Psi \label{dust3}\\
-k^2\Psi &=&4\pi
 Ga^2\rho\l(D_g+3\l(\Psi+\l(\frac{\da}{a}\r)k^{-1}V\r)\r)
   \quad\mr{(Poisson~eqn.)} . \label{dust4}
\eea
In a pure dust universe $\rho\propto a^{-3} \Rightarrow (\da/a)^2 \propto a^{-1}$,
which is solved by $a\propto\eta^2$. The Einstein equations then give
immediately $4\pi G\rho a^2=\nicefrac3/2(\da/a)^2=6/\eta^2$.
Setting $k\eta=x$ and $'=d/dx$, the system (\ref{dust1}-\ref{dust4})
then becomes
\bea
D'_g&=&-V\label{dustb1} \\
V'+\frac{2}{x}V&=&\Psi\label{dustb2}\\
\frac{6}{x^2}\l(D_g+3\l(\Psi+\frac{2}{x}V\r)\r)&=& -\Psi. \label{dustb3}
\eea

We use (\ref{dustb3}) to eliminate $\Psi$ and (\ref{dustb1})
to eliminate $D_g$, leading to
\be
\l(18+x^2\r)V''+\l(\frac{72}{x}+4 x\r)V'-\l(\frac{72}{x^2}+4\r)V=0.
\label{Vpp}\ee
The general solution of Eq.~(\ref{Vpp}) is
\be
V=V_0 x+\frac{V_1}{x^4}
\ee
with arbitrary constants $V_0$ and $V_1$.
Since the perturbations are supposed to be small initially,
they cannot diverge for $x\ra 0$, and we have therefore
to choose $V_1=0$ (the growing mode). Another way to argue is as
follows: If the 
mode $V_1$ has to be small already at some early initial time
$\eta_\mr{in}$, it will be even much smaller at later times and may
hence be neglected. The perturbation variables are then
given by
\bea
V &=& V_0 x \\
D_g &=& -15 V_0 -\frac{1}{2}V_0 x^2\\
\Psi &=& 3V_0 .
\eea

The constancy of the gravitational potential $\Psi$ in a matter
dominated universe and the growth of the density perturbations
like the scale factor $a$ led Lifshitz to conclude 1946~\cite{Lifshitz}
that pure gravitational
instability cannot be the cause for structure formation:
If we start from tiny thermal fluctuations of the order
of $10^{-35}$, they can only grow to about
$10^{-30}$ through this process during the matter dominated
regime. Or, to put it differently,
if we do not want to modify the process of structure formation,
we need initial fluctuations of the order of at least $10^{-5}$.
One possibility to create such fluctuations is due to quantum particle
production in the classical gravitational field during inflation.
 The rapid expansion of the universe during inflation quickly transforms 
microscopic scales at which quantum fluctuations are important into 
cosmological scales where these fluctuations are then ``frozen in'' as 
classical perturbations in the energy density and the geometry.

We distinguish two regimes:\\
{\bf \em i}) super-horizon, $x\ll1$ where we have
\bea
D_g&=&-15V_0\\
\Psi&=&3V_0\\
V&=&V_0x
\eea
%{\bf maybe offer some words of interpretation, i.e. V seems to be growing,
%but we are super-horizon, etc.. What does this value mean, anyway? }
and {\bf \em ii}) sub-horizon, $x\gg1$ where the solution is  dominated by 
the terms
\bea
V&=&V_0 x\\
D_g&=&-\frac{1}{2}V_0 x^2\\
\Psi&=&3V_0=\mr{const} \label{eq48}
\eea
%The potential $\Psi$ remains therefore constant {\bf what does
%this mean? Why is not $V(x)\propto x^4$ taken as dominant solution?
%And I have $D_0=-6V_0$...}.

Note that for dust
\[ D =D_g + 3\Psi +{6\over x}V = -\half V_0x^2 ~. \]
In the variable $D$ the constant term has disappeared and we have $D\ll \Psi$ 
on super horizon scales, $x\ll 1$.

\section{The pure radiation fluid, $\ka=0, \La=0$}

In this limit we set $w=c_s^2=\nicefrac1/3$ and $\Pi=0$. We conclude from
$\rho\propto a^{-4}$ that $a\propto\eta$ and $\Phi=-\Psi$,
and the perturbation
equations become (with the same notation as above):
\bea
D'_g&=&-\frac{4}{3} V\\
V' &=& 2\Psi+\frac{1}{4}D_g\\
-2x^2\Psi &=& 3D_g+12\Psi+\frac{12}{x}V \label{radflu3}
\eea
The general solution of this system is
\bea
D_g &=& D_2 \l[\cos\l(\frac{x}{\sqrt{3}}\r)
	-2\frac{\sqrt{3}}{x}\sin\l(\frac{x}{\sqrt{3}}\r)\r]
	\nonumber\\ &&
+ D_1 \l[\sin\l(\frac{x}{\sqrt{3}}\r)
	+2\frac{\sqrt{3}}{x}\cos\l(\frac{x}{\sqrt{3}}\r)\r]\\
V&=&-\frac{3}{4}D'_g \\
\Psi &=& {-3D_g-(12/x)V\over 12+2x^2}~.
\eea
Again, regularity at $x=0$ requires $D_1=0$.

In the  {\bf 
%{\em i}) 
super-horizon, $x\ll1$} regime we obtain
%\be
%V''+\frac{2V'}{x}-\frac{2V}{x^2}=0
%\ee
%{\bf I must have a sign error somewhere, I'll check that
%tomorrow.}
%Ansatz:
\be
\Psi=\Psi_0, \quad D_g=D_0-\frac{2}{3} V_0 x^2, \quad V=V_0 x
\ee
with
\bea
D_0 &=& -6 \Psi_0 = -D_2\\
V_0 &=& \frac{1}{2} \Psi_0 = -\frac{1}{12} D_0.
\eea
%{\bf and this means...}
On {\bf 
%{\em ii}) 
sub-horizon, $x\gg1$} scales we find oscillating 
 solutions with constant amplitude
with a frequency of $1/\sqrt{3}$:
\bea
V&=&	V_2 \sin\l(\frac{x}{\sqrt{3}}\r) \\
D_g&=&	D_2 \cos\l(\frac{x}{\sqrt{3}}\r)~,~~~ \Psi=-{3\over 2}x^{-2}D_g \\
&& D_2=\frac{4V_2}{\sqrt{3}}.
\eea
Note that also for radiation perturbations 
\[ D=-{2\over 3}V_0x^2 \ll \Psi \]
is small on super horizon scales, $x\ll 1$. The perturbation amplitude is 
given by the largest gauge invariant perturbation variable. 
We conclude therefore that perturbations outside the Hubble
horizon are frozen to first order. Once they enter the horizon
they start to collapse, but pressure resists the gravitational
force and the radiation fluid starts to oscillate. The perturbations
of the gravitational potential oscillate and decay like $1/a^2$
inside the horizon.

\section{Adiabatic and isocurvature initial conditions
for a matter \& radiation fluid}

In this section we want to investigate a system with a matter
and a radiation component that are coupled only by gravity.
The matter component acts therefore as {\em dark matter}, since
it does not interact directly with the radiation.

Since the matter and radiation perturbations behave in the
same way on super-horizon scales,
\be
D_g^{(r)}=A+Bx^2, \quad
D_g^{(m)}=A'+B'x^2, \quad
V^{(r)} \propto V^{(m)} \propto x,
\ee
we may require a constant relation between matter and radiation
perturbations. As we have seen in the previous section, 
inside the horizon ($x>1$) radiation
perturbations start to oscillate while matter perturbations keep
following a power law. On sub-horizon scales a constant ratio
can thus no longer be maintained. There are two interesting
possibilities:

\subsection{ Adiabatic initial conditions}

Adiabaticity requires that matter and radiation perturbations are
initially in perfect thermal equilibrium. This implies that their
velocity fields agree (see below, section on Boltzmann eqn.!)
\be
V^{(r)}=V^{(m)} ,
\ee
so that the energy flux in the two fluids is coupled initially.

Let us investigate the radiation solution in the
{\em matter dominated era}, when the corresponding scale is already
sub-horizon. Since $\Psi$ is dominated by the matter contribution, we
have  $\Psi\simeq \mr{const}=\Psi_0$. We neglect the (decaying) 
contribution from the sub-dominant radiation 
to $\Psi$. Energy--momentum conservation for radiation then
gives
\bea
D_g^{(r)\prime} &=& -\frac{4}{3} V^{(r)}\\
V^{(r)\prime}    &=& 2\Psi + \frac{1}{4} D_g^{(r)} .
\eea

Now $\Psi$ is just a constant given by the matter
perturbations, and it acts like a constant source term.
The full solution of this system is then 
\bea
D_g^{(r)} &=& A \cos\l(\frac{x}{\sqrt{3}}\r)
	- \frac{4}{\sqrt{3}}B \sin\l(\frac{x}{\sqrt{3}}\r)
	- 8 \Psi \l[\cos\l(\frac{x}{\sqrt{3}}\r)-1\r]\\
V^{(r)} &=& B \cos\l(\frac{x}{\sqrt{3}}\r)
	+ \frac{\sqrt{3}}{4}A \sin\l(\frac{x}{\sqrt{3}}\r)
	- 2\sqrt{3} \Psi \sin\l(\frac{x}{\sqrt{3}}\r) .
\eea

Our adiabatic initial conditions require
\be
\lim_{x\rightarrow 0} \frac{V^{(r)}}{x}=V_0
=\lim_{x\rightarrow 0} \frac{V^{(m)}}{x} < \infty .
\ee
Therefore $B=0$ and $A=4V_0-8\Psi$. Using in addition $\Psi=3V_0$
(see (\ref{eq48})) we obtain
\bea
D_g^{(r)} &=& -\frac{44}{3} \Psi \cos\l(\frac{x}{\sqrt{3}}\r)
	+8 \Psi \label{Dad}\\
V^{(r)} &=& {1\over \sqrt{3}} \Psi \sin\l(\frac{x}{\sqrt{3}}\r)\\
D_g^{(m)} &=& -\Psi( 5 + \frac{1}{6} x^2) \\
V^{(m)} &=& {1\over 3} \Psi x\\
\Psi &=& 3 V_0 .
\eea
On super-horizon scales, $x\ll 1$ we have 
\be 
	D_g^{(r)}  \simeq -\frac{20}{3} \Psi ~\mbox{ and }~~~ V^{(r)}
	\simeq \frac{1}{3}x \Psi~, \label{Dads}
\ee 
note that $D_g^{(r)} = (4/3)D^{(m)}_g$ and $V^{(r)}=V^{(m)}$ for adiabatic 
initial conditions.
 
\subsection{ Isocurvature initial conditions}

Here we want to solve the system (\ref{scalcntr})
and (\ref{scalcons}) for dark matter and radiation under
the condition that the metric perturbations vanish initially,
\ie, $\Psi=0$,
\be
\Psi = -\frac{3}{2} \l(\frac{\da}{a}\r)^2 k^{-2} \l[D_g
	+3(1+w)\Psi+3(1+w)\l(\frac{\da}{a}\r)k^{-1} V\r]=0. \label{isocond}
\ee

In principle, we have four evolution and one constraint equations.
We therefore have four constants  to adjust. Condition (\ref{isocond}),
however, requires an entire function to vanish. This may be
impossible. Let us nevertheless try:

If $\Psi=0$ the solutions of the radiation dominated
equations are simply
\bea
D_g^{(r)} &=& A \cos\l(\frac{x}{\sqrt{3}}\r)
	+ B \sin\l(\frac{x}{\sqrt{3}}\r)\\
V^{(r)} &=& \frac{\sqrt{3}}{4} A \sin\l(\frac{x}{\sqrt{3}}\r)
	-\frac{\sqrt{3}}{4} B \cos\l(\frac{x}{\sqrt{3}}\r).
\eea
For the matter perturbations we find
\bea
V^{(m)} &=& -\frac{V_0}{a}, \quad a\propto x^\b, \quad 1\leq\b\leq2 \\
D_g^{(m)} &=& C^{(m)} -\frac{V_0}{\b-1} \frac{x}{a} \quad \mbox{ if } 
 \b\neq 1\\
D_g^{(m)} &=& C^{(m)} -{V_0}\log(x)  \quad  \mbox{ if } \b= 1
\eea
Here $\b$ is the exponent of the scale factor $a\propto \eta^\b$, hence 
$\b=1$ in the radiation era and $\b=2$ in the matter era.

$\Psi=0$ implies with
\bea
D_g &=& \frac{1}{\rho}\l(\rho_r D_g^{(r)} 
	+ \rho_m D_g^{(m)} \r) \quad \mr{and} \\
V &=& \frac{1}{\rho+p}\l(\l(\rho_r+p_r\r)V^{(r)}+\rho_m V^{(m)}\r)
\eea
that
\be
0=\frac{\rho_r}{\rho_m}D_g^{(r)}+D_g^{(m)}+
\l(\frac{\da}{a}\r)k^{-1}\l[ \frac{4\rho_r}{\rho_m}V^{(r)}+3V^{(m)}\r] .
\ee

Since $V^{(m)}\propto 1/a$ it can compensate, for small values
of $x$, the term $\propto \cos(x/\sqrt{3})$ of $V^{(r)}$, which
behaves like $1/a$ as well, due to the pre-factor $\rho_r/\rho_m$.
This term can also be compensated in $D_g^{(r)}$ by the term
$V_0 x/a$ of $D_g^{(m)}$ In the purely radiation dominated universe, the 
$\log$-dependence of $D_g^{(m)}$ renders this compensation imperfect. 
However, there is no way to compensate
$C^{(m)}$ or the term proportional to $A$. We therefore have to choose
 $A=C^{(m)}=0$ and
\be
a\frac{\rho_r}{\rho_m}\frac{\sqrt{3}}{3}B=V_0, \quad
B=\frac{\rho_m}{a\rho_r}\sqrt{3}V_0 .
\ee
(The compensation of the smaller terms in $D_g^{(r)}$ and $D_g^{(m)}$
is only complete if $\b\simeq2$.)

With $c_s=1/\sqrt{3}$ we find
\bea
D_g^{(r)} &\simeq& \frac{\rho_m}{a\rho_r c_s} V_0 \sin\l(c_s x\r)
	\quad \mr{(isocurvature)}\\
D_g^{(r)} &\simeq& \Psi \l(8-\frac{44}{3}\cos\l(c_s x\r)\r)
	\quad \mr{(adiabatic)}~.
\eea
The CMB anisotropies, which we are going to determine in the next chapter,
contain a term
\be
\frac{\De T}{T}\l(\bk,\eta_0,\bn\r)=\cdots 
+\frac{1}{4}D_r^{(g)}\l(\bk,\eta_\mr{dec}\r)
e^{i\bk\bn\l(\eta_0-\eta_\mr{dec}\r)}\cdots ~.
\ee
On scales where this term dominates, the peaks
in $D_g$ translate into peaks in the angular power spectrum of CMB
anisotropies.

For isocurvature initial conditions, we find a first peak in $D_g$ at
\be
x_i^{(1)}=k_i^{(1)}\eta_\mr{dec}=\frac{1}{c_s}\frac{\pi}{2}, \quad
\la_i^{(1)} = \frac{\pi}{k_i^{(1)}}=2c_s\eta_\mr{dec}, \quad
\vth_i^{(1)}\simeq \frac{2c_s\eta_\mr{dec}}{\chi\l(\eta_0-\eta_\mr{dec}\r)} ,
\ee
Here $\vth_i^{(1)}$ is the angle under which the comoving scale 
$\la_i^{(1)}$ at comoving distance $\eta_0-\eta_\mr{dec}$ is seen. 
In the next chapter, we will expand the temperature fluctuations in
terms of spherical harmonics. An fluctuation on the angular scale
$\vth$ then shows up around the harmonic $\ell \sim \pi/\vth$. As an
indication, we note that 
for $\La=\ka=0$, the  harmonic of the first isocurvature peak is
\[\ell_i^{(1)}\sim\pi/\vth_i^{(1)}\sim 110 ~,\] 
In the adiabatic case the first ``peak'' is at $k_a^{(1)}=0$.

Since $D^{(r)}_g$  is negative for small $x$, the first peaks are
``expansion peaks'', and due to the gravitational attraction of the baryons
(which we have neglected in this simple argument) they
are less pronounced than the second (``compression'') peaks.

These second peaks are usually called the ``first acoustic peak''. (It is
the first compression peak and we shall adopt the convention to call
it the ``first  peak'' mainly for consistency with the literature.) 
They correspond to  wavelengths and angular scales
\bea
\la_i^{(2)} &=& \frac{2}{3} c_s \eta_\mr{dec}~, ~~
\vth_i^{(2)}\simeq
\frac{(2/3)c_s\eta_\mr{dec}}{\chi\l(\eta_0-\eta_\mr{dec}\r)}~,
~~~\ell_i^{(2)} \sim 350 ~~~~ \mr{(isocurvature)}\\
\la_a^{(2)} &=& c_s\eta_\mr{dec}~,~~
\vth_a^{(2)}\simeq
\frac{c_s\eta_\mr{dec}}{\chi\l(\eta_0-\eta_\mr{dec}\r)}~,
~~~\ell_a^{(2)} \sim 220 ~~~~\mr{(adiabatic)}.
\eea
Here the indicated harmonic is the one obtained in the case
$\La=\ka=0$, for a typical baryon density inferred from nucleosynthesis.

It is interesting to note that the distance between consecutive peaks 
is {\bf independent of the
initial condition}. It is given by 
\be
\De k_i=k_i^{(2)}-k_i^{(1)}=\pi/(c_s\eta_\mr{dec})=\De k_a~,~~~ \De\vth=
\frac{c_s\eta_\mr{dec}}{\chi\l(\eta_0-\eta_\mr{dec}\r)}~,~~
 \De\ell \sim 220~.
\ee
Again, the numerical value indicated for $\De\ell$ corresponds to a
universe with $\La=\ka=0$. The result is strongly dependent especially
on $\kappa$. This is the reason why the measurement of the peak
position (or better of the inter-peak distance) allows an accurate
determination of curvature.

From our analysis we can draw the following important conclusions:
For scales where the $D_g^{(r)}$-term dominates, the CMB anisotropies
show a series of acoustic oscillations with spacing $\De k$,
the position of the first significant peaks is  at $k=k_{a/i}^{(2)}$, 
depending on the initial condition.

The spacing $\De k$ is {\em independent} of initial conditions.
The angle $\De\vth$ onto which this scale is projected in the sky is
determined entirely by the matter content and the geometry of
the universe. According to our findings in Chapter I, $\vth$
will be larger if $\Om_\ka<0$ (positive curvature) and smaller
if $\Om_\ka>0$ (see Fig.~\ref{theta}).

In our analysis we have neglected the presence of baryons, in
order to obtain simple analytical results. Baryons have two effects:
They lead to $(\rho+3p)_\mr{rad+bar}>0$, and therefore to an enhancement
of the {\em compression} peaks
%{\bf don't you want to use ``compression''? 
%Then should change my use earlier to match use here} 
(the first, third, etc. acoustic peak). In addition, the baryons 
slightly {\em decrease} the sound speed $c_s$, increasing thereby 
$\De k$ and decreasing $\De\vth$.

Another point which we have neglected is the fact that the
universe became matter dominated at $\eta_\mr{eq}$, only shortly
before decoupling: $\eta_\mr{dec}\simeq4\eta_\mr{eq}$, for $\Om_m=1$.
As we have seen, the gravitational potential on sub-horizon scales is
decaying in the radiation dominated era. If the radiation dominated
era is not very long before decoupling, the gravitational potential is
still slightly decaying and free streaming photons fall into a deeper
gravitational potential than they have to climb out of. This effect,
called ``early integrated Sachs--Wolfe effect'' adds to the photon
temperature fluctuations at scales which are only slightly larger than
the position of the first acoustic peak for adiabatic
perturbations. It therefore 'boosts' this peak and, at the same time,
moves it to lightly larger scales (smaller angles).
Since $\eta_\mr{eq}\propto h^{-2}$,
the first acoustic peak is larger  if $h$ is smaller.

A small Hubble parameter {\em increases} therefore the acoustic
peaks. A similar effect is observed if a cosmological constant
or  negative curvature are present, since $\eta_\mr{eq}$ is retarded
in those cases.

The real universe contains not only photons and dark matter, but also 
neutrinos and baryons. It has actually be found recently~\cite{bucher1}
that this 4 fluid mixture allows five different modes which grow or
stay constant. The adiabatic mode, the dark matter isocurvature mode which 
we have just discussed, a similar baryon isocurvature mode and two neutrino 
isocurvature modes. The most generic initial conditions which allow for
arbitrary correlations between the different modes are very unpredictable.
We can maybe just say that they lead to a first acoustic peak in the range of
$150\le \ell^{(2)} \le 350$ for a spatially flat universe.
In the rest of this review, we only discuss adiabatic perturbations, which are
by far the most studied, but it is important to keep in mind that all the
results especially concerning the estimation of cosmological parameters
is not valid if we allow for more generic initial 
conditions~\cite{bucher,trotta}.

\subsection{Vector perturbations of perfect fluids}

If $\Pi^{(V)}=0$ equation (\ref{veccons}) implies
\be
\Om \propto a^{3c_s^2-1} .
\ee
For $\dot{p}/\dot{\rho}=c_s^2\leq\nicefrac1/3$, this leads to a
non--growing vorticity. The dynamical Einstein equation implies
\be
\si^{(V)}\propto a^{-2}~, 
\ee
and the constraint (\ref{veccntr})
reads (at early times,
so we can neglect curvature)
\be
\Om \sim x^2\si^{(V)}.
\ee

If perturbations are created in the very early universe on
super--horizon scales (\eg~during an inflationary period), vector
perturbations of the metric decay and become soon entirely
negligible. Even if $\Om_i$ remains constant in a radiation
dominated universe, it has to be so small on relevant scales
at formation ($x_{in}\ll1$) that we may safely neglect it.

\subsection{Tensor perturbations}

The situation is different for tensor perturbations. Again we
consider the perfect fluid case, $\Pi_{ij}^{(T)}=0$. There
(\ref{tensdyn}) implies (if $\ka$ is negligible)
\be
H''_{ij}+\frac{2\b}{x}H'_{ij}+H_{ij}=0~, \label{tens}
\ee
with $\b=1$ in the radiation dominated era and $\b=2$ in the matter
dominated era. The less decaying mode solution to Eq.~(\ref{tens}) is
$H_{ij} =e_{ij}x^{1/2-\b}J_{1/2-\b}(x)$, where $J_\nu$ denotes the
Bessel function of order $\nu$ and $e_{ij}$ is a transverse traceless
polarization tensor. This leads to
\bea
H_{ij}&=&\mr{const}\quad\mr{for}\;x\ll1\\
H_{ij}&=&\frac{1}{a}\quad\mr{for}\;x\gsim1 .
\eea

\chapter{CMB anisotropies}

\section{Light-like geodesics}

After decoupling, $\eta>\eta_\mr{dec}$, photons follow to a good approximation
light-like geodesics. The temperature shift is then given by the energy
shift of a given photon.

The unperturbed photon trajectory follows
$(x^\mu)\equiv(\eta,\bn(\eta-\eta_0)+\bx_0)$, where $\bx_0$ is the
photon position at time $\eta_0$ and $\bn$ is the (parallel transported)
photon direction. With respect to a geodesic basis $\l({\bf e}\r)_{i=1}^3$,
the components of $\bn$ are constant. If $\ka=0$ we may choose
${\bf e}_i=\dd/\dd x^i$; if $\ka\neq0$ these vector fields
are no longer parallel transported and therefore do not form a
geodesic basis ($\nabla_{{\bf e}_i}{\bf e}_j=0$).

Our metric is of the form
\bea
d\bar{s}^2 &=& a^2 ds^2 \quad \mr{,with} \\
ds^2 &=& \l(\ga_{\mu\nu}+h_{\mu\nu}\r)dx^\mu dx^\nu ,
\quad \ga_{00}=-1, \, \ga_{i0}=0,\,\ga_{ij}=\ga_{ji}
\eea
as before.

We make use of the fact that light-like geodesics are conformally
invariant. More precisely $ds^2$ and $d\bar{s}^2$ have the same
light-like geodesics, only the corresponding affine parameters
are different. Let us denote the two
 affine parameters
by $\bar{\la}$ and $\la$ respectively, and the tangent vectors to the
geodesic by
\be n = \frac{dx}{d\la}, \,\,\,\, \;  \bar{n} =
\frac{dx}{d\bar{\la}} \;\;, \;\;\; n^2 = \bar{n}^2 = 0 \;, \;\;  n^0 =1
\;,\;\; {\bf n}^2 =1.
\ee
We set $n^0 = 1 +\de n^0$. The geodesic
equation for the perturbed metric
\be ds^2 =
(\ga_{\mu\nu}+h_{\mu\nu})dx^{\mu}dx^{\nu}  \ee
 yields, to first order,
\be
{d\over d\la}\de{n}^\mu = -\de\Ga^\mu_{\al\b}n^\al n^\b .
\ee
For the energy shift, we have to determine $\de n^0$. Since
$g^{0\mu}=-1\cdot \de_{0\mu}+\mr{first~order}$, we obtain
$\de\Ga^0_{\al\b}=-\nicefrac1/2 (h_{\al 0|\b}+h_{\b 0|\al}-\dot{h}_{\al\b})$,
so that
\be
{d\over d\la}\de{n}^0=h_{\al 0|\b}n^\b n^\al-
	\frac{1}{2}\dot{h}_{\al\b}n^\al n^\b .
\ee
Integrating this equation we use 
  $h_{\al 0|\b} n^\b={d\over d\la} (h_{\al 0}n^\al)$,
so that the change of $n^0$ between some initial time $\eta_i$ and some
final time $\eta_f$ is given by
\be \de n^0 |_i^f = \left[h_{00} + h_{0j}n^j\right]_i^f -
   {1\over 2}\int_i^f\dot{h}_{\mu\nu}n^{\mu}n^{\nu}d\la  \; .
\label{2deltan}
\ee
On the other hand, the ratio of the energy of a photon measured by
some observer at $t_f$ to the energy emitted at $t_i$ is
\be
{E_f\over E_i} = \frac{(\bar{n}\cdot u)_f}{(\bar{n}\cdot u)_i}
	= {T_f\over T_i}
     \frac{(n\cdot u)_f}{(n\cdot u)_i}  \; , \label{Ef/Ei}
\ee
where $u_f$ and $u_i$ are the four-velocities of the observer and
emitter
respectively, and the factor $T_f/T_i$ is the usual (unperturbed)
redshift, which relates
 $n$ and $\bar{n}$.
The velocity field of observer and emitter is given by
 \be u = (1-A)\dd_\eta +v^i\dd_i \; . \ee

An observer measuring
a temperature  $T_0$ receives photons that were emitted at the
time $\eta_{dec}$ of decoupling of matter and radiation, at the fixed
temperature $T_{dec}$. In first-order perturbation theory, we find the
following relation between the unperturbed temperatures $T_f$,
$T_i$,  the measurable temperatures $T_0$, $T_{dec}$, and the photon
density perturbation:
\be {T_f \over T_i} =
	{T_0\over T_{dec}}\left(1 - {\de T_f\over T_f} + {\de T_i
\over T_i}\right) =
    {T_0\over T_{dec}}\left(1 - {1\over 4}\de^{(r)}|_i^f\right) \; ,
\ee
where $\de^{(r)}$ is the intrinsic density perturbation in the
radiation and we used $\rho^{(r)}\propto T^4$ in the last
equality. Inserting the above equation 
and Eq.~(\ref{2deltan}) into
Eq.~(\ref{Ef/Ei}),
and  using Eq.~(\ref{defpert}) for the definition of $h_{\mu\nu}$,
one finds, after  integration by parts \cite{Rfund} the following
result for scalar perturbations:
\be
 {E_f\over E_i} = {T_0\over T_{dec}}\left\{1-\left[ {1\over
4}D^{(r)}_g +
	  V_j^{(b)}n^j  +\Psi-\Phi\right]_i^f +
   	\int_i^f(\dot{\Psi}-\dot{\Phi})d\la\right\}  \; .
\label{2deltaE}  \ee
Here $D_g^{(r)}$ denotes the density perturbation in the radiation
fluid, and  $V^{(b)}$ is the peculiar velocity of the baryonic matter
component (the emitter and observer of radiation).
The final time values in the square bracket of Eq. (\ref{2deltaE}) give
rise only to monopole contributions and to the dipole due to our
motion with respect to the CMB, and will be neglected in what
follows.

Evaluating Eq.~(\ref{2deltaE}) at final time $\eta_0$ (today) and
initial time $\eta_{dec}$, we obtain the temperature difference
of photons coming from different directions $\bf n$ and ${\bf n}'$
 \be {\De T\over T} \equiv
{\de T({\bf n})\over T}- {\de T({\bf n}')\over T},
\ee
with  temperature perturbation
\be
{\Delta T({\bf n})\over T} =\left[ {1\over 4}D^{(r)}_g +
V_{j}^{(b)}n^j
+\Psi -\Phi\right](\eta_{dec},{\bf x}_{dec})
   + \int_{\eta_{dec}}^{\eta_0}(\dot{\Psi}-\dot{\Phi})(\eta,{\bf
	x}(\eta))d\eta~, \label{dT0}
\ee
where ${\bf x}(\eta)$ is the unperturbed photon position at time $\eta$ for 
an observer at ${\bf x}_0$, and ${\bf x}_{dec}={\bf x}(\eta_{dec})$ (If 
$\ka=0$ we simply have ${\bf x}(\eta)={\bf x}_0-(\eta_0-\eta){\bf n}$.). h
The first term in Eq.~(\ref{dT0}) describes the intrinsic
inhomogeneities on the surface of  last scattering, due to acoustic
oscillations prior to decoupling. Depending on the initial conditions,
it can contribute significantly on super-horizon scales.
This is especially important in the case of adiabatic initial
conditions. As we have seen in Eq.~(\ref{Dads}), in a dust $+$ radiation
universe with $\Om=1$, adiabatic initial conditions imply $D_g^{(r)}(k,\eta)
=-20/3\Psi(k,\eta)$ and $V^{(b)}=V^{(r)}\ll D_g^{(r)}$ for 
$k\eta\ll 1$. With $\Phi=-\Psi$ the the square bracket of 
Eq.~(\ref{dT0}) gives 
\[
 \l({\Delta T({\bf n})\over T}\r)^{(OSW)}_{\rm adiabatic} = 
	{1\over 3}\Psi(\eta_{dec},{\bf x}_{dec})
\]
on super-horizon scales. The contribution to ${\delta T\over T}$ from
the last scattering surface on very large scales is called the
'ordinary Sachs--Wolfe effect' (OSW). It has been derived for the first
time by Sachs and Wolfe~\cite{SW}. 
For isocurvature perturbations,  the
initial condition $D_g^{(r)}(k,\eta)\ra 0$ for ${\eta\ra 0}$ is satisfied
and  the contribution of $D_g$ to the ordinary Sachs--Wolfe effect 
can be neglected. 
\[
 \l({\Delta T({\bf n})\over T}\r)^{(OSW)}_{\rm isocurvature} 
	= 2\Psi(\eta_{dec},{\bf x}_{dec})
\]
The second term in (\ref{dT0}) describes
the relative motions of emitter and  observer. This is the Doppler
contribution to the  CMB anisotropies. It appears on the same
angular scales as the acoustic term, and we thus call the sum of
the acoustic and Doppler contributions ``acoustic peaks''.

The last two terms are due to the inhomogeneities in the spacetime
geometry; the first contribution determines the change in the photon
energy due to the difference of the gravitational potential at the
position of emitter and observer. Together with the part contained in
$D_g^{(r)}$ they represent the ``ordinary'' Sachs-Wolfe  effect. The
integral accounts for red-shift or blue-shift caused by the
time dependence of the gravitational field along the  path of the
photon, and represents the so-called integrated Sachs-Wolfe (ISW)
effect. In a $\Om=1$, pure dust universe, the Bardeen potentials are
constant and there is no integrated Sachs-Wolfe effect; the blue-shift
which the photons acquire by falling into a gravitational potential is
exactly canceled by the redshift induced by climbing out of it. This
is no longer true in a universe with substantial radiation
contribution, curvature or a cosmological constant.

The sum of the ordinary Sachs--Wolfe term and the integral is the full
Sachs-Wolfe contribution (SW).

For {\bf vector} perturbations $\de^{(r)}$ and $A$ vanish and Eq.~(\ref{Ef/Ei})
leads to
\be (E_f/E_i)^{(V)} = (a_i/a_f)[1 - V_j^{(m)}n^j|_i^f +
  \int_i^f\dot{\si}_jn^jd\la]   ~. \label{2dev} \ee
We obtain a Doppler term and a gravitational contribution.
For {\bf tensor} perturbations, i.e. gravitational waves,  only the
gravitational part remains:
\be (E_f/E_i)^{(T)} = (a_i/a_f)[1 -   \int_i^f\dot{H}_{lj}n^ln^jd\la]
   ~. \label{2det} \ee
Equations (\ref{2deltaE}), (\ref{2dev}) and (\ref{2det}) are the
manifestly gauge invariant results for the Sachs--Wolfe effect for
scalar vector and tensor perturbations. Disregarding again the dipole
contribution due to our proper motion, Eqs.~(\ref{2dev},\ref{2det}) 
 imply the vector and tensor temperature fluctuations
\bea
\left({\Delta T({\bf n})\over T}\right)^{(V)} &=& 
  V_j^{(m)}(\eta_{dec},{\bf x}_{dec})n^j +
  \int_i^f\dot{\si}_j(\eta,{\bf x}(\eta))n^jd\la  \label{dTV}\\
\left({\Delta T({\bf n})\over T}\right)^{(T)} &=&
-   \int_i^f\dot{H}_{lj}(\eta,{\bf x}(\eta))n^ln^jd\la \label{dTT}
   ~.  \eea
Note that for models where initial fluctuations have been led down in
the very early universe, vector perturbations are irrelevant as we
have already pointed out. In this sense Eq.~(\ref{dTV}) is here
mainly for completeness. However, in models where perturbations are 
sourced by some inherently inhomogeneous component ({\em e.g.} topological
defects) vector perturbation can be important.

\section{Power spectra}

One of the basic tools to compare models of large scale structure
with observations are power spectra. They are the ``harmonic
transforms'' of the two point correlation functions. If the perturbations
of the model under consideration are Gaussian (a relatively generic
prediction from inflationary models), then the power spectra
contain the full statistical information of the model.

One  important power spectrum is the dark matter power
spectrum,
\be
P_D(k)=\l\lan\l| D_g^{(m)}\l(\bk,\eta_0\r)\r|^2\r\ran ,
\ee
where $\lan~\ran$ indicates a statistical average over ``initial
conditions'' in a given model. $P_D(k)$ is usually compared with
the observed power spectrum of the galaxy distribution.

Another power spectrum is given by the velocity perturbations,
\be
P_V(k)=\l\lan\l| {\bf V}\l(\bk,\eta_0\r)\r|^2\r\ran
\simeq H_0^2 \Om^{1.2}P_D(k)k^{-2}~.
\ee
For $\simeq$ we have used that 
$|kV|(\eta_0) = \dot{D}_g^{(m)}(\eta_0) \sim H_0\Om^{0.6}D_g$
on sub-horizon scales (see \eg~\cite{peebles}).

The power spectrum we are most interested in 
is the CMB anisotropy power spectrum. It is defined as follows:
$\De T/T$ is a function of position $\bx_0$, time $\eta_0$ and
photon direction $\bn$. We develop the $\bn$--dependence in
terms of spherical harmonics. We will suppress the argument $\eta_0$
and often also $\bx_0$ in the
following calculations. All results are for today ($\eta_0$) and here
($\bx_0$). By statistical homogeneity expectation values are supposed
to be independent of position. Furthermore, we assume that the process 
generating the initial perturbations is statistically isotropic. Then,
the off-diagonal correlators of the expansion coefficients $a_{\ell m}$
vanish and we have
\be
\frac{\De T}{T}\l(\bx_0,\bn,\eta_0\r)
=\sum_{\ell,m} a_{\ell m}(\bx_0) Y_{\ell m}(\bn), \quad
\l\lan a_{\ell m}\cdot a_{\ell'm'}^*\r\ran 
= \de_{\ell\ell'}\de_{mm'}C_\ell
\ee

The $C_\ell$'s are the CMB power spectrum. We assume that the
perturbations are generated by a {\em homogeneous} and {\em isotropic}
process, so that $C_\ell$ depends neither on $\bx_0$ nor on
$m$, and that $\l\lan a_{\ell m}\cdot a_{\ell'm'}^*\r\ran$
vanishes for $\ell\neq\ell'$ or $m\neq m'$.

Let us, at this point insert a comment on the problem of {\bf cosmic
variance}: Even if our 'ergodic hypothesis' is correct and we may
interchange ensemble and spatial averages, we cannot obtain very
precise averages for measurements of large scale characteristics, due
to the fact that we can observe only the universe around a given
position. For example, let us assume that temperature fluctuations are
Gaussian, as they are in most inflationary models. The functions
$a_{\ell m}$ are then also Gaussian distributed, and we have  a variance
of
\[ \left|{1\over 2\ell+1}\sum_{m=-\ell}^\ell |a_{\ell m}|^2
 -C_\ell\right| = |C_\ell^{obs}-C_\ell|={C_\ell\over 2\ell+1} ~,\]
on the average of the $2\ell+1$ values $a_{\ell m}$ which can in principle be 
measured from one point with full sky coverage. For simplicity, we neglect 
the increase of the variance due to the fact that our own milky ways
blocks a portion of sky of about 20\%. Wick's theorem now gives
\[ \lan C_{\ell}^2\ran -  \lan C_{\ell}\ran^2= 
 \lan |a_{\ell m}|^4\ran -  \lan |a_{\ell m}|^2\ran^2= 
   2\lan |a_{\ell m}|^2\ran^2 \] 
For a given multipole $\ell$ we
then expect a variance of the $C_\ell$'s
\be
  {\sqrt{(C_\ell^{obs})^2 -C_\ell^2} \over C_\ell} =\sqrt{2\over 2\ell+1}~,
\ee
in real experiments, this 'cosmic variance' is in general much larger due to 
the limited sky coverage.

The two point correlation function is related to the $C_\ell$'s by
\bea
\l\lan \frac{\De T}{T}(\bn)\frac{\De T}{T}(\bn')\r\ran_{\bn\cdot\bn'=\mu}
=  \sum_{\ell,\ell',m,m'}\l\lan a_{\ell m}\cdot a_{\ell'm'}^*\r\ran
	 Y_{\ell m}(\bn) Y_{\ell m}^*(\bn') =  \nonumber \\
 \sum_\ell C_\ell \underbrace{\sum_{m=-\ell}^\ell 
   Y_{\ell m}(\bn) Y_{\ell m}^*(\bn')}_{\frac{2\ell+1}{4\pi}
   P_\ell(\bn\cdot\bn')}
= \frac{1}{4\pi}\sum_\ell (2\ell+1)C_\ell P_\ell(\mu) ,\label{correl}
\eea
where we have used the addition theorem of spherical harmonics
for the last equality.  The $P_\ell$'s are the Legendre polynomials.

Clearly the $a_{lm}$'s from scalar, vector and tensor perturbations
are uncorrelated,
\be
\l\lan a_{\ell m}^{(S)} a_{\ell'm'}^{(V)} \r\ran
=\l\lan a_{\ell m}^{(S)} a_{\ell'm'}^{(T)} \r\ran
=\l\lan a_{\ell m}^{(V)} a_{\ell'm'}^{(T)} \r\ran
=0 .
\ee

Since vector perturbations decay, their contributions, the $C_\ell^{(V)}$,
are negligible in models where initial perturbations have been
laid down very early, \eg, after an inflationary period. Tensor
perturbations are constant on super-horizon scales and perform damped
oscillations once they enter the horizon.

Let us first discuss in somewhat more detail scalar perturbations.
We specialize to the case $\ka=0$ for simplicity.
We suppose the initial perturbations to be given by a spectrum,
\be
\l\lan\l|\Psi\r|^2\r\ran k^3=A^2 k^{n-1}\eta_0^{n-1} . \label{inspec}
\ee
We multiply by the constant $\eta_0^{n-1}$, the actual comoving size of the 
horizon, in order to keep $A$
dimensionless for all values of $n$. $A$ then represents the amplitude of 
metric perturbations at horizon scale today, $k=1/\eta_0$.

On {\em super-horizon scales} we have, for {\em adiabatic} perturbations:
\be
\frac{1}{4}D_g^{(r)} = -\frac{5}{3} \Psi +\OO(x^2), \quad
V^{(b)}=V^{(r)}=\OO(x)
\ee

The dominant contribution on super-horizon scales (neglecting the 
integrated Sachs--Wolfe effect $\int \dot{\Phi}-\dot{\Psi}$~) is then
\be
\frac{\De T}{T}(\bx_0,\bn,\eta_0) = \frac{1}{3} \Psi(x_\mr{dec}, 
\eta_\mr{dec}).		\label{sw}
\ee

The Fourier transform of (\ref{sw}) gives
\be
\frac{\De T}{T}(\bk,\bn,\eta_0)  = \frac{1}{3} \Psi(k, \eta_\mr{dec}) \cdot
	e^{i\bk\bn\l(\eta_0-\eta_\mr{dec}\r)}~.
\ee

Using the decomposition
\[
e^{i\bk\bn\l(\eta_0-\eta_\mr{dec}\r)} =
   {\sum_{\ell=0}^\infty (2\ell+1)i^\ell
   j_\ell(k(\eta_0-\eta_\mr{dec})) P_\ell( \widehat{\bk}\cd\bn)}~,
\]
where $j_\ell$ are the spherical 
Bessel functions, we obtain
\bea
\lefteqn{\l\lan \frac{\De T}{T}(\bx_0,\bn,\eta_0)
 \frac{\De T}{T}(\bx_0,\bn',\eta_0) \r\ran}\\
 &=& \frac{1}{V} \int d^3x_0 \l\lan\frac{\De T}{T}(\bx_0,\bn,\eta_0)
 \frac{\De T}{T}(\bx_0,\bn',\eta_0) \r\ran \nonumber \\
&=&\frac{1}{(2\pi)^3}\int d^3k \l\lan\frac{\De T}{T}(\bk,\bn,\eta_0)
 \l(\frac{\De T}{T}\r)^*(\bk,\bn',\eta_0) \r\ran  \nonumber\\
& =& \frac{1}{(2\pi)^3 9}\int d^3k \l\lan\l|\Psi\r|^2\r\ran 
	\sum_{\ell,\ell'=0}^\infty
 (2\ell+1)(2\ell'+1)j_\ell(k(\eta_0-\eta_\mr{dec}))
 j_\ell'(k(\eta_0-\eta_\mr{dec})) i^{\ell-\ell'} \nonumber \\ &&  \cdot
 P_\ell(\hat{\bk}\cd\bn) \cdot
P_\ell'(\hat{\bk}\cd\bn')~.
\eea
Inserting $P_\ell(\hat{\bk}\bn) =
 \frac{4\pi}{2\ell+1} \sum_m Y_{\ell m}^*(\hat{\bk})Y_{\ell m}(\bn)$
and  $P_\ell'(\hat{\bk}\bn')=
 \frac{4\pi}{2\ell'+1}\sum_{m'} Y_{\ell' m'}^*(\hat{\bk})Y_{\ell' m'}(\bn')$,
integration over the directions $d\Om_{\hat{k}}$ 
 gives $\de_{\ell\ell'}\de_{mm'}\sum_m Y_{\ell m}^*(\bn)Y_{\ell m}(\bn')$.
Using as well $\sum_m Y_{\ell m}^*(\bn)Y_{\ell m}(\bn')=\frac{2\ell+1}{4\pi}
P_\ell(\mu)$, where  $\mu=\bn\cdot\bn'$, we find
\bea
\lefteqn{\l\lan \frac{\De T}{T}(\bx_0,\bn,\eta_0)
 \frac{\De T}{T}(\bx_0,\bn',\eta_0) \r\ran_{\bn\bn'=\mu}
 =} \nonumber \\ &&
\sum_\ell \frac{2\ell+1}{4\pi} P_\ell(\mu) \frac{2}{\pi}
\int\frac{dk}{k} \l\lan\frac{1}{9}|\Psi|^2\r\ran k^3
j_\ell^2(k(\eta_0-\eta_\mr{dec})) .
\eea

Comparing this equation with~Eq.~(\ref{correl})
we obtain for {\em adiabatic perturbations}
on scales $2\le \ell$ $\ll$ 
$\chi(\eta_0-\eta_\mr{dec})/\eta_\mr{dec}$ $\sim 100$
\be
C_\ell^{(SW)} \simeq C_\ell^{(OSW)} \simeq 
 \frac{2}{\pi} \int_0^\infty \frac{dk}{k}
 \l\lan\l|\frac{1}{3}\Psi\r|^2
 \r\ran k^3 j_\ell^2 \l(k\l(\eta_0-\eta_\mr{dec}\r)\r).
\label{clsw}
\ee

If $\Psi$ is a pure power law and we set $k(\eta_0-\eta_\mr{dec})\sim
k\eta_0$,  the integral (\ref{clsw}) can be performed
analytically. For the ansatz (\ref{inspec}) one finds for $-3<n<3$
\be
C_\ell^{(SW)} = \frac{A^2}{9} \frac{\Ga(3-n)\Ga(\ell-\frac{1}{2}+\frac{n}{2})}{
2^{3-n}\Ga^2(2-\frac{n}{2})\Ga(\ell+\frac{5}{2}-\frac{n}{2})} . \label{clswsol}
\ee

Of special interest is the {\em scale invariant} spectrum, $n=1$. This spectrum
with a time and scale independent gravitational potential has first been
investigated by Harrison and by Zel'dovich \cite{HZ}. It is therefore
called the Harrison--Zel'dovich spectrum. It leads to
\be
\ell(\ell+1)C_\ell^{(SW)} = \mr{const.} \simeq
\l\lan\l(\frac{\De T}{T}(\vth_\ell)\r)^2\r\ran~,~~~~ 
	\vth_\ell\equiv \pi/\ell~.
\ee
This  is precisely (within the accuracy of the experiment) the
behavior  observed by the DMR experiment aboard
COBE \cite{DMR}.

Inflationary models predict very generically a HZ spectrum (up to
small corrections). The DMR discovery has therefore been
regarded as a great success, if not a proof, of inflation.
There are however other models like topological defects \cite{PST,ZD,Aetal}
or certain string
cosmology models \cite{dgsv} which also predict scale--invariant,
\ie, Harrison Zel'dovich spectra of fluctuations. These models do
however not belong to the class investigated here, since in these
models perturbations are induced by seeds which evolve non--linearly
in time.

For isocurvature perturbations, the main contribution on large
scales comes from the integrated Sachs--Wolfe effect and (\ref{clsw})
is replaced by
\be
C_\ell^{(ISW)} \simeq \frac{8}{\pi} \int \frac{dk}{k}
k^3 \l\lan \l| \int_{\eta_\mr{dec}}^{\eta_0}  \dot{\Psi}(k,\eta)
j_\ell^2(k(\eta_0-\eta))d\eta\r|^2\r\ran . \label{clisw}
\ee
Inside the horizon $\Psi$ is roughly constant (matter dominated).
Using the ansatz (\ref{inspec}) for $\Psi$ inside the horizon
and setting the integral in (\ref{clisw})
$\sim2\Psi(k,\eta=1/k) j_\ell^2(k\eta_0)$, we obtain again
(\ref{clswsol}), but with $A^2/9$ replaced by $4A^2$. The Sachs--Wolfe
temperature anisotropies coming from isocurvature perturbations are
therefore about a factor of $6$ times larger than those coming from
adiabatic perturbations.

On smaller scales, $\ell\gsim 100$ the contribution to $\De T/T$
is usually dominated by acoustic oscillations, the first two terms in
Eq.~(\ref{dT0}).  Instead of 
(\ref{clisw}) we then obtain
\bea
\lefteqn{C_\ell^{(AC)}\simeq} \nonumber \\
&& \frac{2}{\pi} \int_0^\infty \frac{dk}{k}
k^3\l\lan\l|\frac{1}{4}D_g^{(r)}(k,\eta_\mr{dec})j_\ell(k\eta_0)
+V^{(r)}(k,\eta_\dec)j_\ell'(k\eta_0)\r|^2\r\ran~.
\eea

On sub-horizon scales $D_g^{(r)}$ and $V^{(r)}$ are oscillating
like sine or cosine waves depending on the initial conditions.
Correspondingly the $C_\ell^{(AC)}$ will show peaks and
minima. On very small scales they are damped by the photon
diffusion which takes place during the recombination process (see next
section).

\begin{figure}[ht]
\centerline{\epsfig{file=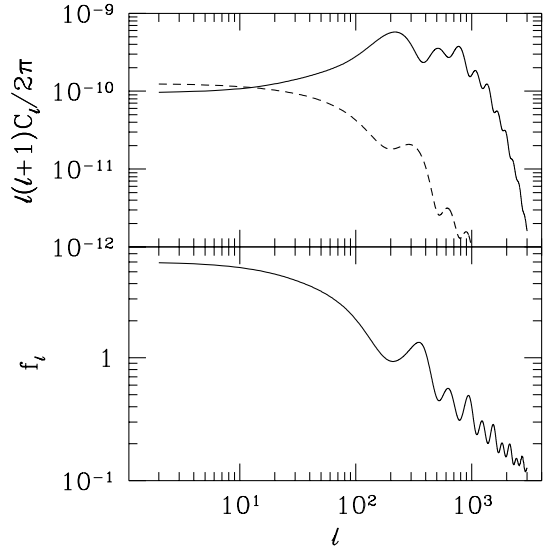,width=8cm}}
\caption{A COBE normalized sample adiabatic( solid line) and isocurvature 
(dashed line) CMB anisotropy spectrum,  $\ell(\ell+1)C_\ell$, are shown on the 
top panel. The quantity
shown in the bottom panel is the ratio of temperature fluctuations for fixed 
value of $A$ (from Kanazawa et al.~\protect\cite{Kanaza}).}\label{adisofig}
\end{figure}
%\begin{figure}[ht]
%\centerline{\epsfig{file=Peefig1.eps,width=8cm}}
%\caption{\label{isofig}Some sample isocurvature 
% CMB anisotropy spectra are shown. The variable
%$T_\ell= T_0\sqrt{\ell(\ell+1)C_\ell/(2\pi)}$ is plotted for the Peebles model
%(from \protect\cite{Peemod}).}\label{Isoc}
%\end{figure}

For gravitational waves (tensor fluctuations), a formula analogous to
(\ref{clswsol}) can be derived (see appendix),
\be
C_\ell^{(T)}=\frac{2}{\pi}\int dk k^2 \l\lan\l|\int_{\eta_\dec}^{\eta_0}
d\eta \dot{H}(\eta,k) \frac{j_\ell(k(\eta_0-\eta))}{(k(\eta_0-\eta))^2}
\r|^2\r\ran\frac{(\ell+2)!}{(\ell-2)!} .
\ee

To a very crude approximation we may assume $\dot{H}=0$ on super-horizon
scales and $\int d\eta \dot{H}j_\ell(k(\eta_0-\eta)) 
	\sim H(\eta=1/k)j_\ell(k\eta_0)$. For a pure power
law,
\be
k^3\l\lan\l|H(k,\eta=1/k)\r|^2\r\ran=A_T^2 k^{n_T}\eta_0^{-n_T} ,
\ee
this gives
\bea
C_\ell^{(T)} &\simeq & \frac{2}{\pi} \frac{(\ell+2)!}{(\ell-2)!} A_T^2
\int \frac{dx}{x} x^{n_T} \frac{j_\ell^2(x)}{x^4} \nonumber \\
&=& \frac{(\ell+2)!}{(\ell-2)!} A_T^2 
	\frac{\Ga(6-n_T)\Ga(\ell-2+\frac{n_T}{2})}{
    2^{6-n_T}\Ga^2(\frac{7}{2}-n_T)\Ga(\ell+4-\frac{n_T}{2})} .
\eea
For a scale invariant spectrum ($n_T=0$) this results in
\be
\ell(\ell+1)C_\ell^{(T)} \simeq \frac{\ell(\ell+1)}{(\ell+3)(\ell-2)}
	A_T^2{8\over 15\pi}~ .
\ee
The singularity at $\ell=2$ in this crude approximation is not real,
but there is some enhancement of $\ell(\ell+1)C_\ell^{(T)}$
at $\ell\sim2$.

Since tensor perturbations decay on sub-horizon scales, $\ell \gsim 60$,
they are not very sensitive to cosmological parameters.

Again, inflationary models (and topological defects) predict a scale
invariant spectrum of tensor fluctuations ($n_T \sim 0$).

On very small angular scales, $\ell\gsim 800$, fluctuations are damped
by collisional damping (Silk damping). This effect has to be discussed
with the Boltzmann equation for photons derived in the next section.

\begin{figure}[ht]
\centerline{\epsfig{file=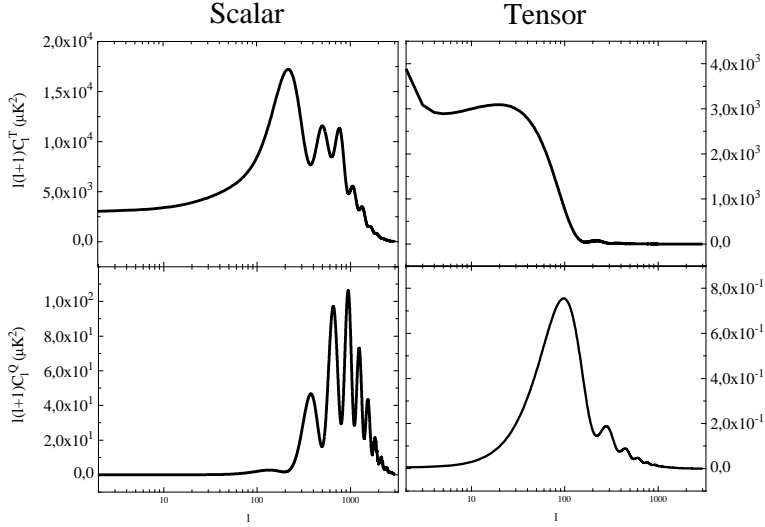,width=8cm,angle=-90}}
\caption{\label{STfig}Adiabatic scalar and tensor 
 CMB anisotropy spectra are shown (top panels).
The bottom panels show the corresponding polarization spectra (see
Section~\ref{pola}). 
(from \protect\cite{melvit}).}
\end{figure}

\section{The Boltzmann equation}
\subsection{Elements of the derivation}
When particles are not very tightly coupled, the fluid approximation
breaks down and they have to be described by a Boltzmann equation,
\be
p^\mu \dd_\mu f- \Ga^i_{\al\b} p^\al p^\b \frac{\dd f}{\dd p^i}
= C[f] ~.\label{beq}
\ee
 $C[f]$ is a collision integral which describes the interactions
with other matter components. The left hand side of (\ref{beq}) just
requires the particles to move along geodesics in the absence of
collisions.

Let us first consider the situation where collisions are negligible,
$C[f]=0$. The unperturbed Boltzmann equation implies that
$f$ be a function of $v=a p$ only. Setting
$f=\bar{f}(v)+F(\eta,\bx,v,\bn)$, where $\bn$ denotes the momentum
directions, leads then to the perturbation equation
\be
\dd_\eta F - n^i\dd_i F-\Ga^{(S)\,i}_{jk} n^j n^k \frac{\dd F}{\dd n^i}
= v \frac{d\bar{f}}{dv}\l[n^i A_{,i}-n^i n^j\l(B_{i|j}-\dot{H}_{ij}\r)
+H_L\r]. \label{bpertu}
\ee
Here $\Ga^{(S)\,i}_{jk}$ are the Christoffel symbols of the space
of constant curvature $\ka$.

To derive (\ref{bpertu}), we have used $p^2=0$. The Liouville
equation for particles with non--vanishing mass can be found
in Ref.~\cite{Rfund}.

The ansatz
\be
f(x,\bp) = \bar{f}\l(\frac{g^{(3)}(\bp,\bp)^\frac{1}{2}}{T(x,\bn)}\r)
=\bar{f}\l(\frac{\overline{T}v}{T(x,\bn)}\r)  \label{fdistr}
\ee
with $T(x,\bn)=\overline{T}(\eta)+\De T(x,\bn)$ leads to
\be
f=\bar{f}-v \frac{d\bar{f}}{dv}\frac{\De T}{\overline{T}}.  \label{fT}
\ee

Integrating (\ref{fdistr}) over photon energies, we can also write
\be
\frac{\De T}{T} = \frac{1}{4} \imath ,
\ee
where $\imath$ is the brightness perturbation,
\be
\imath=\frac{4\pi}{\bar{\rho}a^4}\int_0^\infty F v^3 dv .
\ee

Comparing Eq.~(\ref{fT}) with  (\ref{bpertu}), we find
\be
\dd_\eta \l(\frac{\De T}{T}\r) + n^i \dd_i \l(\frac{\De T}{T}\r)
-\Ga^{(S)\,i}_{jk} n^j n^k \frac{\dd \l(\frac{\De T}{T}\r)}{\dd n^i}
= -\l[n^i A,_i-\l(B_{i|j}-\dot{H}_{ij}\r)n^i n^j+H_L\r]. \label{tpertu}
\ee

The fact that gravitational perturbations of Liouville's equation can
be cast purely in {\em temperature perturbations} of the original distribution
is not astonishing. This is just an expression of gravity being ``achromatic'',
\ie~independent of the photon energy.

We now decompose (\ref{tpertu}) into scalar, vector and tensor
components. Even though $\De T/T$ is just a function, it can be
represented in the form
\be
\frac{\De T}{T}(\bx,\bn) = \sum_{\ell=0}^\infty \al_{i_1,\ldots,i_\ell}(\bx)
n^{i_1} \cdots n^{i_\ell} ,
\ee
where the $\al_{i_1,\ldots,i_\ell}$ are symmetric traceless tensor
fields that contain scalar, vector, 2--tensor and in principle also
higher tensor components. Since spin components larger than 2 are
not sourced by the right hand side of equation (\ref{tpertu}) and
since they are suppressed at early times, when collisions
are important, we neglect them here.

For the {\em scalar} contribution to $\De T/T$ we obtain from (\ref{tpertu})

\bea
\dd_\eta \l(\frac{\De T}{T}\r)^{(S)} + 
	 k\mu \l(\frac{\De T}{T}\r)^{(S)}
	-\Ga^{(S)\,i}_{jk} n^j n^k 
	\frac{\dd \l(\frac{\De T}{T}\r)^{(S)}}{\dd n^i} = \nonumber \\
-\l[k\mu A+ \mu^2k^2 \l(B-\dot{H}_T\r)+H_L+
	\frac{1}{3} k^2\dot{H}_T\r],
\eea
where we have introduced the ``direction cosine'' $\mu$ defined by
$n^iY,_i=k\mu Y$. Note that in flat space, $\ka=0$, we have just
$\mu=i\hat{\bm{k}}\cd \bm{n}$.

This equation is not manifestly gauge--invariant. However, setting
\be
\MM^{(S)}=\l(\frac{\De T}{T}\r)^{(S)}+H_L+\frac{1}{3}k^2 H_T
+k\mu\l(\dot{H}-B\r) ,
\ee
it reduces to
\be
\dd_\eta\MM^{(S)}+k \mu\MM^{(S)}
-\Ga^{(S)\,i}_{jk} n^j n^k \frac{\dd \MM^{(S)}}{\dd n^i}
=k \mu\l(\Phi-\Psi\r) , \label{scalbright}
\ee
where $\Phi$ and $\Psi$ are the Bardeen potentials. If $n^j$ 
are components with respect to a geodesic basis (or $\ka=0$), the third
term on the left hand side of Eq.~(\ref{scalbright}) vanishes. 
For simplicity we now concentrate on the case $\ka=0$. We can
then integrate the equation and obtain
\bea
\MM^{(S)}(\eta_0,\bn,\bk) &=&
   \exp[i\bk\cd\bn(\eta_{in}-\eta_0)]\MM^{(S)}(\eta_{in},\bn,\bk) 
 \nonumber \\ &&  +
   \int_{\eta_{in}}^{\eta_0} i\exp[i\bk\cd\bn(\eta-\eta_0)] \bn\cd\bk
   \l(\Phi-\Psi\r) d\eta ~.
\eea
Integration by parts and neglecting the monopole term
$\l(\Phi-\Psi\r)(\eta_0)$, leads to
\bea
\lefteqn{\MM^{(S)}(\eta_0,\bn,\bk) =}  \nonumber \\
&&   \exp[i\bk\cd\bn(\eta_{in}-\eta_0)] \l[
\MM^{(S)}(\eta_{in},\bn,\bk) 
  + \l(\Phi-\Psi\r)(\eta_{in},\bk)\r] \nonumber \\ &&
 - \int_{\eta_{in}}^{\eta_0} \exp[i\bk\cd\bn(\eta-\eta_0)]
  \l(\dot\Phi-\dot\Psi\r) d\eta ~. \label{dTLiou}
\eea
Comparing this equation with (\ref{dT0}), we see again that
$\MM^{(S)}=\l(\frac{\De T}{T}\r)^{(S)}$ (up to gauge dependent
monopole and dipole contributions) if the initial condition is
\[ \MM^{(S)}(\eta_{in},\bn,\bk) = {1\over 4}D_g^{(r)}(\eta_{in},\bk)
	+\bn\cd\bk V^{(b)}(\eta_{in},\bk) ~,\]
which is equivalent to require that $\MM^{(S)}(\eta_{in})$ has no
higher than first moments. As we shall see below, this assumption is
quite reasonable since collisions suppress the build up of higher moments
before recombination.

Since the right hand side of (\ref{scalbright}) is gauge invariant,
the left hand side must be so as well and we conclude that
$\MM^{(S)}$ is a gauge--invariant variable (a direct proof of this,
analyzing the gauge transformation properties of the distribution
function, can be found in Ref.~\cite{Rfund}).

$\MM^{(S)}$ used in this work coincides with the scalar temperature fluctuations
up a to a gauge dependent monopole and dipole contribution. In other
work~\cite{HS} the gauge invariant variable $\Th \equiv \MM^{(S)}+\Phi$ has
been used.  Since $\Phi$ is independent of the photon direction $\bn$
this difference in the definition shows up only in the monopole,
$C_0$. 

The vector and tensor parts of  $\De T/T$ are gauge--invariant
by themselves and we denote them by $\MM^{(V)}$ and $\MM^{(T)}$ for
consistency. 
In the absence of collisions and with vanishing spatial curvature, they 
satisfy the equations
\bea
\dot{\MM}^{(V)} +i\mu k\MM^{(V)}
&=& -i n^\ell n^m k_\ell \si_m^{(V)}\\
\dot{\MM}^{(T)} +i\mu k\MM^{(T)}
&=& -i n^\ell n^m \dot{H}_{m\ell} .
\eea

The components of the energy momentum tensor are obtained by
integrating the second moments of the distribution function 
over the mass shell,
\be
T^{\mu\nu} = \int_{P_m(x)} p^\mu p^\nu
f(p,x){p^2dpd\Om_{\hat\mr{p}}\over p^0}~,
\ee
where $\Om_{\hat\mr{p}}$ denotes the angular integration over momentum
directions. 
One finds for $\ka=0$ and setting $\mu=\bn\cdot\hat{\bk}$:
\bea
D_g^{(r)} &=& \frac{1}{\pi}\int\MM^{(S)} d\Om \\
V^{(r)} &=& \frac{3 i}{4\pi}\int \mu \MM^{(S)} d\Om \label{Vrad}\\
\Pi^{(r)} &=& \frac{9}{2\pi}\int \l(\mu^2-\frac{1}{3}\r) \MM^{(S)} d\Om \\
\Ga^{(r)} &=& 0 \\
V^{(V)}_i &=& \frac{1}{4\pi} \int n_i \MM^{(V)} d\Om \\
\Pi_j^{(V)} &=& \frac{6}{\pi} \int \mu n_j \MM^{(V)} d\Om \\
\Pi_{ij}^{(T)} &=& \frac{3}{\pi} \int n_i n_j \MM^{(T)} d\Om .
\eea

Let us now turn to the collision term. At recombination (when
the fluid description of radiation breaks down) the temperature
is $\sim 0.4 \mr{~eV}$. The electrons and nuclei are non--relativistic
and the dominant collision process is non--relativistic Thomson
scattering. Since collisions are important only before and during
recombination, where curvature effects are entirely negligible, we set
$\ka=0$ in the reminder of this section.

The collision term is given by
\be
C[f]=\frac{df_+}{d\eta}-\frac{df_-}{d\eta} ,
\ee
where $f_+$ and $f_-$ denote the distribution of photons scattered into
respectively out of the beam due to Compton scattering.

In the matter  (baryon/electron) rest frame,
which we indicate by  a prime, we know
\[ {df'_+\over dt'}(p,\bn)= {\si_Tn_e\over 4\pi}\int
      f'(p',\bn')\om(\bm{ n, n}')d\Om'  \; , \]
where $n_e$ denotes the  number density of free electrons, $\si_T$ is 
the Thomson cross section,  and $\om$ is the normalized
angular dependence of the Thomson cross section:
\[ \om(\bm{ n, n}') = 3/4[1 + (\bm{ n\cd n}')^2] =
  1 + {3\over 4} n_{ij} n'_{ij} ~~~\mbox{ with }~~
   n_{ij} \equiv  n_i n_j - {1\over 3}\de_{ij}  \; . \]
In the baryon rest frame which moves with four velocity $u$, the photon
energy is given by
\[ p' = p_\mu u^\mu \; . \]
We denote by $p$ the photon energy with respect to a tetrad adapted to
the slicing of spacetime into $\{ \eta=$ constant$\}$  hyper--surfaces:
\[ p =  p_\mu n^\mu \; ,~~~\mbox{ with }~~
  n = a^{-1}[(1-A)\dd_\eta +B^i\dd_i] ~.\]
The unit vector $n$ is the normal the the hyper-surfaces of constant time.
The lapse function and  the shift vector of the slicing are given by
  $\al= a(1+A)$ and  $\bm{\beta}=
	-B^{i}\dd_i$ .  In first order,
\[ p_0 = ap(1+A) - ap n_iB^i~~,\]
and in zeroth order, clearly,
\[ p_i = ap n_i ~ .\]
Furthermore, the baryon four velocity has the form
$ u^0 = a^{-1}(1-A)~~,~~~ u^i = u^0v^i $ up to first order.
This yields
\[ p' = p_\mu u^\mu = p(1+ n_i(v^i-B^i)) \; . \]
Using this identity and performing the integration over photon energies,
we find
\[ \rho_r{d\io_+( n)\over dt'} = \rho_r\si_Tn_e[1+4 n_i(v^i-B^i) +
 {1\over 4\pi}\int\io( n')\om( n, n')d\Om'] ~ .\]
The distribution of photons scattered out of the beam, has the well known
form \\ (see e.g. Lifshitz and Pitajewski [1983])
\[ {df_-\over dt'} = \si_Tn_ef'(p',\bn) ~, \]
so that we finally obtain
\[ C' = {4\pi\over\rho_r a^4} \int dp({df_+\over dt'} -{df_-\over dt'})p^3 =
\si_Tn_e[
   \de_r -\io + 4 n_i(v^i-B^i) + {3\over 16\pi} n_{ij}\int
  \io( n') n'_{ij}d\Om'] ~ , \]
where $\de_r = (1/4\pi)\int\io( n)d\Om$ is the photon energy density
perturbation.\\
Using the definitions of the gauge--invariant variables $\MM^{(S)}$ ~and
  $V^{(b)}$ for the photon brightness perturbation and the baryon
  velocity potential, we
can write $C'$ in gauge--invariant form.
\be C'  = 4\si_Tn_e[{1\over 4}D_g^{(r)} -\MM^{(S)} +  n^iV^{(b)}_i +
              {1\over 2} n_{ij}M^{ij}]~,\label{BC} \ee
with $D_g^{(r)} = (1/\pi)\int\MM^{(S)} d\Om$
and  \[ M^{ij}   \equiv {3\over 8\pi}\int\MM^{(S)}( n') n'_{ij}d\Om'~. \]
Since the term in square brackets of (\ref{BC}) is already first order
we can set $dt'=dt$ which yields
$C = {dt'\over d\eta}C' = {dt\over d\eta}C'= aC'$.
The  Boltzmann equation for scalar perturbations expressed in terms of
the gauge invariant variable $\MM^{(S)}$ then becomes
\be \dot{\cal M}^{(S)} + n^i\dd_i\MM^{(S)} =  n^i\dd_i(\Phi-\Psi) +
    a\si_Tn_e[{1\over 4}D_g^{(r)}-\MM^{(S)} -n^i\dd_iV^{(b)} + 
   {1\over 2} n_{ij}M^{ij}]
	 \; . \label{3B} \ee 

For vector and tensor perturbations we obtain in the same way
\bea
\hspace{-5mm} \dot{\MM}^{(V)}+i\mu k\MM^{(V)} &=& -n^i n^j \si_{i|j}
	+a \si_T n_e\l[n^i V^{(Vb)}_i 
	+\frac{1}{2} n^{ij} M_{ij}^{(V)}-\MM^{(V)}\r]  \label{3BV}\\
\hspace{-5mm} \dot{\MM}^{(T)}+i\mu k\MM^{(T)} &=& -n^i n^j \dot{H}_{ij}
	+a \si_T n_e\l[n^{ij} M_{ij}^{(T)} 
	-\MM^{(T)}\r]~.\label{3BT}
\eea

\subsection{The tight coupling limit}
Before recombination, when $n_e\simeq \rho_b/m_p$,
\bea
\eta_T \equiv \frac{1}{a\si_T n_e} &\sim & 
\frac{10}{\Om_b h}(1+z)^{-\frac{3}{2}} \eta \ll \eta, \quad 
	z_{\rm eq}\gsim z\gsim z_\dec, \\
 &\sim & \frac{10}{\Om_b h}(1+z_{\rm eq})^{-\frac{1}{2}}(1+z)^{-1}\eta
 \quad  z\gsim z_{\rm eq}~.
\eea

To lowest order in $\eta_T$, collisions force the photon distribution 
to be of the form
\be
\MM^{(S)} = \frac{1}{4}D_g+n^i V^{(b)}_i +\frac{1}{2}n^{ij}M_{ij} ,\label{3r1}
\ee
the building up of higher moments is strongly suppressed by collisions.

During recombination, the number density of free electrons, $n_e$,
decreases rapidly and the collision term becomes less and less
important. Higher moments in the photon distribution develop by
free streaming.

The collision term $C[\MM^{(S)}]$ of equation (\ref{3B}) also appears in the
equation of motion of the
baryons as a drag. The Thomson drag force is given by
\be F_j = {\rho_r\over 4\pi}\int C[\MM^{(S)}] n_jd\Om =
	 {-4a\si_Tn_e\rho_r\over 3}(M_j +V^{(b)}_i) ~,\label{3drag}\ee
\[  \mbox{ with }~~~ M_j ={3i\over 4\pi}\int n_j\MM^{(S)} d\Om ~.\]
This yields the following  scalar baryon equation of motion  in an 
ionized plasma
\be \dot{V}^{(b)} + (\dot{a}/a)V^{(b)} = k\Psi -
  {4a\si_Tn_e\rho_r\over 3\rho_b}(-\hat k^jM_j +V^{(b)}) ~, \label{3bar}
\ee
where we have added the drag force to 
the second eq. of (\ref{scalcons}) with $w=c_s^2=0$.

We now want to discuss equations (\ref{3B},\ref{3bar}) in the limit of very
many collisions. The comoving photon mean free path  is given by
$\eta_T =l_T = (a\si_Tn_e)^{-1}$. In lowest order $\eta_T/ \eta$ and $l_T/\la$,
\footnote{Here $\la$ is a typical size of a perturbation. For a given
Fourier mode $k$, it is $\la \sim \pi/k$.}
$\MM^{(S)}$ is given by (\ref{3r1}), and Eq.~(\ref{3bar}) implies
\[ -\hat k^jM_j + V^{(b)} = 0~.\]
Inserting the solution (\ref{3r1}) in the Boltzmann equation~(\ref{3B})
and integrating over directions this implies
\be 
kV^{(b)} = k^jM_j = k V^{(r)} = {-3\over 4}\dot{D}_g^{(r)}
~, \label{3r2} \ee
Implying especially $ V^{(b)} = V^{(r)} \equiv V$. 
Eq. (\ref{3r2}) is equivalent to the energy conservation equation 
(\ref{scalcons}) for radiation.
Using also (\ref{scalcons}) for baryons, $w=0$, we obtain
\[ \dot{D}_g^{(r)} ={-4k\over 3} V^{(b)} = {4\over 3}\dot{D}_g^{(b)} .\]
This shows that entropy per baryon is conserved, $\Ga=0$. Before
recombination, when the collisions are sufficiently frequent, baryons
and photons are adiabatically coupled.
Inserting (\ref{3r1}) in (\ref{3B}) we find up to first order in $\eta_T$
\bea \MM^{(S)} &=& D_g^{(r)} -4i n^jk_jV +{1\over 2} n_{ij}M^{ij}
	-\eta_T[\dot{D}_g^{(r)}
	-4i n^jk_j\dot{V} +{1\over 2} n_{ij}\dot{M}^{ij} \nonumber\\
	&&  + in^jk_jD_g^{(r)}
	+4 n^i n^jk_ik_jV +{i\over 2} n^i n_{mj}k_iM^{mj}
	-i4 n^jk_j(\Phi-\Psi)]~ .   \label{3mm1} \eea
Using (\ref{3mm1}) to calculate the drag force yields
\[ F_j = ik_j(\rho_r/3)[4\dot{V} -D_g^{(r)} + 4(\Phi-\Psi)] ~.\]
Inserting $F_j$ in (\ref{3bar}), we obtain
\[ (\rho_b +(4/3)\rho_r)\dot{V} +\rho_b(\dot{a}/a)V =
   (\rho_r/3)D_g^{(r)} + (\rho_b+(4/3)\rho_r)\Psi -
	(4\rho_r/3)\Phi ~.\]
This is equivalent to momentum conservation, the second equation of
(\ref{scalcons}) for $\rho=\rho_b+\rho_r$,
$p=\rho_r/3$  and $\Ga=\Pi =0$, if we use
\[ D_g^{(r)}= (4/3)D_g^{(b)} ~~~\mbox{ and }~~~~ D_g =
	{\rho_rD_g^{(r)} +\rho_bD_g^{(b)}\over \rho_b+\rho_r}   ~ . \]
In this limit therefore, baryons and photons behave like a single fluid
with density $\rho =\rho_r+\rho_b$ and pressure $p=\rho_r/3$.

  From (\ref{scalcons}) we can derive a second order
equation for $D_g$. This equation can be simplified if expressed in
terms of the variable $D$ related by (\ref{DDgrel}).  We obtain
\[ \ddot{D} +c_s^2k^2D + (1+3c_s^2 - 6w)(\dot{a}/a)\dot{D} -
  3[w(\ddot{a}/a) - (\dot{a}/a)^2(3(c_s^2-w) -(1/2)(1+w))]D = 0 ~.\]
For small wavelengths (sub-horizon), which are however sufficiently 
large for the fluid approximation to be valid, $1/\eta_T \gg c_sk\gg 1/\eta$,
we may drop the term in square brackets. The ansatz
$ D(t) = A(t)\exp(-i\int kc_sdt)$
then eliminates the term of order $c_s^2k^2$. For the terms of order
$c_sk/\eta$ we obtain the equation
\be 2\dot{A}/A + (1+3c_s^2 -6w)(\dot{a}/a) +\dot{c_s}/c_s = 0 \label{3dA}~.\ee
For the case $c_s^2 =w=$const. , this equation is solved by
$ A\propto (k\eta)^{1-\nu}$ with $\nu =2/(3w+1)$, i.e., the short wave limit.
% (\ref{3fs}). {\bf I don't think we have that anywhere}
In our situation we have
\bean
 w &=& {\rho_r \over 3(\rho_r+\rho_b)}  \\
 c_s^2 &=& {\dot{\rho_r} \over 3(\dot{\rho_r}+\dot{\rho_b})} ~
    = ~ {(4/3)\rho_r \over 4\rho_r+3\rho_b}~~\mbox{ and} \\
 \dot{c}_s/c_s &=& -3/2(\dot{a}/a){\rho_b \over 4\rho_r+3\rho_b} ~.\eean
Using all this, one finds that
\[ A = \left({\rho_b+(4/3)\rho_r \over c_s(\rho_r+\rho_b)^2a^4}\right)^{1/2}
     =  \left({\rho+p \over c_s\rho^2a^4}\right)^{1/2} \]
solves (\ref{3dA}) exactly, so that we finally obtain the
approximate solution for the, tightly coupled matter radiation fluid
on sub-horizon scales
\be D(t) \propto  \left({\rho+p \over c_s\rho^2a^4}\right)^{1/2}
	\exp(-ik\int c_sd\eta) ~.\label{3mrt} \ee
Note that this short wave approximation is only valid in the limit
$\eta\gg 1/(c_sk)$, thus the limit to the matter dominated regime, $c_s\ra 0$
cannot be performed. In the limit to the radiation dominated regime,
$c_s^2\ra 1/3$ and $\rho\propto a^{-4}$ we recover the acoustic waves
with constant amplitude which we have already found in the last subsection.
But also in this limit, we still need matter to ensure $\eta_T =1/(a\si_Tn_e)
\ll \eta$. In the opposite case, $\eta_T\gg \eta$, radiation does not behave like
an ideal fluid but it becomes collisionless and has to be treated with
Liouville's equation ((\ref{3B}) without the collision term).

\subsection{Damping by photon diffusion}
In this subsection we discuss the Boltzmann equation in the next
order, $(\eta_T/\eta)^2$. In this order we will obtain the damping of
 fluctuations in an ionized plasma due to the finiteness of the mean
free path; the non-perfect coupling. We follow the treatment by 
Peebles~\cite{Pee80} (using our gauge-invariant approach instead of
synchronous  gauge). Again we  consider Eqs. (\ref{3B}) and
(\ref{3bar}), but since we are mainly interested in collisions which
take place on time scales $\eta_T\ll \eta$, we neglect gravitational 
effects and the time dependence of the coefficients. We can then look
for solutions of the form
\[ V \propto \MM^{(S)} \propto \exp(i(\bm{k\cd x}-\om \eta)) ~. \]
In (\ref{3B}) and (\ref{3bar}) this yields (neglecting also the
angular dependence of Compton scattering, i.e., the term $ n_{ij}M^{ij}$)
\be \MM^{(S)} = {1\over 4}{D_g^{(r)} -i\bm{k n}V\over 1 -i\eta_T(\om -\bm{k\cd  n})}
	\label{3md}  \ee
and
\be \eta_T\bm{k}\om V = (4\rho_r/3\rho_b)(i\bm{k}V +\bm{M}) ~, \label{3Vd}\ee
with $\bm{M} = (3/4\pi)\int\bm{ n}\MM^{(S)} d\Om$. Integrating (\ref{3B}) over
angles, one obtains $\dot{D}_g^{(r)} +(i/3)k_jM^i = 0$. With our ansatz
therefore $\bm{k\cd M} =3\om D_g^{(r)}$. Using this after scalar
multiplication
of (\ref{3Vd}) with $\bm{k}$, we find, setting $R =3\rho_b/4\rho_r$,
\[ V = {(3/4)\om  D_g^{(r)} \over \eta_Tk^2R\om -ik^2} ~. \]
Inserting this result for $V$ in (\ref{3md}) leads to
\[ \MM^{(S)} = {D_g^{(r)}\over 4}{1+{3\mu\om/k\over1-i\eta_T\om R}\over 
   1-i\eta_T(\om -k\mu)}~,\]
where we have set $\mu =\hat{\bk}\cd \bn$. This is  the  result of
 Peebles~\cite{Pee80}, where the same calculation is performed 
in synchronous
gauge. Like there (\S 92), one obtains in lowest non-vanishing
order $\om \eta_T$ the following dispersion relation: Using 
\[ {1\over 2}\int_{-1}^{1}\MM^{(S)} d\mu=  {D_g^{(r)}\over 4}~, ~\mbox{ which
yields }~ 1= {1\over 2}\int_{-1}^{1}{1+{3\mu\om/k\over1-i\eta_T\om R}\over 
   1-i\eta_T(\om -k\mu)} d\mu \]
one finds
\be \om =\om_0 -i\ga~ \mbox{ with }~~ \om_0 =k/[3(1+R)]^{1/2} ~ \mbox{ and }
  ~~ \ga = (k^2\eta_T/6){R^2+{4\over 5}(R+1)\over (R+1)^2} ~ .
	\label{3disp}\ee
In the baryon dominated regime, $R\ge 1$, therefore
\be \ga \approx k^2\eta_T/6 ~.\ee
(If the angular dependence of Thompson scattering is not neglected,
the term ${4\over 5}(R+1)$ in Eq.~(\ref{3disp}) becomes ${8\over 9}(R+1)$. 
If also polarization is taken into account, one obtains ${16\over 15}(R+1)$.)

Posing $ k_{\rm damp}\eta_T/6 =1$, this leads to a damping scale
$\lambda_{\rm damp} \sim \eta_T(\eta_{dec})/2$, which is projected in the
microwave sky to an angle 
\[ \vth_{\rm damp} \sim  {\eta_T(\eta_{\rm dec})\over 2\chi(\eta_0-\eta_{\rm
dec})}~. \]
For $\ka =0$ this corresponds to a few arc minutes and to the
harmonic number 
\be \ell_{\rm damp} =\pi/\vth_{\rm damp} \simeq  
{\pi \eta_0 \over 20\eta_T(\eta_{\rm dec})} \simeq 
	{(1 +z_{\rm dec})^2\over 20}\Om_bh~. \ee 
This estimate is very crude since we are using the the approximation
for $\eta_T$ from the tight coupling regime just where
coupling stops to be tight, and we assume an arbitrary value of 
$ n_e	\sim 0.1n_b$ at the moment of decoupling.
Both these errors enhance the value of $\ell_{\rm damp}$ somewhat.
Numerical results give
\[\ell_{\rm damp} \sim 800 - 1000 \]
in a $\ka=0$ universe. In open (closed) universes, this scale (which
of course also depends on $\Om_b$) is moved to larger (lower) $\ell$
due to projection. A reasonable approximation for the damping harmonic
is
\[\ell_{\rm damp} \sim 1000 \l({\Om_bh\over 0.02(1-\Om_{\ka})^{1/2}}\r) ~.\]
Temperature fluctuations on smaller scales, $\ell>\ell_{\rm damp}$ are
exponentially damped by photon diffusion.

\section{Polarization and moment expansion \label{pola}}

Thomson scattering is not isotropic. And what is more, for a 
non--isotropic photon
distribution it depends on the polarization: Even if the incident
photon beam is unpolarized, the scattered beam will be, unless the
incident distribution is perfectly isotropic. In the previous section
we have neglected this effect by summing over initial
polarizations and averaging over final polarizations. Now we want
to derive the difference in the Boltzmann equation taking into account
polarization.

Polarization is usually characterized by means of  the
{\em Stokes parameters} \cite{jackson,chandra,koso}.

For a harmonic electromagnetic wave with electric field
\be
{\bf E} (\bx,t)=\l({\bm\vep}_1 E_1+{\bm\vep}_2 E_2\r)
e^{i\om( \bn \bx - t)} ~,
\ee
where $\bn$, ${\bm \vep}_1$ and ${\bm \vep}_2$ form an orthonormal
basis and the complex field amplitudes are parameterized
as $E_j=a_j e^{i\de_j}$, the Stokes parameters are given by
\bea
I &=& a_1^2 + a_2^2\\
Q &=& a_1^2 - a_2^2\\
U &=& 2 a_1 a_2 \cos(\de_2-\de_1)\\
V &=& 2 a_1 a_2 \sin(\de_2-\de_1) .
\eea
$I$ is the intensity of the wave (whose perturbation
$\imath$ has been introduced in the previous section), 
while $Q$ is a measure of
the strength of linear polarization (the ratio of the principal
axis of the polarization ellipse). $V$ measures circular polarization
which is not generated by Thomson scattering and therefore $V$ vanishes if
the initial circular polarization vanishes (which we assume).
$U$ is then determined via the identity $I^2=Q^2+U^2$. For scalar
perturbations also $U$ vanishes.

Since $Q$ vanishes in the background, to first order it obeys the unperturbed
Boltzmann equation,
\be
\dd_\eta Q+i n^j k_jQ
-\Ga^{(S)\,i}_{jk} n^j n^k \frac{\dd Q}{\dd n^i}
= C[Q], 
\ee
where $C$ is the collision integral. The same type of equation, with a somewhat
different collision integral is satisfied by $U$. The collision integral for 
$V$ does not couple to $I,Q$ or $U$  and hence $V\equiv 0$ is a consistent 
solution.

 An explicit derivation of the following Boltzmann hierarchy
including polarization is presented in Appendix~\ref{AppBoltz}. Here we just 
repeat the necessary definitions and the results.

The brightness anisotropy $\MM$ and the non-vanishing Stokes parameters $Q$ 
and $U$ can be expanded as
\bea
 \MM(\ct,\bk,\bn) &=& \sum_{\ell}\sum_{m=-2}^2\MM_{\ell}^{(m)}(\ct,k)
	{ _0 G^m_{\ell}}(\bn),\\
 Q(\ct,\bk,\bn)\pm iU(\ct,\bk,\bn) &=& \sum_{\ell}\sum_{m=-2}^2
	(E_{\ell}^{(m)} \pm iB_{\ell}^{(m)}){ _2G^m_{\ell}}(\bn).
\eea
The special functions $ _sG^m_{\ell}$ are described in 
Appendix~\ref{AppBoltz}. The coefficients $m=0,m=\pm1$ and $m=\pm2$ describe 
the scalar $(S)$, vector $(V)$ and tensor $(T)$ components respectively. 
The Boltzmann equation for the coefficients $X^{(m)}_{\ell}$ is given by
\bea
\lefteqn{ \dot\MM_{\ell}^{(m)} -k\l[{_0\ka^m_{\ell}\over 2\ell-1}
	\MM_{\ell-1}^{(m)}
-{_0\ka^m_{\ell+1}\over 2\ell+3}\MM_{\ell+1}^{(m)}\r] = } &&\nonumber \\
  &&	-n_e\si_Ta\MM_{\ell}^{(m)}  +S_{\ell}^{(m)} ~~~ (\ell\ge m)\\
\lefteqn{ \dot E_{\ell}^{(m)} -k\l[{_2\ka^m_{\ell}\over 2\ell-1}
E_{\ell-1}^{(m)}  -{2m\over \ell(\ell+1)}B_{\ell}^{(m)}
-{_2\ka^m_{\ell+1}\over 2\ell+3}E_{\ell+1}^{(m)}\r] = } && \nonumber \\ && 
  -n_e\si_Ta[E_{\ell}^{(m)} + \sqrt{6}C^{(m)}\de_{\ell,2} \\
\lefteqn{ \dot B_{\ell}^{(m)} - k\l[{_2\ka^m_{\ell}\over 2\ell-1}
  B_{\ell-1}^{(m)}   +{2m\over \ell(\ell+1)}E_{\ell}^{(m)}
-{_2\ka^m_{\ell+1}\over 2\ell+3}B_{\ell+1}^{(m)}\r] = } && \nonumber \\ &&  
  -n_e\si_TaB_{\ell}^{(m)}~.
\eea
where we  set
\be \begin{array}{cc}
 S_0^{(0)} =n_e\si_Ta\MM^{(0)}_0, & S^{(0)}_1 =n_e\si_Ta4V_b +4k(\Psi-\Phi), \\
 S^{(0)}_2 =n_e\si_TaC^{(0)}, & S^{(1)}_1 =n_e\si_Ta4\om_b,  \\
  S^{(1)}_2 =n_e\si_TaC^{(1)} +4\Si, &   S^{(2)}_2 =n_e\si_TaC^{(2)} +4\dot H 
\end{array}
\ee
and $C^{(m)}= {1\over 10}[\MM^{(m)}_2 -\sqrt{6}E^{(m)}_2] $.
The coupling coefficients are
\[ 
	_s\ka^m_{\ell} =\sqrt{{(\ell^2-m^2)(\ell^2-s^2)\over \ell^2}}.
\]

The CMB temperature and polarization power spectra
are given in terms of the expansion coefficients $\MM_{\ell}^{(m)}$,
 $E_{\ell}^{(m)}$ and $B_{\ell}^{(m)}$ as
\be
 (2\ell+1)^2C_\ell^{XY(m)} = {n_m\over 8\pi}\int k^2dkX_{\ell}^{(m)}
	Y_{\ell}^{(m)*}~,  \label{PSpecC2}
\ee
where $n_m=1$ for $m=0$ and $n_m=2$ for $m=1,2$, accounting for the number 
of modes. Since $B$ is parity odd, the only non-vanishing cross correlation 
spectrum is $C^{TE}$.

\begin{figure}[ht]
\centerline{\epsfig{file=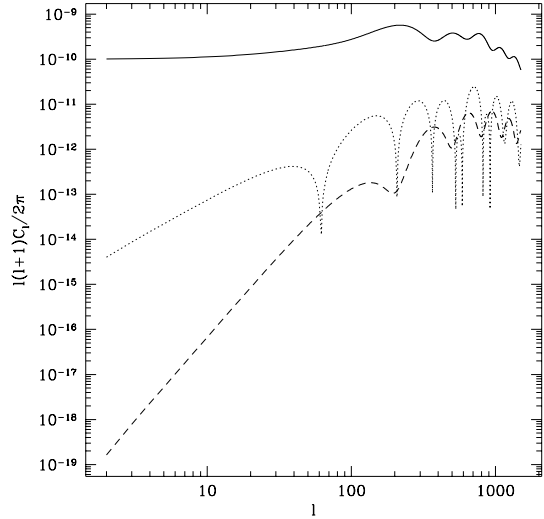,width=7.5cm}}
\caption{\label{pol} The temperature anisotropy (solid), the
polarization (dashed) and their correlation (dotted) are shown for
a purely scalar standard CMD model.}
\end{figure}

The Boltzmann hierarchy presented above can be solved numerically with 
publicly available fast numerical codes like CMBfast~\cite{SZ} or
CAMCODE~\cite{AAA}. This enables us to compute the CMB anisotropy and 
polarization spectra for many different values of cosmological parameters, and 
compare them with present data.

\section{Parameter estimation}
In the last section 
of this chapter we make some general remark about the dependence of the 
CMB anisotropy spectrum on different parameters and about degeneracies.
We start by enumerating the relevant physical processes.

\subsection{Physical processes}

\begin{itemize}
\item Before recombination, photons and baryons form a tightly coupled
fluid which performs acoustic oscillations on sub-horizon scales.
\item Depending on the initial conditions, these oscillations are
sine waves (isocurvature case) or cosine waves (adiabatic case).
\item After recombination, the photons move along perturbed geodesics,
only influenced by the metric perturbations.
\item Vector perturbations of the metric decay as $a^{-2}$ after
creation and their effects on CMB anisotropies are negligible for
models where initial fluctuations are created early, {\it e.g.} during
an inflationary phase. This is different for models which constantly
seed fluctuations in the geometry,  {\it e.g.} topological defects.
\item Tensor perturbations of the metric have constant amplitude
on super-horizon scales and perform damped oscillations
$\propto a^{-1}$ once they enter the horizon.
\item Scalar perturbations of the metric are roughly constant if
they enter the horizon only after the time of matter and radiation
equality. On scales which enter the horizon before equality they
are damped by a factor $(z_\mr{eq}/z_\mr{in})^2$, where $z_\mr{eq}$
and $z_\mr{in}$ are the redshift of equality and of horizon crossing,
respectively. 
\item Perturbations on small scales, $k\gsim k_T\simeq
(\Om_b h/20) (z_\dec+1)^2 H_0$ are exponentially damped by collisional
damping during recombination (Silk damping).
\end{itemize}

\subsection{Scale dependence}

\begin{itemize}
\item On large scales (larger than the horizon scale at recombination,
$\ell \lsim \ell_H \simeq \pi/\vth_H$, with
$\vth_H=\eta_\dec/\chi(\eta_0-\eta_\dec)$, perturbations are dominated
by gravitational effects: Inflationary models typically lead to
$k^3\l\lan|\Psi-\Phi|^2(k,\eta_\dec)\r\ran \simeq \mr{const.}$ and
$k^3\l\lan H^2\r\ran\simeq\mr{const.}$ on these scales. This
implies a flat ``Harrison-Zel'dovich'' spectrum,
\be
\l(\frac{\De T}{T}\r)^2(\vth_\ell) \simeq \ell(\ell+1)C_\ell \simeq
 \mr{const.} , \quad
\vth_\ell = \frac{\pi}{\ell} .
\ee
\item On intermediate scales, $\ell_H<\ell<\ell_\mr{damp}\sim 800$,
CMB anisotropies mainly reflect the acoustic oscillations of
the photon/baryon plasma prior to recombination. The position of the first
peak is severely affected by initial conditions (adiabatic or
isocurvature). For $\ka=0$, the first contraction
%{\bf what does positive mean here?}
 peak is at about $\ell_1^{(a)}\sim 220$ if the initial conditions
 are adiabatic, while the first
 contraction peak is at $\ell_1^{(i)}\sim350$ for  isocurvature
 initial conditions. The
amplitude of and the distance between the peaks depend strongly
on cosmological parameters.
\item On small scales, $\ell_\mr{damp}<\ell$, fluctuations are
collisionally damped during recombination (``Silk damping'').
The damping scale depends mainly on $\Om_b h$ and $\Om$.
\end{itemize}

\subsection{The main influence of cosmological parameters}

\begin{itemize}
\item {\bf Curvature, $h^2 \Om_\ka$:}
\begin{itemize}
\item Mainly affects the inter--peak distance, $\De\ell$, and,
for given initial conditions, the position of the first peak.
Positive curvature lowers $\De\ell$ while negative curvature
enhances it.
\item Curvature also leads to an integrated Sachs--Wolfe contribution
which is especially important for $\ka>0$ at very low $\ell$.
 Overall, this leads to some enhancement of the
Sachs--Wolfe contribution and therefore (after normalization to
the COBE measurements) to somewhat lower acoustic peaks.
\end{itemize}

\item {\bf Baryon density, $\rho_b=\Om_b h^2\cdot 10^{-29}
\mr{g/cm^3}$:}
\begin{itemize}
\item A high baryonic density enhances the compression peaks
and decreases the expansion peaks via the self--gravity of the
baryons.
\item It also reduces the damping scale,
$\la_T=1/(a_\dec\si_T n_e(\eta_\dec))$, leading to an increase in
$\ell_\mr{damp}$.
\item Baryons  decrease the plasma sound velocity,
$c_s=\nicefrac1/3 (1+\dot{\rho}_b/\dot{\rho}_\ga)^{-1}$, and hence
prolongs the oscillation period. This increases the spacing between 
acoustic peaks.
\end{itemize}

\item{\bf Cosmological Constant, $\La=\frac{\Om_\La h^2}{8\pi G}
\cdot 10^{-29}\mr{g/cm^3}$:}

The presence of a cosmological constant at fixed
$\Om_\mr{tot}=\Om_m+\Om_\La$ delays the epoch of equal matter
and radiation. During the radiation dominated era, the
gravitational potential is not constant, but decays as soon
as a given scale enters the horizon. If $\eta_\mr{eq}\sim\eta_\dec$
this induces an integrated Sachs--Wolfe (ISW) contribution which
boosts mainly the first acoustic peak. $\Om_\La$ 
 also boosts the late integrated  Sachs--Wolfe
contribution.

\item{\bf Hubble Parameter, $H_0=100 h \mr{~km/(s~Mpc)}$:}
The influence of the Hubble parameter is complicated and depends sensitively
on the variables which are kept fixed during its variation ($\Om_\bullet$ 
or $\om_\bullet =h^2\Om_\bullet$). As one example of its influence:
for fixed curvature and cosmological constant, lowering
the Hubble parameter also delays the epoch of equal matter and
radiation, $\eta_\mr{eq}\rightarrow\eta_\dec$, since
\be
z_\mr{eq}+1=\frac{\Om_m}{\Om_\mr{rad}}\simeq2.4\cdot10^4\Om_m h^2 .
\ee
Therefore the same type of ISW contribution as for $\La$--models
boosts the first acoustic peak.

\item{\bf Initial conditions:}
\begin{itemize}
\item A tensor contribution enhances the large scales fluctuations
but not the acoustic peaks, thereby lowering their relative
amplitude.
\item A ``blue'' fluctuation spectrum, $n>1$, enhances fluctuations
on smaller scales and raises thereby the acoustic peaks.
\end{itemize}

\end{itemize}

\subsection{Degeneracy}
\begin{figure}[ht]
\centerline{\epsfig{file=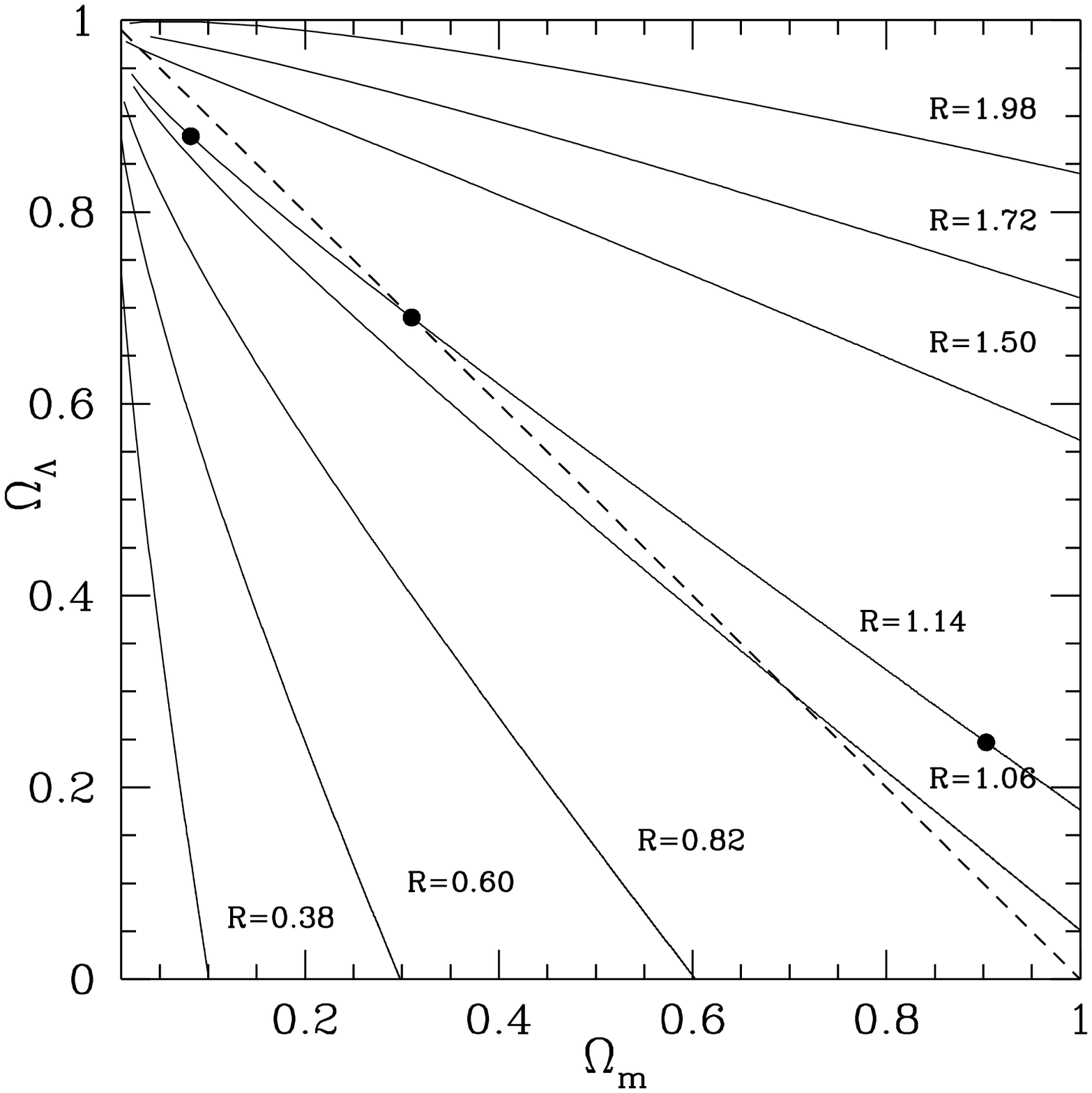,width=0.49\textwidth}\epsfig{file=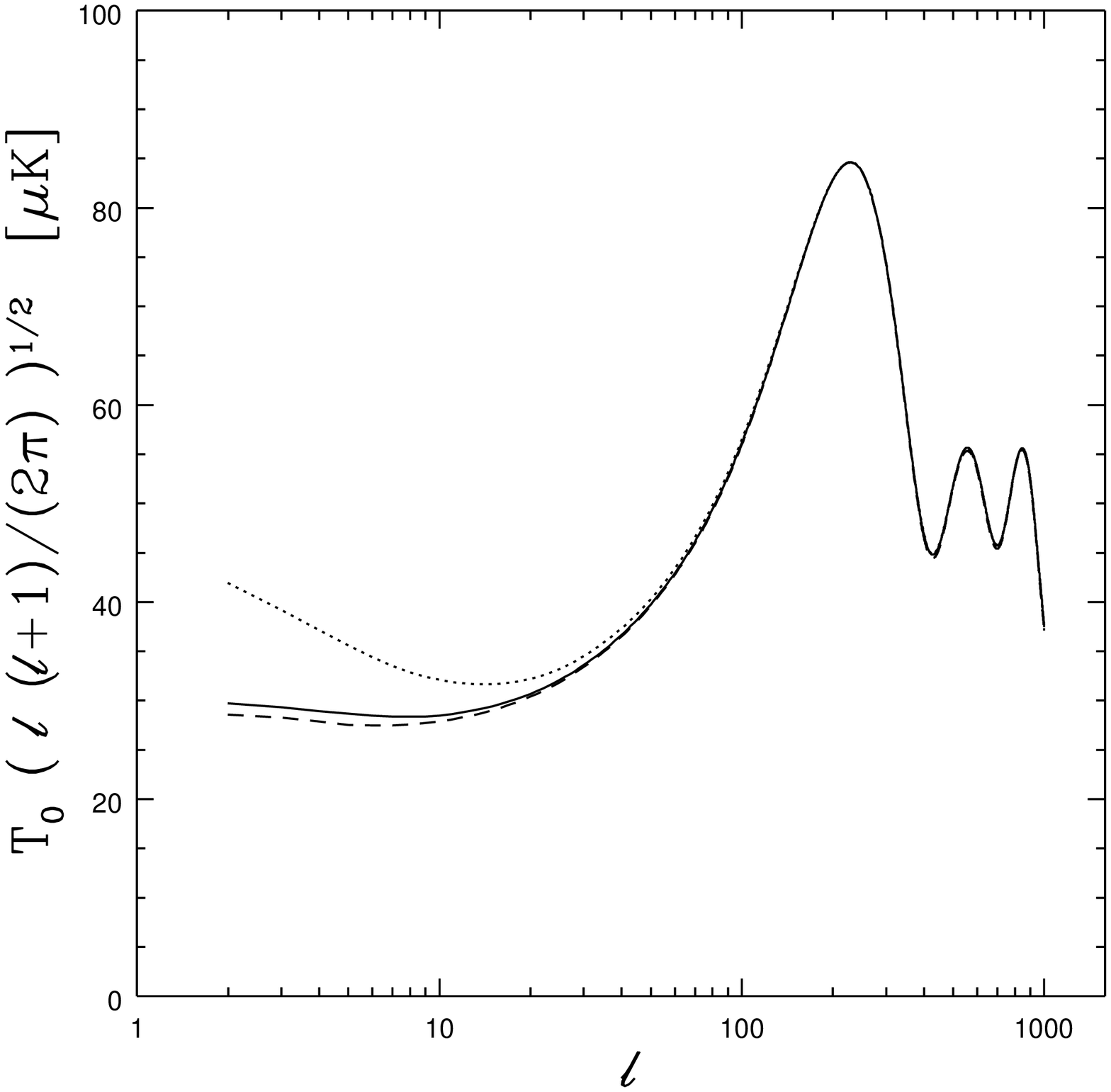,width=0.49\textwidth}}
\caption{{\bf Left:} The lines of constant $R$ are shown in the 
$\Om_\La$--$\Om_m$
plane. The values $\Om_\La,\Om_m$ for which the CMB anisotropy spectra are
shown right are indicated as black dots.
\hspace{0.3cm}
{\bf Right:} Three CMB anisotropy spectra with different values of 
$\Om_\La$ and $\Om_m$ but fixed $R$ are shown. For $\ell\gsim 50$ these 
spectra are clearly degenerate.The solid line represents a flat model, while 
the dotted line corresponds to a closed and the dashed line to an open 
universe.}
\label{fig:degen}
\end{figure}

One important issue in determining cosmological parameters from CMB 
anisotropy measurements is the choise of good variables, which requires
physical insight in how anisotropies are influenced. As we have argued before,
the Hubble parameter, $h$ is not a good variable since its influence is 
very complicated. It enters the cosmic densities 
$\rho_\bullet \propto \Om_\bullet h^2$ and the length scales like 
$\eta_{\rm eq}$ or $\eta_\dec$. Another limitation for parameter estimation 
from CMB anisotropies is degeneracy. We illustrate here just
one example. As we have discussed in Chapter~3, the position of the first 
acoustic peak only depends on the sound horizon, $\tau_s=\int^{\eta_\dec}
c_sd\eta$ and the angular diameter distance to the last scattering surface,
$\chi(\eta_0-\eta_\dec)$. The distance between subsequent peaks in the CMB 
power spectrum is proportional to 
$$\De\ell= {\chi(\eta_0-\eta_\dec)\over\tau_s} $$
In Fig.~\ref{fig:degen} (left panel) we show lines of constant 
$R=\De\ell/\De\ell_0$ in the $\Om_m$ -- $\Om_\La$ plane. Here $\De\ell_0 =
 \De\ell_0(\Om_\La=\Om_\ka=0)$ is the value of $\De\ell$ in a spatially flat 
universe with vanishing cosmological constant. To the right the
CMB anisotropy spectra for scalar perturbations with fixed index $n=1$ and 
fixed values of the matter density $\om_m$ and the baryon density $\om_b$.
But the cosmological constant and $h$ vary, so that $\Om_\La$ and $\Om_m$ 
correspond to the values indicated by bullets on the left panel. Clearly, for 
$\ell>50$ these spectra are perfectly degenerate. On the other hand, due to
cosmic variance, the low $\ell$ CMB anisotropies will never be known to 
very good accuracy so that this degeneracy cannot be lifted by CMB anisotropy 
observations alone. Additional data like the supernova type Ia measurements,
observations of the galaxy distribution (large scale structure) or CMB 
polarization are needed.

There are also other degeneracies like the optical depth to reionization and 
the tensor contribution or the scalar spectral index and the tensor 
contribution. The important lesson to learn is that even if the very
stringent model assumptions are correct, we still need other data
to measure cosmological parameters and especially we will only feel 
comfortable with a sufficient amount of redundancy. 

\chapter{Observations and Results}

In this short, final chapter we want to discuss briefly the experimental
situation which is very much in flow and may have changed considerably 
already at the moment when this review appears. It has been clear for a 
long time that, if initial
fluctuations have led to the formation of large scale structure
by gravitational instability, they should have induced fluctuations
in the cosmic microwave background \cite{SW,Dor}. Before
 spring 1992, however, only the dipole anisotropy had been detected
\cite{Conklin,Henry}. Its value is~\cite{PDG}
\[ \l\langle\l({\De T\over T}\r)^2\r\rangle^\mr{dipole} = 
 (1.528\pm 0.004)\times10^{-6} ~.\]

\begin{figure}[ht]
\centerline{\epsfig{file=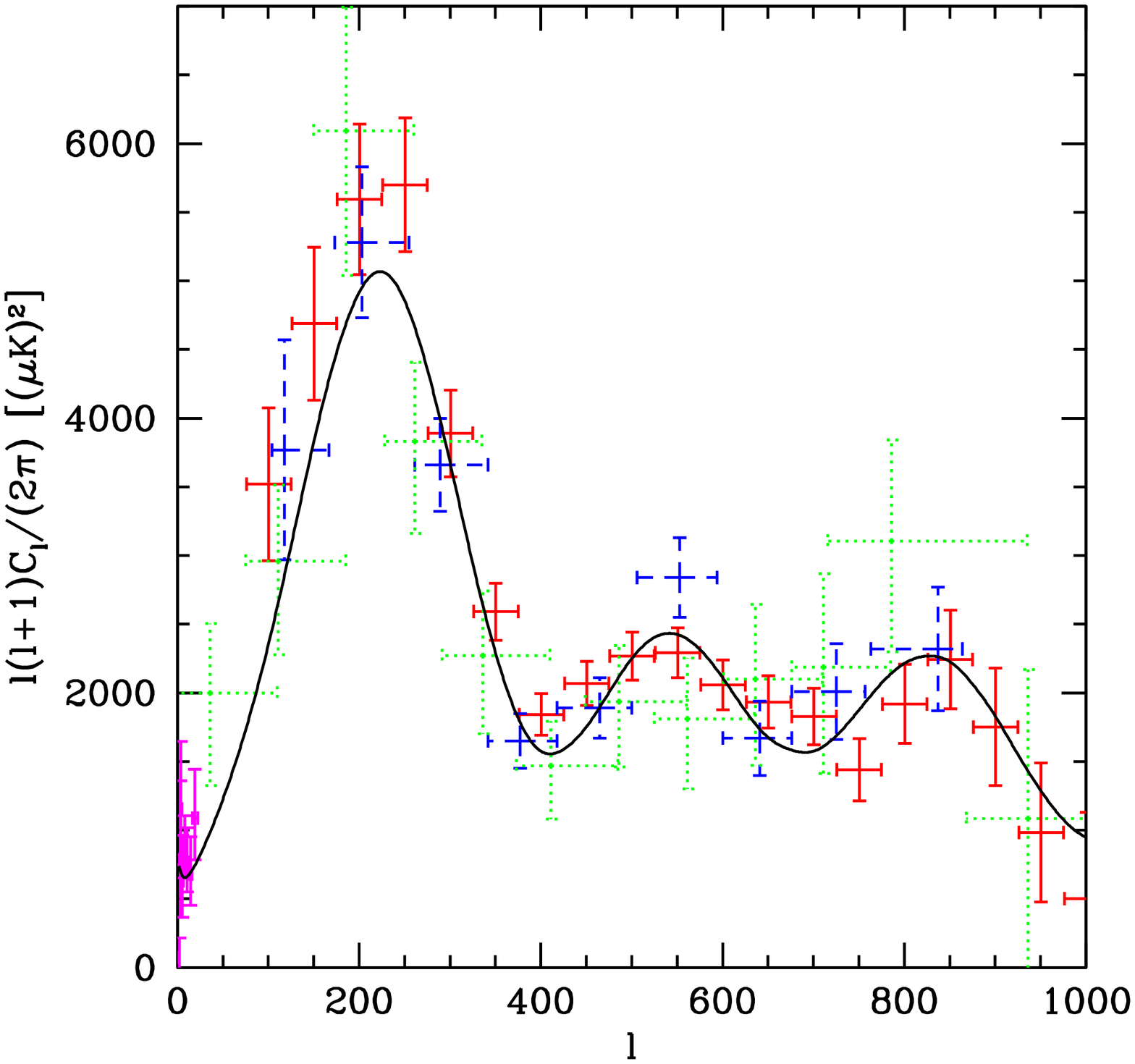,width=7cm} \quad
  \epsfig{file=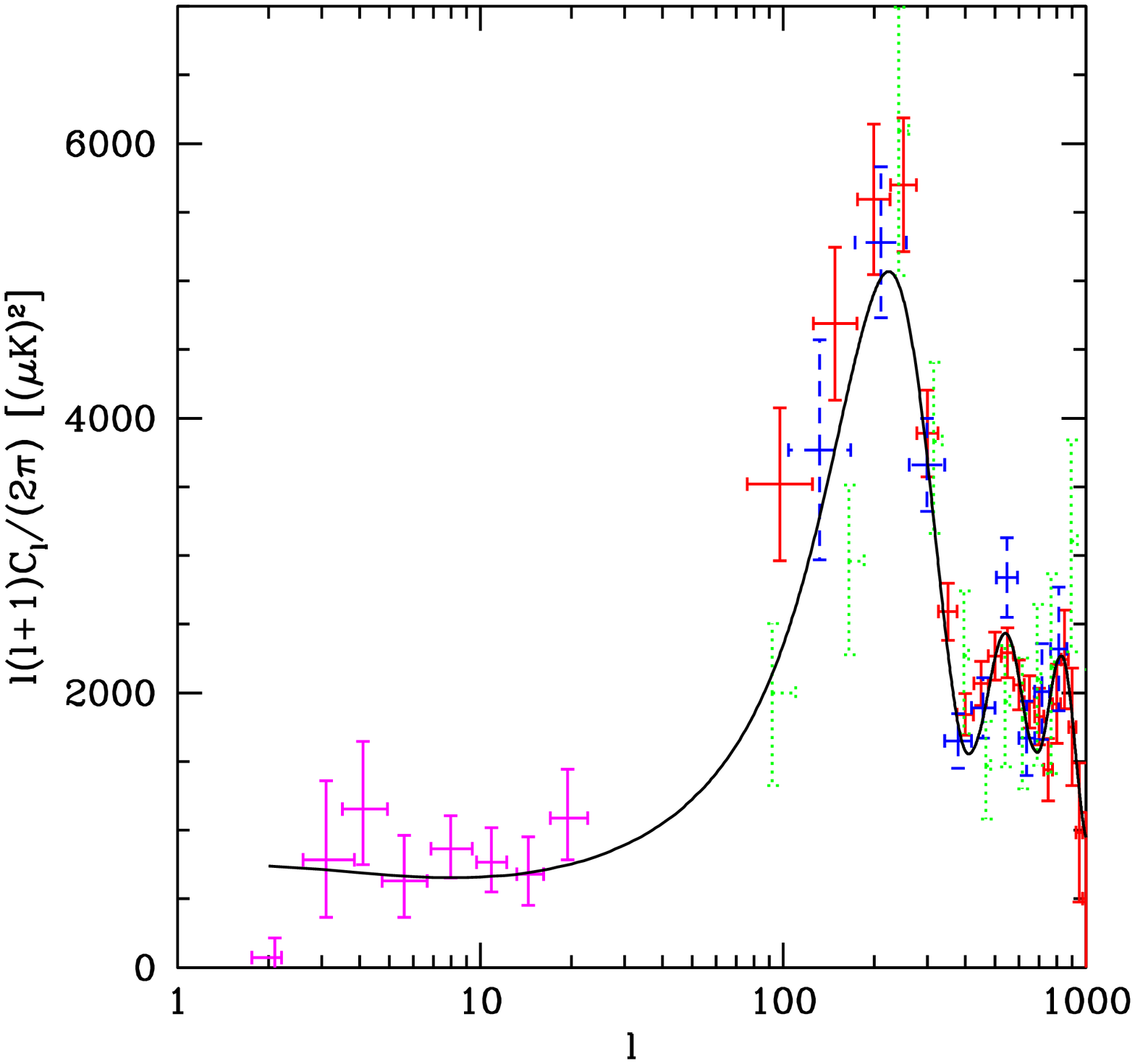,width=7cm}}
\caption{\label{fig:CMBdata} The measured temperature anisotropies,
$\ell(\ell+1)C_{\ell}$ are shown in a
lin-lin plot (left) and in a log-lin plot (right)  with the theoretical curve
from a standard, adiabatic cold dark matter model. The data points shown 
are those from COBE DMR (solid, magenta, low $\ell$), BOOMERANG (solid, red), 
DASI (dashed, blue) and MAXIMA-1 (dotted, green). }
\end{figure}

After many upper limits, the DMR experiment aboard the COBE satellite
measured for the first time convincingly positive anisotropies \cite{DMR}.
It found
\be
\l\lan\l(\frac{\De T}{T}\r)^2\r\ran (\th) \sim (30 \mu\mr{K})^2
\ee
on all angular scales $\th \geq 7^\circ$. Many more positive measurements
have been performed since then. In Fig.~\ref{fig:CMBdata} we just show the
COBE DMR results~\cite{cobe} together with the three most recent experiments, 
BOOMERANG~\cite{b01}, MAXIMA-1~\cite{max01} and DASI~\cite{dasi} 

As one sees in this figure, present data, agrees very well with a
simple flat model of purely scalar, scale invariant, $n_s=1$, adiabatic 
fluctuations with cosmological parameter $\om_b=0.02$, $\Om_\ka=0$,
$\Om_\La=0.7$, $h=0.65$ which are also preferred from other cosmological 
data. However, the error-bars are still considerable.

The experiments can be split into three classes: Satellite experiments,
balloon--borne experiments and ground based experiments. The technical
and economical advantages of ground based experiments are obvious.
Their main problem is atmospheric fluctuation. This can be reduced by
two methods:
\begin{itemize}
\item Choose a very high altitude and very cold site, \eg~the south
pole. Several experiments like SP, Python and White Dish have chosen
this site.
\item Measure anisotropies on small scales, preferably by interferometry
(DASI, CAT, VSA, Jodrell Bank).
\end{itemize}

Balloon--borne experiments flying at about 40km altitude have less 
problems with the Earths atmosphere but they face the following difficulties:
\begin{itemize}
\item They are limited in weight.
\item They cannot be manipulated at will in flight.
\item They have a rather short duration.
\item To secure all the data taken on the balloon, they have to be 
	recovered intact.
\end{itemize}

Yet the advantages of overcoming the atmosphere are so significant 
that many groups have chosen this approach, like \eg~MAXIMA-1, TopHat, etc.
The BOOMERanG experiment combines the two advantages of a cold site and 
balloon altitude. It has performed a long--duration flight (10 days) on the 
south pole in December 1998.

The third possibility are satellite experiments. They avoid atmospheric
problems altogether, but this solution is very expensive. So far two
satellite experiments have been launched: COBE in 1989 (NASA mission) and 
 MAP in June 2001 (Microwave Anisotropy Probe, a NASA MIDEX mission),
one more is planned: PLANCK, an ESA medium size mission of the
``Horizon 2000'' program, to be launched in 2007.

As I am writing this lines, MAP has safely arrived at its destination, 
the Lagrange point L2 of the sun-earth system. It will perform measurements 
at five frequencies in the range
from 22 to 90 GHz. The instruments of PLANCK will operate
at nine frequencies, covering 20 to 800 GHz. At low frequencies
(below 100 GHz) {\em radio receivers} are used (so called ``HEMTs'', high
electron mobility transistors) while the
high frequency instruments are {\em bolometers}. Recent progress in
detector technologies should enable the two new satellites to
perform significantly better than COBE -- the radio receivers of
PLANCK, \eg, are supposed to be 1000 times more sensitive than the
ones used for COBE, and the angular resolution has improved from
seven degrees to four arc minutes. For more details also on other experiments
see
\begin{itemize}
\item {\tt http://astro.estec.esa.nl/PLANCK}
\item {\tt http://map.gsfc.nasa.gov}
\item {\tt http://www.gsfc.nasa.gov/astro/cobe/cobe\_home.html}
\item {\tt http://spectrum.lbl.gov/www/max.html }
\item {\tt http://oberon.roma1.infn.it/boomerang/ }
\vspace{1cm}
\end{itemize}

I finish this short chapter with Table~\ref{tab:parms} which shows the ranges 
for the 
cosmological parameters $\Om_{tot}=1-\Om_{\ka}$, $h^2\Om_b$ and $n_s$
as determined {\bf purely by CMB anisotropies}. Except for the last 
reference, a purely scalar spectrum of adiabatic fluctuations is assumed. 
The parameter estimation process also assumes 
'weak priors' on the values of other cosmological observables, like e.g. 
that the age of the Universe be larger than 10Gyrs. or $0.4<h<0.9$. 
I do not comment this table much further but refer the reader to the original
literature and many improved papers on this subject which will appear
shortly.
\begin{table}
\begin{center}\footnotesize
\begin{tabular}{||c|c|c|c|c|c||}
 \hline\hline
Ref. & Data & $\Om_{tot}$ & $\Om_bh^2$ & $n_s$  & errors \\
\hline 
 & & & & & \\
\protect\cite{deBe} & BOOM and DMR data & $1.02^{+0.06}_{-0.05}$  & 
$0.022^{+0.004}_{-0.003} $ & $0.96^{+0.1}_{-0.09} $ & $1-\si$ errors \\
 & & & & & \\
\protect\cite{Pryke} & DASI and DMR data & $1.05^{+0.06}_{-0.06} $  &
$0.022^{+0.004}_{-0.004} $ & $1.01^{+0.09}_{-0.07} $ & $1-\si$ errors \\
 & & & & & \\
\protect\cite{Stompor} & MAX and DMR data & $0.90^{+0.18}_{-0.16}  $ &
 $0.0325^{+0.0125}_{-0.0125} $ & $0.99^{+0.14}_{-0.14} $ & $2-\si$ errors \\
 & & & & & \\
\protect\cite{Wang} & all data, no priors & $1.06^{+0.59}_{-0.13}  $  &
$0.02^{+0.06}_{-0.01} $ & $0.93^{+0.75}_{-0.16} $  & $2-\si$ errors \\
 & allows also tensor mode & & & & \\
\hline \hline
\end{tabular}
\end{center}
\vspace{0.3cm}
\caption{Some results from parameter estimations from recent CMB data alone.
The errors given are formal $1$ or $2$-$\si$ errors which assume the 
underlying model to be correct and no systematic problems in the data. 
They are obtained by marginalization or maximization over all other model 
parameters. \label{tab:parms}
}
\end{table}

Clearly, the results shown in Table~\ref{tab:parms} are very consistent. It
is interesting to note, how the upper limit on the scalar spectral index 
deteriorates if one allows for a tensor component. This is one of the 
degeneracies in the CMB data which can be broken by including large scale 
structure data in the analysis (see~\cite{Wang}). Other 
cosmological parameters are not well constrained by CMB data alone. However,
if CMB data is combined with SN1a and large scale structure data,
the error bars are significantly reduced and evidence for a non-vanishing 
cosmological constant $\Om_\La\sim 0.7$ becomes very strong 
(see~\cite{deBe,Pryke,Wang}).
\vspace{1cm}

{\bf Acknowledgment} It is a pleasure to thank Pedro Ferreira, 
Roman Juszkiewicz. Martin Kunz, Alessandro Melchiorri, Bohdan Novosyadlyj 
and Norbert Straumann for stimulating
discussions. I'm also very grateful to Martin Kunz for his help with the
preparation of this manuscript. Finally, thanks for hospitality go to the 
Institute for Advanced Study in Princeton, where this review was terminated.

% LocalWords: ij dx dec const lj ln jd mm dk jk dv rad df df spacetime Tanaka
% LocalWords: al im

\newpage
\appendix
\chapter{The $C_\ell$'s from gravitational waves}

We consider metric perturbations which are produced by some isotropic
random process (for example during inflation). After production, they
evolve according to a deterministic equation of motion. 
By reasons of isotropy and due to symmetry, the
correlation functions of $h_{ij}({\bf k},\eta)$  have to be of the form
\bea
\langle h_{ij}({\bf k},\eta) h^*_{lm}({\bf k},\eta')\rangle &=& 
  [k_ik_jk_lk_mH_1(k,\eta,\eta') + \nonumber\\  &&
  (k_ik_l\de_{jm}+k_ik_m\de_{jl}+k_jk_l\de_{im}
 +k_jk_m\de_{il})H_2(k,\eta,\eta') + \nonumber \\ &&
k_ik_j\de_{lm}H_3(k,\eta,\eta') +k_lk_m\de_{ij}H_3^*(k,\eta',\eta) +
\nonumber\\ &&
+\de_{ij}\de_{lm}H_4(k,\eta,\eta') +
(\de_{il}\de_{jm}+\de_{im}\de_{jl})H_5(k,\eta,\eta')] ~.
\label{hijlmansatz}
\eea
Here the functions $H_a$ are functions of the modulus $k=|{\bf k}|$ only.
Furthermore, all of them except $H_3$ are hermitian in $t$ and $t'$.
This is the most general ansatz for an isotropic correlation tensor
satisfying the required symmetries. To project out the tensorial part
of this correlation tensor we act on $h_{ij}$ it with the tensor projection
operator,
\be
 T_{ij}^{~~mn}= P_i^mP_j^n-(1/2)P_{ij}P^{mn}   ~~
	\mbox{ with }~~~
P_{ij} =  \de_{ij}-\hat{k}_i\hat{k}_j~.
\ee
This yields
\bea
\lefteqn{\langle h^{(T)}_{ij}({\bf k},\eta)h^{(T)*}_{lm}({\bf
	k},\eta')\rangle =} \nonumber \\
&&	H_5(k,\eta,\eta')[\de_{il}\de_{jm}+\de_{im}\de_{jl}   -\de_{ij}\de_{lm} + 
	k^{-2}(\de_{ij}k_lk_m + \nonumber \\ 
&&	\de_{lm}k_ik_j -\de_{il}k_jk_m - \de_{im}k_lk_j -\de_{jl}k_ik_m
	-\de_{jm}k_lk_i) + \nonumber \\
&&	k^{-4}k_ik_jk_lk_m] \label{Ctau}
    ~.\eea
From Eq.~(\ref{dTT}), we then obtain
\bea
&& \left\langle{\Delta T \over T}( {\bf n}){\Delta T \over T}( {\bf
n}')\right\rangle \equiv
 {1\over V}\int d^3x \left({\Delta T \over T}({\bf n,x})
	{\Delta T \over T}( {\bf n',x})\right) =
 \nonumber \\  && \left({1\over 2\pi}\right)^3
\int k^2dkd\Om_{\hat{\bf k}}\int_{\eta_{dec}}^{\eta_0}d\eta  
\int_{\eta_{dec}}^{\eta_0}d\eta'
\exp(i{\bf k}\cdot{\bf n}(\eta_0-\eta))
\exp(-i{\bf k}\cdot{\bf n}(\eta_0-\eta')) \cd \nonumber\\ &&
\left[\langle{\dot h}^{(T)}_{i j}(\eta,{\bf k}){\dot h}_{lm }^{(T)*}
	(\eta',{\bf k})
	\rangle n_in_jn'_ln'_m \right] ~.   \label{dT2}
\eea
Here $d\Om_{\hat{\bf k}}$ denotes the integral over directions in $\bf
k$ space. We use the normalization of the Fourier
transform 
\[ \hat f({\bf k})={1\over \sqrt{V}}\int d^3x \exp(i{\bf x\cd k})f({\bf x}) 
	~,~~~~
  f({\bf x})={1\over (2\pi)^3}\int d^3k \exp(-i{\bf x\cd k})
	\hat f({\bf k}) ~,\]
where $V$ is an (arbitrary) normalization volume.

We now introduce the form (\ref{Ctau}) of $<h^{(T)}h^{(T)}>$. 
We further make use
of the assumption that the perturbations have been created at some
early epoch, e.g. during an inflationary phase, after which they
evolved deterministically. The function $H_5(k,\eta,\eta')$ is thus a
product of the form
\be  H_5(k,\eta,\eta') = H(k,\eta)\cd H^*(k,\eta') ~.\ee
Introducing this in Eq.~(\ref{dT2})  yields
\bea
\lefteqn{ \left\langle{\Delta T \over T}( {\bf n}){\Delta T \over T}
	( {\bf n}')\right\rangle=}
 \nonumber \\ && \left({1\over 2\pi}\right)^3
\int k^2dkd\Om_{\hat{\bf k}}
\left[({\bf n}\cd{\bf n}')^2 -1+\mu'^2+\mu^2-
	4\mu\mu'({\bf n}\cd{\bf n}')+\mu^2\mu'^2\right] \cd \nonumber\\
&&	\int_{\eta_{dec}}^{\eta_0}d\eta  
\int_{\eta_{dec}}^{\eta_0}d\eta' \left[\dot H(k,\eta)\dot H^*(k,\eta')
\exp(ik\mu(\eta_0-\eta))\exp(-ik\mu'(\eta_0-\eta')) \right]~, \label{dT2.5}
\eea
where $\mu=(n\cd\hat{\bf{k}})$ and  $\mu'=(n'\cd\hat{\bf{k}})$.
To proceed, we  use the identity \cite{AbSt}
\be
\exp((ik\mu(\eta_0-\eta))=\sum_{r=0}^{\infty}(2r+1)i^rj_r(k(\eta_0-\eta))
	P_r(\mu)~.
\ee
Here $j_r$ denotes the spherical Bessel function of order $r$ and
$P_r$ is the Legendre polynomial of degree $r$.

Furthermore, we replace each factor of $\mu$ in Eq.~(\ref{dT2.5})
by a derivative of the
exponential $\exp(ik\mu(\eta_0-\eta))$ with respect to $k(\eta_0-\eta)$ and
correspondingly with $\mu'$. We then obtain
\bea
\lefteqn{\left\langle{\Delta T \over T}( {\bf n}){\Delta T \over T}
	( {\bf n}')\right\rangle=}
 \nonumber \\ && \left(1\over 2\pi\right)^3 
\sum_{r,r'}(2r+1)(2r'+1)i^{(r-r')}\int k^2dkd\Om_{\hat{\bf k}}
P_r(\mu)P_{r'}(\mu') \times
 \nonumber \\ && 
\Big[2 ({\bf n}\cd{\bf n}')^2 
      \int d\eta d\eta'j_r(k(\eta_0-\eta))j_{r'}(k(\eta_0-\eta'))
	\dot H(k,\eta)\dot H^*(k,\eta')
 \nonumber \\ && 
-\int d\eta d\eta'[j_r(k(\eta_0-\eta))j_{r'}(k(\eta_0-\eta'))+
	j_r''(k(\eta_0-\eta))j_{r'}(k(\eta_0-\eta')) +
 \nonumber \\ && 
	j_r(k(\eta_0-\eta))j_{r'}''(k(\eta_0-\eta'))-  
	j_r''(k(\eta_0-\eta))j_{r'}''(k(\eta_0-\eta'))
]\dot H(k,\eta)\dot H^*(k,\eta')
 \nonumber \\ && 
-4({\bf n}\cd{\bf n}') \int d\eta d\eta'
	j_r'(k(\eta_0-\eta))j_{r'}'(k(\eta_0-\eta'))
	\dot H(k,\eta)\dot H^*(k,\eta')\Big]~. \label{dT3}
\eea
Here only the Legendre polynomials, $P_r(\mu)$ and  $P_{r'}(\mu')$  depend
 on the direction $\hat{\bf k}$. To perform the integration
$d\Om_{\hat{\bf k}}$, we use the addition theorem for the spherical
harmonics $Y_{rs}$,
\be
P_r(\mu)={4\pi\over (2r+1)}   \label{add}
	\sum_{s=-r}^rY_{rs}({\bf n})Y^*_{rs}(\hat{\bf k})~.
\ee
 The orthogonality of the spherical harmonics then yields
\bea
\lefteqn{(2r+1)(2r'+1)\int d\Om_{\hat{\bf k}}P_r(\mu)P_{r'}(\mu')=}
\nonumber \\ &&
 16\pi^2\de_{rr'}\sum_{s=-r}^rY_{rs}({\bf n})Y^*_{rs}({\bf n}') =
\nonumber \\ &&
 4\pi\de_{rr'}P_r({\bf n}\cd{\bf n}') ~. \label{Yorth}
\eea
In Eq.~(\ref{dT3}) the integration over $d\Om_{\hat{\bf k}}$ then
leads to  terms of the form
$({\bf n}\cd{\bf n}')P_r({\bf n}\cd{\bf n}')$ and 
$({\bf n}\cd{\bf n}')^2P_r({\bf n}\cd{\bf n}')$. To reduce them, we use
\be
xP_r(x)={r+1\over 2r+1}P_{r+1} +{r\over 2r+1}P_{r-1}~.
\ee
%\bea
%\lefteqn{x^2 P_r(x)=}
%\nonumber \\ &&
%{(r+1)(r+2)\over (2r+1)(2r+3)}P_{r+2} + {2r^2+2r-1 \over 
%(2r-1)(2r+3)}P_r +{r(r-1)\over (2r-1)(2r+1)}P_{r-2}~.
%\eea
Applying this and its iteration for $x^2 P_r(x)$, we obtain

\bea
\lefteqn{\langle{\Delta T \over T}( {\bf n}){\Delta T \over T}^* ({\bf
	n}') \rangle=}
 \nonumber \\ && {1\over 2\pi^2} 
 \sum_{r}(2r+1)\int k^2dk \int d\eta d\eta' \dot H(k,\eta)
	\dot H^*(k,\eta')\Big\{
 \nonumber \\ &&  \left[ 
{2(r+1)(r+2)\over (2r+1)(2r+3)} P_{r+2}+{1 \over (2r-1)(2r+3)}P_{r}
+ {2r(r-1)\over (2r-1)(2r+1)}P_{r-2} \right]\times
\nonumber \\ &&
       j_r(k(\eta_0-\eta))j_{r}(k(\eta_0-\eta')) - 
	P_r [j_r(k(\eta_0-\eta)j_r''(k(\eta_0-\eta'))
\nonumber \\ &&  + j_r(k(\eta_0-\eta'))j_r''(k(\eta_0-\eta))
                 -j_r''(k(\eta_0-\eta))j_{r'}''(k(\eta_0-\eta'))]
 \nonumber \\ && 
-4 \left[{r+1 \over 2r+1}P_{r+1} + {r \over 2r+1}P_{r-1}\right]
	j_r'(k(\eta_0-\eta))j_{r}'(k(\eta_0-\eta'))
 \Big\}~, \label{dT4}
\eea
where the argument of the Legendre polynomials, $\bf n\cd n'$, has
been suppressed.
Using the relations 
\be
j_r'=-{r+1\over 2r+1}j_{r+1} +{r\over 2r+1}j_{r-1}
\ee
%\bea
%\lefteqn{j_{r}'' =}
%\nonumber \\ &&
%{(r+1)(r+2)\over (2r+1)(2r+3)}j_{r+2} - {2r^2+2r-1 \over 
%(2r-1)(2r+3)}j_r +{r(r-1)\over (2r-1)(2r+1)}j_{r-2}~.
%\eea
for Bessel functions,
and its iteration for $j''$, we can rewrite Eq.~(\ref{dT4}) in terms of
the Bessel functions $j_{r-2}$ to $j_{r+2}$.

We now insert  the definition of $C_\ell$:
\be
\left\langle{{\Delta T} \over T}({\bf n}) \cdot
{{\Delta T} \over T}({\bf n}')\right\rangle_{({\bf n} \cdot {\bf n}')
	=\cos\theta}=
{1 \over {4\pi}}\Sigma_\ell(2\ell+1)C_{\ell}P_{\ell}(\cos\theta) ~,
  \label{correlAp}
\ee
and compare the coefficients in Eqs. (\ref{dT4}) and (\ref{correlAp}).
We obtain  the somewhat lengthy expression
\bea \lefteqn{C_\ell =}
\nonumber \\ &&
{2\over \pi}\int dk k^2 
\int d\eta d\eta' \dot H(k,\eta)\dot H^*(k,\eta')\Big\{
   j_l(k(\eta_0-\eta))j_l(k(\eta_0-\eta')) \times  
\nonumber \\ &&
      \left (
        {1 \over (2 \ell-1)(2 \ell +3)} + {2 (2 \ell^2+2 \ell -1) \over 
        (2 \ell-1)(2 \ell +3)} + 
        {(2 \ell^2+2 \ell -1)^2 \over (2 \ell-1)^2(2 \ell +3)^2} \right. 
\nonumber \\ && \left.
       -{4 \ell^3 \over (2 \ell-1)^2(2 \ell +1)} -
       {4 (\ell+1)^3 \over (2 \ell+1)(2 \ell +3)^2}  \right ) 
\nonumber \\ && 
-\left[j_{\ell}(k(\eta_0-\eta))j_{\ell+2}(k(\eta_0-\eta'))
+j_{\ell+2}(k(\eta_0-\eta))j_{\ell}(k(\eta_0-\eta'))\right] \times 
\nonumber \\ && {1\over 2l+1}
     \left ( {2(\ell+2)(\ell+1)(2\ell^2+2\ell-1) 
              \over (2\ell-1)(2\ell+3)^2}
            +{2(\ell+1)(\ell+2) \over (2\ell+3)}
     -{8(\ell+1)^2(\ell+2) \over (2\ell+3)^2} \right )
\nonumber \\ && 
-\left[j_{\ell}(k(\eta_0-\eta))j_{\ell-2}(k(\eta_0-\eta'))
+j_{\ell-2}(k(\eta_0-\eta))j_{\ell}(k(\eta_0-\eta'))\right] \times 
\nonumber \\ &&{1\over 2l+1}
\left ( {2 \ell (\ell-1)(2\ell^2+2\ell-1) \over (2\ell-1)^2(2\ell+3)}
+{2 \ell (\ell-1) \over (2\ell-1)(2}
 -{8 \ell^2 (\ell-1) \over (2\ell-1)^2} \right )
\nonumber \\ &&
+j_{\ell+2}(k(\eta_0-\eta))j_{\ell+2}(k(\eta_0-\eta')) \times
\nonumber \\ && 
\left ( {2(\ell+2)(\ell+1) \over (2\ell+1)(2\ell+3)}-
{4(\ell+1)(\ell+2)^2 \over (2\ell+1)(2\ell+3)^2} + 
{(\ell+1)^2(\ell+2)^2 \over (2\ell+1)^2(2\ell+3)^2} \right )
\nonumber \\ && 
+j_{\ell-2}(k(\eta_0-\eta))j_{\ell-2}(k(\eta_0-\eta')) \times
\nonumber \\ &&
\left. \left ( {2 \ell (\ell-1) \over (2\ell-1)(2\ell+1)}-
{4 \ell (\ell-1)^2 \over (2\ell-1)^2 (2\ell+1)} + 
{\ell^2(\ell-1)^2 \over (2\ell-1)^2(2\ell+1)^2} \right )
\right \}~\label{dT5}
\eea

An analysis of the coefficient of each term reveals that the curly
bracket in this expression is just 
\[ \{\cdots \} = \ell(\ell-1)(\ell+1)(\ell+2)
	\l({j_{\ell}(k(\eta_0-\eta))\over
(k(\eta_0-\eta))^2}\r)^2 \]
 and the result is equivalent to 
\be
C_\ell={2\over \pi}\int dkk^2|I(\ell,k)|^2
\ell(\ell-1)(\ell+1)(\ell+2) ~,
\label{Cl}\ee
with
\be
I(\ell,k)=\int_{\eta_{dec}}^{\eta_0}d\eta\dot{H}(\eta,k)
           {j_\ell((k(\eta_0-\eta))\over(k(\eta_0-\eta))^2} ~.
\label{II} \ee

\chapter{Boltzmann equation and polarization}
\label{AppBoltz}
In this appendix we derive the Boltzmann equation taking into account 
polarization, and we write it as a hierarchy of equations using an orthonormal
expansion in the space of photon directions. Up to the collision term, the 
Eqs. (\ref{3B}), (\ref{3BV}) and (\ref{3BT}) are still valid. 
We first re-derive the collision term taking into account the polarization 
dependence of Thomson scattering.

Just before the process of 
recombination during which the fluid description of radiation breaks down,
the temperature
is $\sim 0.4 \mr{~eV}$. The electrons and nuclei are non-relativistic
and the dominant collision process is non-relativistic Thomson
scattering.

Thomson scattering depends on the polarization
of the incoming radiation field. We describe the polarization state of the 
radiation field by the 
{\em Stokes parameters}~\cite{jackson,koso,melvit,chandra}:

For a harmonic electro-magnetic wave with electric field
\be
{\bf E} (\bx,t)=\l({\bm\ep}_1 E_1+{\bm\ep}_2 E_2\r)
e^{ip \bm{\scriptsize n\cd x} - i\om t} ~,
\ee
where $\bn$, $\bm{ \ep}_1$ and $\bm{\ep}_2$ form an orthonormal
basis and the complex field amplitudes are parameterized
as $E_j=a_j e^{i\de_j}$, the Stokes parameters are given by
\bea
I &=& a_1^2 + a_2^2\\
Q &=& a_1^2 - a_2^2\\
U &=& 2 a_1 a_2 \cos(\de_2-\de_1)\\
V &=& 2 a_1 a_2 \sin(\de_2-\de_1) .
\eea
$I$ is the intensity of the wave (whose perturbation
$\MM$ has interested us so far), while $Q$ is a measure of
the strength of linear polarization (the ratio of the principal
axis of the polarization ellipse). $U$ and $V$ give phase
information (the orientation of the ellipse). For non-relativistic Thomson 
scattering $V$ is completely decoupled
and (since it vanishes at early times) is therefore never generated.

As $Q$ and $U$ vanish in the background, perturbations cannot couple
to them (since such terms are 2nd order), and the Liouville equations
 for $Q$ and $U$ become (neglecting scattering
and spatial curvature)
\be
\dd_\eta{(Q;U)}+i n^\ell k_\ell{(Q;U)}  =0 . \label{scalstokes}
\ee

The differential cross section of Thomson scattering for a photon
with incident polarization ${\bm \ep}_{(i)}$ scattering into the
 outgoing polarization ${\bm \ep}_{(s)}\equiv {\bm \ep}'$ is~\cite{jackson}
\be
\frac{d\si}{d\Om} = \frac{3}{8\pi}\si_T 
\l|{\bm\ep}_{(s)}^* {\bm\ep}_{(i)}\r|^2 .
\ee

\begin{figure}[ht]
\centerline{\epsfig{file=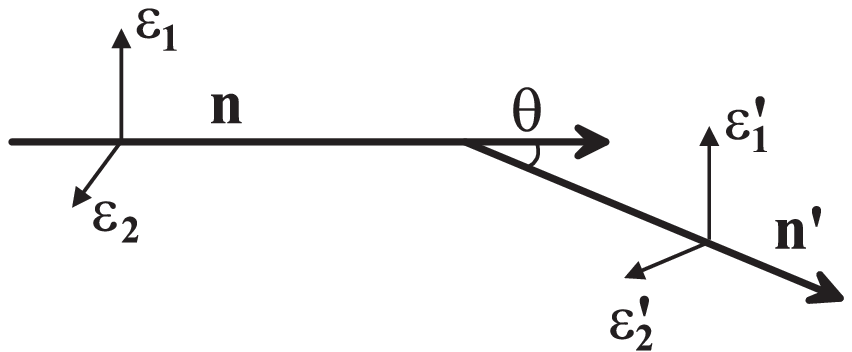,width=6cm}}
\caption{\label{scat1}
Definition of the angles and vectors for Thomson scattering
in the  $(\bn,{\bm \ep}_2)$ plane.}
\end{figure}
It is often convenient to introduce the two `partial' intensities
$I_1\equiv a_1^2 = (I+Q)/2$ and $I_2 \equiv a_2^2 = (I-Q)/2$. A
wave scattered in the $(\bn,{\bm \ep}_2)$ plane (see figure 
\ref{scat1}) by an angle $\th$ has the intensities
\bea
I_1^{(s)} &=& \frac{3\si_T}{8\pi} I_1^{(i)} \nonumber \\
I_2^{(s)} &=& \frac{3\si_T}{8\pi} I_2^{(i)} \cos^2\th \label{thomscat1a} ,
\eea
or, expressed in terms of the Stokes parameters,
\be
\l( \begin{array}{c} I^{(s)} \\ Q^{(s)} \end{array} \r) = \frac{3\si_T}{16\pi}
\l( \begin{array}{cc} 1+\cos^2\th & \sin^2\th \\
	\sin^2\th & 1+\cos^2\th \end{array} \r)
\l( \begin{array}{c} I^{(i)} \\ Q^{(i)} \end{array} \r) . \label{thomscat1b}
\ee

A rotation in the $({\bm\ep}_1,{\bm\ep}_2)$ plane doesn't change
the intensity of the wave, but it changes $Q$ and $U$ to
\bea
Q' &=& Q \cos(2\phi) + U \sin(2\phi) \label{thomrota} \\
U' &=& -U \sin(2\phi) + Q \cos(2\phi)\label{thomrotb} ~.
\eea

\begin{figure}[ht]
\centerline{\epsfig{file=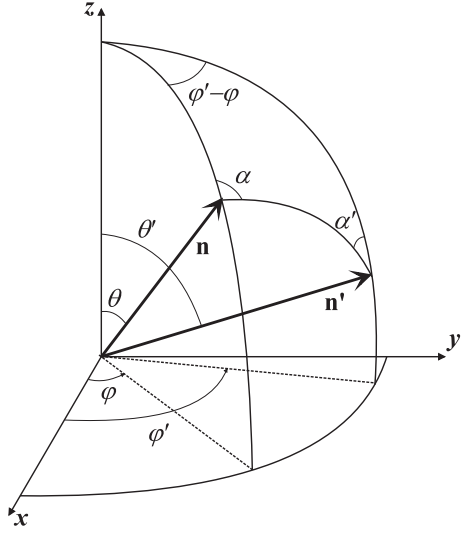,width=6cm}}
\caption{\label{scat2}
Definition of the angles and vectors for Thomson scattering
in the general case. The polarization vectors are oriented
like in figure \ref{scat1}.}
\end{figure}
To determine the cross section that a given 'initial' wave \\
$(I^{(i)},Q^{(i)},U^{(i)})$
propagating in direction $\bn$ be scattered into a wave $(I^{(s)},Q^{(s)},U^{(s)})$
with direction $\bn'$, we need to go through the following steps
(we will use the plane  $({\bm y}, {\bm z})$ as reference
plane, see figure (\ref{scat2}) for definitions of the angles
and vectors):
\begin{enumerate}
\item Rotate around $\bn$ such that the plane $(\bn,\bn')$ turns into 
the plane $(\bn {\bm z})$. One needs to apply
the rotation (\ref{thomrota},\ref{thomrotb}) for $\phi=\al$ to the
Stokes parameters.
\item Rotate the new plane $(\bn,\bn')$ around  ${\bm z}$
into the reference plane  $({\bm y}, {\bm z})$. This operation does not 
influence the incoming Stokes parameters..
\item Now we are in the known case of (\ref{thomscat1a}) 
and (\ref{thomscat1b}). Hence we can apply the
scattering matrix.
\item We then rotate the scattering plane back around ${\bm z}$
into the old $({\bm z},\bn')$ plane. This does  not
change the scattered Stokes parameters.
\item Finally we rotate around $\bn'$ by the angle $\al'$
to reach the original state. To do this, we have to apply
the rotation matrix (\ref{thomrota},\ref{thomrotb}) again, but now for
$\phi=\al'$.
\end{enumerate}

Following the steps outlined above, we recover the scattering matrix in the 
basis $(I_1,I_2,U)$ given in equations (\ref{chandra_eq}) - (\ref{ch_eq_2}) 
(see also~\cite{chandra}). $V$ is completely decoupled from the other
parameters and follows an evolution which is independent of the
rest. Hence by starting with $V(t \ll t_{dec}) = 0$ it will stay
zero and can be neglected. The scattering matrix $P$, which determines
the (non vanishing) scattered Stokes parameters from the initial ones,
\be
 \left( \begin{array}{c} I_1^{(s)} \\ I_2^{(s)} \\ U^{(s)} \end{array} \right)
 = {\si_T\over 4\pi} P
 \left( \begin{array}{c} I_1^{(i)} \\ I_2^{(i)} \\ U^{(i)} \end{array} \right)
\ee
is then given by
\be
  P = \l[P^{(0)} + \sqrt{1-\mu^2}\sqrt{1-\mu^{\prime2}} P^{(1)}
  +P^{(2)}\r] , \label{chandra_eq}
\ee
where
\be
P^{(0)} = \frac{3}{4} \l( \begin{array}{ccc}
 2 (1-\mu^2) (1-\mu^{\prime2})+\mu^2\mu^{\prime2} & \mu^2 & 0  \\
 \mu^{\prime2}                                    & 1     & 0  \\
 0                                                & 0     & 0
\end{array} \r) ,
\ee
\be
P^{(1)}=\frac{3}{4} \l( \begin{array}{ccc}
 4 \mu \mu' \cos(\phi'-\phi) & 0 & 2\mu\sin(\phi'-\phi)  \\
 0                           & 0 & 0                    \\
 -4\mu'\sin(\phi'-\phi)      & 0 & 2\cos(\phi'-\phi)    
\end{array} \r) ,
\ee
\be
P^{(2)} = \frac{3}{4} {\scriptsize\l( \begin{array}{ccc}
 \mu^2\mu^{\prime2}\cos[2(\phi'-\phi)] & -\mu^2\cos[2(\phi'-\phi)] & \mu^2\mu'\sin[2(\phi'-\phi)]  \\
 -\mu^{\prime2}\cos[2(\phi'-\phi)]     & \cos[2(\phi'-\phi)]       & -\mu'\sin[2(\phi'-\phi)]    \\
 -2\mu\mu^{\prime2}\sin[2(\phi'-\phi)] & 2\mu\sin[2(\phi'-\phi)]   & 2\mu\mu'\cos[2(\phi'-\phi)]
\end{array} \r)}~ .   \label{ch_eq_2}
\ee

As we are in an isotropic situation, we will perform all the
calculations in a special coordinate system with $\bk \parallel \hat{z}$
and $\bn, \bn'$ as in Fig.~\ref{scat2}.
Clearly the results are independent of this coordinate choice.

The matrix $R$ connecting $(I_1,I_2,U)$ with $(I,Q,U)$ is
given by
\be \label{matrixR}
\l( \begin{array}{c}
	I_1 \\
	I_2 \\
	U \end{array} \r)
= \l( \begin{array}{c}
	1/2 (I+Q) \\
	1/2 (I-Q) \\
	U \end{array} \r)
= \frac{1}{2} \l( \begin{array}{rrr}
	1 &  1 & 0 \\
       	1 & -1 & 0 \\
	0 &  0 & 2 \end{array} \r) 
\l( \begin{array}{c}
	I \\
	Q \\
	U \end{array} \r)
\equiv R \l( \begin{array}{c}
	I \\
	Q \\
	U \end{array} \r).
\ee
	
To calculate the collision term including polarization , we 
change into the $(I_1,I_2)$ basis .
For each of the two intensities $\la\in\{1,2\}$ we then have
the collision term given by
\be
C[f^{(\la)}]=\frac{df_+^{(\la)}}{d\eta}-\frac{df_-^{(\la)}}{d\eta} ,
\ee
where $f_+^{(\la)}$ and $f_-^{(\la)}$ denote the distribution of photons 
in the polarization state $\la$ scattered into
respectively out of the beam due to Compton scattering.

In the matter  (baryon/electron) rest frame,
which we indicate by  a prime, we know that
\[ {df_+^{(\la)\prime}\over dt'}(p,\bn)= {\si_Tn_e\over 4\pi}\int
      f^{(\de)\prime}(p',\bn')P^\la_\de({\bn,\bn}')d\Om'  \; , \]
where $n_e$ denotes the electron number density
and $P^\la_\de$ is the $2\times2$ upper left corner of the normalized Thomson 
scattering matrix (\ref{chandra_eq}).
In the baryon rest frame which moves with four velocity $u$, the photon
energy is given by
\[ p' = p_\mu u^\mu \; . \]
We denote the photon energy with respect to a tetrad adapted to
the slicing of space-time into $\{ \eta=$ constant$\}$  hyper--surfaces by $p$  :
\[ p =  p_\mu n^\mu \; ,~~~\mbox{ with }~~
  n = a^{-1}[(1-A)\dd_\eta +B^i\dd_i] ~,~ \]
The lapse function and  the shift vector of the slicing are given by
  $\al= a(1+A)$ and  $\bm{\beta}=
	-B^{i}\dd_i$ .  In first order,
\[ p_0 = ap(1+A) - ap n_iB^i~~,\]
and to zeroth order $ p_i = ap n_i$.
Furthermore, the baryon four velocity has the form
$ u^0 = a^{-1}(1-A)~~,~~~ u^i = u^0v^i $ up to first order.
This yields
\[ p' = p_\mu u^\mu = p(1+ n_i(v^i-B^i)) \; . \]
Using this identity and performing the integration over photon energies,
we obtain
\bean 
\rho_\ga{d\io_+^{(\la)}( \bn)\over d\eta'} &=& a\rho_\ga\si_Tn_e\l[1+
4 n_i(v^i- B^i) + \r.  \\  && \l.
 {1\over 4\pi}\int\io^{(\de)}( \bn')P^\la_\de( \bn, \bn')d\Om'\r] ~ .
\eean

Photons which are scattered leave the beam, with the probability
given by the Thomson cross section
 (see \eg~\cite{LP})
\[ 
{df^{(\la)}_-\over dt'} = \si_Tn_ef^{(\la)\prime}(p',\bn) ~, 
\]
so that we finally have
\bea
C^{(\la)\prime} &=& \frac{4\pi}{\rho_\ga a^4} 
	\int dp\l({df^{(\la)}_+\over dt'} -{df^{(\la)}_-\over dt'}\r)p^3 
	=\frac{1}{2}\si_Tn_e\big[4 n_i(v^i-B^i)  -\io^{(\la)}
 \nonumber \\  
&& ~~~ 
 + {1\over 4\pi} \int
  \io^{(\de)}( \bn') P^\la_\de({\bn,\bn}') d\Om'\big] ~ .
\eea

By setting $C^{(\MM)} = C^{(1)}+C^{(2)}$ and $C^{(Q)} = C^{(1)}-C^{(2)}$
we transform the collision integral back to the normal stokes
parameters. The term for $U$ has the same form as the one for
$Q$, just with the corresponding matrix elements in the integral.
The Boltzmann equation then finally becomes (setting 
$\EE\equiv (\MM,Q,U)$ and for the flat case, $\ka = 0$):

\bea
%\dot{\MM} + i \mu k \MM - \Ga^{(S)i}_{jk} n^j n^k \frac{\dd \MM}{\dd n^i}
%&=& 4 i \mu k (\Phi-\Psi+n^m \Si^{(V)}_m) + 4 n^\ell n^m \dot{H}_{m \ell} \nonumber \\
% + a \si_T n_e &&\l[-\MM -4 i \mu V_b+4 n^\ell \om_{b,\ell} +\int d\Om' P^\al_1 \EE'_\al \r] \\
%\dot{Q} + i \mu k Q - \Ga^{(S)i}_{jk} n^j n^k \frac{\dd Q}{\dd n^i}
%&=& a \si_T n_e \l[-Q + \int d\Om' P^\al_2 \EE'_\al \r] \\
%\dot{U} + i \mu k U - \Ga^{(S)i}_{jk} n^j n^k \frac{\dd U}{\dd n^i}
%&=& a \si_T n_e \l[-U + \int d\Om' P^\al_3 \EE'_\al \r] \\
\lefteqn{\dot{\MM} + i \mu k \MM = 4 i \mu k (\Phi-\Psi+n^m \Si^{(V)}_m) 
  + 4 n^\ell n^m \dot{H}_{m \ell}} \nonumber \\
 &&+ a \si_T n_e \l[-\MM -4 i \mu V_b+4 n^\ell \om_{b,\ell} +
	\int d\Om' P^\al_1 \EE'_\al \r] \label{BMAp} \\
\lefteqn{\dot{Q} + i \mu k Q 
= a \si_T n_e \l[-Q + \int d\Om' P^\al_2 \EE'_\al \r]} \label{BVAp} \\
\lefteqn{\dot{U} + i \mu k U 
= a \si_T n_e \l[-U + \int d\Om' P^\al_3 \EE'_\al \r]}, \label{BUAp}
\eea
where we have to use the scattering matrix transformed into the 
$(\MM,Q,U)$ basis,
\bea
 P  &=& P_S +P_V +P_T \quad \mbox{ with } \\ 
P_S &=& R^{-1}P^{(0)} R \\
  &=& \frac{3}{8}\scriptsize{ \l( \begin{array}{ccc}
	   3-\mu^2-\mu^{\prime 2} + 3 \mu^2 \mu^{\prime2} & 
	(1-3\mu^2)(1-\mu^{\prime2})  & 0\\
	 (1-\mu^2)(1-3\mu^{\prime2})&3 (1-\mu^2)(1-\mu^{\prime2}) & 0 \\
	0 & 0 & 0 
	\end{array} \r)} \\
P_V &=& \sqrt{1-\mu^2} \sqrt{1-\mu^{\prime2}} R^{-1} P^{(1)} R \\
  &=& \frac{3}{2} \sqrt{1-\mu^2} \sqrt{1-\mu^{\prime2}}
	\l( \begin{array}{ccc}
	\mu \mu' C & \mu \mu' C & - \mu S \\
	\mu \mu' C & \mu \mu' C & - \mu S \\
	\mu' S    & \mu' S    & C \end{array} \r)  \\
P_T &=& R^{-1} P^{(2)} R \\
  &=& \frac{3}{8}\scriptsize{ \l( \begin{array}{ccc}
	(1-\mu^2)(1-\mu^{\pr2})C_T   & -(1-\mu^2)(1+\mu^{\pr2})C_T 
	& 2 (1-\mu^2)\mu' S_T  \\
	- (1+\mu^2)(1-\mu^{\pr2})C_T & (1+\mu^2)(1+\mu^{\pr2})C_T  
	& -2(1+\mu^2)\mu' S_T \\
	-2\mu(1-\mu^{\pr2})S_T & 2\mu(1+\mu^{\pr2})S_T  & 4\mu\mu' C_T
	\end{array} \r)}
\eea
with $C = \cos(\phi-\phi')$, $S = \sin(\phi-\phi')$ and   \\
 $C_T = \cos(2(\phi-\phi'))$, $S_T = \sin(2(\phi-\phi'))$.
The parts $P_S,~P_V,~P_T$ of $P$ describe the scattering of the scalar,
vector and tensor contribution to $\EE$ respectively.

The functions $\MM$, $Q$ and $U$ depend on the wave vector $\bk$, the
photon direction $\bn$ and conformal time $\eta$.
We choose for each $\bk$-mode a reference system with 
$z$-axis parallel to $\bk$. 
For scalar perturbations we achieve in this way 
azimuthal symmetry --- the right-hand side of the Boltzmann equation and
therefore also the brightness $\MM^{(S)}$ depend only
on $\mu =  (\hat{\bk}\cd \bn)$ and can be expanded
in Legendre polynomials. The right-hand side of
the Boltzmann equation also determines the azimuthal
dependence of vector and tensor perturbations.
One can continue with this approach, but the resulting equations for $Q$ 
and $U$ and especially also their power spectra depend explicitly on the 
coordinate system. Therefore, we prefer an approach which is inherently 
covariant under rotation.

\section{Electric and magnetic polarization}
Since the Stokes parameters explicitly depend on the 
coordinate system, and Eqs.~(\ref{BVAp}) and (\ref{BUAp}) transform rather 
complicated under rotations of the coordinate system.
A  method to characterize CMB polarization due to non-relativistic 
Thomson scattering which is more convenient than 
the Stokes parameters since its transformation properties are very simple
has been developed a couple years ago~\cite{Kam1,Kam2,HSZ}. A detailed
derivation of this method goes beyond the scope of this report. Here 
we just repeat the definitions and the main results. We set
\be
	\bm{\cal T} = (\MM, Q+ iU,Q-iU) \label{TTAp}
\ee

In terms of this vector the collision integral above can 
we written (in vector form) as
\bea
 \bm{C}[\bm{\cal T}] &= & a\si_Tn_e\big[-\bm{\cal T} +
	\l({1\over 4\pi}\int {\rm d}\Om' \MM' 
 +(\bn\cd{\bf v}_b),0,0\r)  \nonumber \\
 &&  +{1\over 10}\sum_{m=-2}^2\int {\rm d}\Om' 
	P^{(m)}(\bn,\bn')\bm{\cal T}' \big]\label{colA} 
\eea
From Eqs.~(\ref{chandra_eq}) to (\ref{matrixR}) one can determine the 
scattering matrix for the vector $\bm{\cal T}$
\be
  P^{(m)} = \left( \begin{array}{ccc}
  Y_2^{m\prime}Y_2^m & -\sqrt{{3\over 2}} {_2Y_2^{m\prime}}Y_2^m & 
	-\sqrt{{3\over 2}}\,{_{-2}Y_2^{m\prime}}Y_2^m \\
  -\sqrt{6}Y_2^{m\prime}{_2Y_2^m} & 3\, {_2Y_2^{m\prime}}{_2Y_2^m} & 
	3\, {_{-2}Y_2^{m\prime}}{_2Y_2^m} \\
 -\sqrt{6} Y_2^{m\prime}{_{-2}Y_2^m} & 3\, {_2Y_2^{m\prime}}{_{-2}Y_2^m} & 
	3\, {_{-2}Y_2^{m\prime}}{_{-2}Y_2^m} 
\end{array}\right) \label{TscatA}
\ee
where $_sY_l^{m\prime} =~ _sY_l^{m*}(n')$ and $_sY_l^{m}$ are the spin-weighted
spherical harmonics~\cite{NP,Kam2}. 

We now decompose the Fourier components of the 
temperature anisotropy $\MM$ and the polarization variables $E$ and $B$ as

\bea
 \MM &=& \sum_{\ell}\sum_{m=-2}^2\MM_{\ell}^{(m)}
	{ _0 G^m_{\ell}},\\
 Q \pm iU  &=& \sum_{\ell}\sum_{m=-2}^2
	(E_{\ell}^{(m)} \pm iB_{\ell}^{(m)}){ _2G^m_{\ell}}(\bn).
\eea
Here $m=0$ is the scalar mode, $m=\pm1$ are the vector and $m=\pm2$ are
the tensor modes. The functions $_sG^m_{\ell}$ are closely related to the spin 
weighted harmonics $_sY^m_{\ell}$:
\[
 _sG^m_{\ell}(\bn) = (-i)^{\ell}\sqrt{{4\pi\over 2\ell +1}} {_sY^m_{\ell}}(\bn)
\]

The evolution equations in term of these variables can be given in the 
following compact form~\cite{HSZ}
\bea
\lefteqn{ \dot\MM_{\ell}^{(m)} -k\l[{_0\ka^m_{\ell}\over 2\ell-1}
	\MM_{\ell-1}^{(m)}
-{_0\ka^m_{\ell+1}\over 2\ell+3}\MM_{\ell+1}^{(m)}\r] = }\nonumber \\
  &&	-n_e\si_Ta\MM_{\ell}^{(m)}  +S_{\ell}^{(m)} ~~~ (\ell\ge m)\\
\lefteqn{ \dot E_{\ell}^{(m)} -k\l[{_2\ka^m_{\ell}\over 2\ell-1}
E_{\ell-1}^{(m)}  -{2m\over \ell(\ell+1)}B_{\ell}^{(m)}
-{_2\ka^m_{\ell+1}\over 2\ell+3}E_{\ell+1}^{(m)}\r] = } \nonumber \\ && 
  -n_e\si_Ta[E_{\ell}^{(m)} + \sqrt{6}C^{(m)}\de_{\ell,2} \\
\lefteqn{ \dot B_{\ell}^{(m)} - k\l[{_2\ka^m_{\ell}\over 2\ell-1}
  B_{\ell-1}^{(m)}   +{2m\over \ell(\ell+1)}E_{\ell}^{(m)}
-{_2\ka^m_{\ell+1}\over 2\ell+3}B_{\ell+1}^{(m)}\r] = } \nonumber \\ &&  
  -n_e\si_TaB_{\ell}^{(m)}~.
\eea
where we have set
\be \begin{array}{cc}
 S_0^{(0)} =n_e\si_Ta\MM^{(0)}_0, & S^{(0)}_1 =n_e\si_Ta4V_b +4k(\Psi-\Phi), \\
 S^{(0)}_2 =n_e\si_TaC^{(0)}, & S^{(1)}_1 =n_e\si_Ta4\om_b,  \\
  S^{(1)}_2 =n_e\si_TaC^{(1)} +4k\Si, &   S^{(2)}_2 =n_e\si_TaC^{(2)} +4\dot H 
\end{array}
\ee
and $C^{(m)}= {1\over 10}[\MM^{(m)}_2 -\sqrt{6}E^{(m)}_2] $.
The coupling coefficients are
\[ 
	_s\ka^m_{\ell} =\sqrt{{(\ell^2-m^2)(\ell^2-s^2)\over \ell^2}}.
\]
Note that for scalar perturbations, $m=0$, $B$-polarization is not sourced 
and we have $B_{\ell}^{(0)}\equiv 0$.

Finally we want to connect the intensities $\MM^{(m)}_{\ell}$ with the 
more familiar expansion of the 
scalar $(S)$, vector $(V)$ and tensor $(T)$ contributions to the 
brightness function in terms of Legendre polynomials. Usually one sets
\[ \MM = \MM^{(S)} +  \MM^{(V)} +  \MM^{(T)} ~.\]
Here $\MM^{(S)}$ only depends on $\mu=(\bn\cdot\bk)/k$ and the 
$\bn$-dependence of $\MM^{(V)}$ and $\MM^{(T)}$ can be written as  
\bea 
\MM^{(V)}(\mu,\phi) &=& \sqrt{1\!-\!\mu^2}\l[ \MM_1^{(V)}(\mu) \cos\phi 
	\!+\! \MM_2^{(V)}(\mu) \sin\phi\r] \label{v1}\\
\MM^{(T)}(\mu,\phi) &=& (1-\mu^2) \l[ \MM_+^{(T)} \cos(2\phi) + 
	\MM_\times^{(T)} \sin(2\phi)\r], \label{t1}
\eea
where $\phi$ is the azimuthal angle in the plane normal to $\bk$.
 Each of the functions $\MM_{\bullet}^{(S,V,T)}(\mu)$ is now expanded 
in Legendre polynomials
\be
\MM_{\bullet}^{(S,V,T)} =\sum_{\ell}(-i)^{\ell}(2\ell+1)
	\si_{\bullet,\ell}^{(S,V,T)}P_{\ell}(\mu)~.  \label{devel}
\ee
The coefficients $\si_{\bullet,\ell}^{(S,V,T)}$ are then related to 
$\MM_{\ell}^{(m)}$ via  the identities
\bea
  \MM_{\ell}^{(0)} &=&  (2\ell+1) \si_{\ell}^{(S)}  \label{Asi0} \\
  \MM_{\ell}^{(\pm 1)} &=& {i\over 2}\sqrt{\ell(\ell+1)} 
  [\si_{1,\ell+1}^{(V)}\mp i\si_{2,\ell+1}^{(V)} +
	\si_{1,\ell-1}^{(V)}\mp i\si_{2,\ell-1}^{(V)}] \\
\MM_{\ell}^{(\pm 2)} &=& -\sqrt{(\ell+2)!\over (\ell-2)!} \big[ 
{1\over 2\ell+3} \si_{\uparrow\downarrow,\ell+2}^{(T)} +
{2(\ell+1)\over (2\ell-1)(2\ell+3)}\si_{\uparrow\downarrow,\ell}^{(T)}
 \nonumber \\ && \quad  +
{1\over 2\ell-1} \si_{\uparrow\downarrow,\ell-2}^{(T)}]~, \label{Asi2}
\eea
where 
\[  \si_{\uparrow\downarrow,\ell}= {1\over 2}[\si_{+\ell} \mp 
		i\si_{\times\ell}] ~.\]

We do not repeat this correspondence for the Stokes parameters $Q$ and $U$ 
since it is rather complicated and not very useful as it depends on the 
coordinate system chosen. 

\section{Power spectra}
In the previous appendix and in Chapter~4 we have derived the expression
for the CMB anisotropy power spectrum for scalar and tensor perturbations.
Here we give the general expression for scalar, vector and tensor fluctuations,
polarizations and cross correlations. To make contact with the results 
derived before, one has to use Eqs.~(\ref{devel},\ref{Asi0}) and 
(\ref{Asi2}) and neglect the collision term in the Boltzmann equation.

We expand the present CMB anisotropies and polarization in spherical
harmonics: $\Delta T (\bn,\eta_0)/T_0 = \sum_{\ell m} a_{\ell m} Y_{\ell}^m
 (\bn)$.
The coefficients $a_{\ell m}$ are random variables with zero mean and
rotationally invariant
variances, $C_\ell \equiv \langle \mid a_{\ell m} \mid ^2 \rangle$.
The  correlation function of the anisotropy pattern then has the
standard expression:
\be
 \left\langle {\delta T\over T_0}(\bn_1){\delta T\over
T_0}(\bn_2)\right\rangle = {1\over 4\pi} \sum_\ell (2\ell+1) {C_\ell} P_\ell
(\cos\theta)
\ee
 where $\cos\theta = \bn_1 \cdot \bn_2$ and $\lan\cdots\ran$ denotes 
 ensemble average. We use the Fourier transform normalization
\be
  \hat{f}(\bk) = {1\over V}\int f({\bf x})\exp(i\bk\cd{\bf x})d^3x~,
\ee
with some normalization volume $V$. Using statistical homogeneity 
we have
\bea  \lefteqn{
\left\langle {\delta T\over T_0}(\bn_1)
 {\delta T\over T_0}(\bn_2)\right\rangle  = 
{1\over V}\int d^3x \left\langle {\delta T\over T_0}({\bf x},\bn_1)
	{\delta T\over T_0}({\bf x}, \bn_2)\right\rangle} \nonumber \\
&=& {1\over (2\pi)^3}\int d^3k \left\langle  {\delta T\over T_0}(\bk,\bn_1)
	{\delta T\over T_0}(\bk, \bn_2)\right\rangle~.
\eea
Inserting our ansatz (\ref{devel}) for  ${\delta T\over T_0} ={1\over 4}\MM$,
and using the addition theorem for spherical harmonics, \\
$P_\ell(\bn_1\cdot\bn_2) = \frac{4\pi}{2\ell+1} 
	\sum_m Y_{\ell m}^*(\bn_1) Y_{\ell m}(\bn_2)$,
we find
\bea   \lefteqn{
\left\langle {\delta T\over T_0}(\bn_1){\delta T\over T_0}(\bn_2)\right\rangle
 = {1\over 8\pi}\sum_{\ell,\ell',m,m'}(-1)^{(\ell-\ell')}Y_{\ell m}(n_1)
    Y^*_{\ell' m'}(n_2)} \nonumber\\
~~~~~~  & ~~ &\times\int k^2dkd\Om_{\hat{\bk}}Y_{\ell m}^*(\hat{\bk})
	Y_{\ell' m'}(\hat{\bk})
  \langle\si_\ell\si^*_{\ell'}\rangle(k) \nonumber \\
~~~~~~   &=&{1\over 32\pi^2}\sum_\ell(2\ell+1)P_\ell(\bn_1\cd\bn_2) \int
	k^2dk \langle\si_\ell\si^*_\ell\rangle(k)~, 
\label{ClSa}\eea
from which we conclude

\be
C_\ell^{\MM\MM,(S)} =  {1 \over {8 \pi}}\int
 k^2dk\langle|\si^{(S)}_\ell(t_0,k)|^2\rangle ~,
\label{ClSapp}\ee
where the superscript $^{(S)}$ indicates that Eq.~(\ref{ClSapp}) gives
the contribution from {\em scalar} perturbations and $^{\MM\MM}$ means
that it is the contribution to the intensity perturbation.

The $QQ$, $UU$, $\MM Q$, $\MM U$ and $QU$ correlators depend with the Stokes 
parameters on the particular coordinate system chosen.
It is much more convenient to express the polarization power spectra in 
terms of the variables $E$ and $B$ which are independent of the coordinate 
system. Furthermore, since $B$ is parity odd, its correlators with $\MM$ 
and $E$ vanishes. One finds the simple general expression~\cite{HSZ}
\be
 (2\ell+1)^2C_\ell^{XY(m)} = {n_m\over 8\pi}\int k^2dkX_{\ell}^{(m)}
	Y_{\ell}^{(m)*}~,  \label{PSpec}
\ee
where $n_m=1$ for $m=0$ and $n_m=2$ for $m=1,2$, accounting for the number 
of modes.

\end{document}